\documentclass[%
 aip,
apl,
 amsmath,amssymb,
 reprint,%  
]{revtex4-1}
\usepackage[utf8]{inputenc}
\usepackage{graphicx}% Include figure files
\usepackage{dcolumn}% Align table columns on decimal point
\usepackage{bm}% bold math
\usepackage{epstopdf}        % TCM Converts eps images to PDF to use in graphicx package.  Must be loaded after graphicx package
\usepackage{moresize}        % TCM Adds \HUGE and \ssmall font sizes
\usepackage{multirow}        % TCM Allows multirow cells in tables (similar to merge and center in Excel)
\usepackage{siunitx}		 % TCM package required to ensure units are typeset properly with numbers e.g. \SI{10}{\kg\m\per\square\s}
\usepackage{tabularx}
\usepackage{xspace} 
\usepackage{bookmark}		 		  % TCM Eliminates Warning Bookmark level greater than one.  Must be loaded after Hyperref

\newcommand\Tstrut{\rule{0pt}{2.6ex}}

\newcommand{\FDvalSolo}{$\left(6.08\pm0.17\right)$\,\%\xspace}
\newcommand{\detSeven}{$\left(5.52 \pm 0.10\right)$\,\%\xspace}
\newcommand{\detNine}{$\left(6.67 \pm 0.11\right)$~\%\xspace}

\newcommand{\FDvalFin}{$\left(6.08\pm0.18\right)$\,\%\xspace}
\newcommand{\GeThval}{$\left(19.7^{+0.6}_{-0.5}\right)$\,eV\xspace}

\usepackage{color,soul}
\definecolor{orange}{rgb}{0.75,0.25,0}

\begin{document}
	
\title{Energy Loss Due to Defect Formation from $^{206}$Pb Recoils in SuperCDMS Germanium Detectors}

\author{R.~Agnese} \affiliation{Department of Physics, University of Florida, Gainesville, FL 32611, USA}
\author{T.~Aralis} \affiliation{\mbox{Division of Physics, Mathematics, \& Astronomy, California Institute of Technology, Pasadena, CA 91125, USA}}
\author{T.~Aramaki} \affiliation{SLAC National Accelerator Laboratory/Kavli Institute for Particle Astrophysics and Cosmology, Menlo Park, CA 94025, USA}
\author{I.J.~Arnquist} \affiliation{Pacific Northwest National Laboratory, Richland, WA 99352, USA}
\author{E.~Azadbakht} \affiliation{Department of Physics and Astronomy, and the Mitchell Institute for Fundamental Physics and Astronomy, Texas A\&M University, College Station, TX 77843, USA}
\author{W.~Baker} \affiliation{Department of Physics and Astronomy, and the Mitchell Institute for Fundamental Physics and Astronomy, Texas A\&M University, College Station, TX 77843, USA}
\author{S.~Banik} \affiliation{\mbox{School of Physical Sciences, National Institute of Science Education and Research, HBNI, Jatni - 752050, India}}
\author{D.~Barker} \affiliation{\mbox{School of Physics \& Astronomy, University of Minnesota, Minneapolis, MN 55455, USA}}
\author{D.A.~Bauer} \affiliation{Fermi National Accelerator Laboratory, Batavia, IL 60510, USA}
%\author{D.~Balakishiyeva} \affiliation{Department of Physics, Southern Methodist University, Dallas, TX 75275, USA}
\author{T.~Binder} \affiliation{\mbox{Department of Physics, University of South Dakota, Vermillion, SD 57069, USA}}
%\author{S.~Banik} \affiliation{School of Physical Sciences, National Institute of Science Education and Research, HBNI, Jatni - 752050, India}
%\author{D.~Barker} \affiliation{School of Physics \& Astronomy, University of Minnesota, Minneapolis, MN 55455, USA}
%\author{R.~Basu~Thakur} \affiliation{Fermi National Accelerator Laboratory, Batavia, IL 60510, USA}\affiliation{Department of Physics, University of Illinois at Urbana-Champaign, Urbana, IL 61801, USA}
%\author{D.A.~Bauer} \affiliation{Fermi National Accelerator Laboratory, Batavia, IL 60510, USA}
%\author{T.~Binder} \affiliation{Department of Physics, University of South Dakota, Vermillion, SD 57069, USA}
\author{M.A.~Bowles} \affiliation{\mbox{Department of Physics, South Dakota School of Mines and Technology, Rapid City, SD 57701, USA}}
\author{P.L.~Brink} \affiliation{SLAC National Accelerator Laboratory/Kavli Institute for Particle Astrophysics and Cosmology, Menlo Park, CA 94025, USA}
\author{R.~Bunker} \affiliation{Pacific Northwest National Laboratory, Richland, WA 99352, USA}
\author{B.~Cabrera} \affiliation{Department of Physics, Stanford University, Stanford, CA 94305, USA}
%\author{D.O.~Caldwell} \affiliation{Department of Physics, University of California, Santa Barbara, CA 93106, USA}
\author{R.~Calkins} \affiliation{\mbox{Department of Physics, Southern Methodist University, Dallas, TX 75275, USA}}
\author{C.~Cartaro} \affiliation{SLAC National Accelerator Laboratory/Kavli Institute for Particle Astrophysics and Cosmology, Menlo Park, CA 94025, USA}
\author{D.G.~Cerde\~no} \affiliation{Department of Physics, Durham University, Durham DH1 3LE, UK}\affiliation{\mbox{Instituto de F\'{\i}sica Te\'orica UAM/CSIC, Universidad Aut\'onoma de Madrid, 28049 Madrid, Spain}}
\author{Y.-Y.~Chang} \affiliation{\mbox{Division of Physics, Mathematics, \& Astronomy, California Institute of Technology, Pasadena, CA 91125, USA}}
%\author{Y.~Chen} \affiliation{Department of Physics, Syracuse University, Syracuse, NY 13244, USA}
\author{J.~Cooley} \affiliation{\mbox{Department of Physics, Southern Methodist University, Dallas, TX 75275, USA}}
\author{B.~Cornell} \affiliation{\mbox{Division of Physics, Mathematics, \& Astronomy, California Institute of Technology, Pasadena, CA 91125, USA}}
\author{P.~Cushman} \affiliation{\mbox{School of Physics \& Astronomy, University of Minnesota, Minneapolis, MN 55455, USA}}
%\author{M.~Daal} \affiliation{Department of Physics, University of California, Berkeley, CA 94720, USA}
\author{P.C.F.~Di~Stefano} \affiliation{Department of Physics, Queen's University, Kingston, ON K7L 3N6, Canada}
\author{T.~Doughty} \affiliation{Department of Physics, University of California, Berkeley, CA 94720, USA}
\author{E.~Fascione}\affiliation{Department of Physics, Queen's University, Kingston, ON K7L 3N6, Canada}
\author{E.~Figueroa-Feliciano} \affiliation{\mbox{Department of Physics \& Astronomy, Northwestern University, Evanston, IL 60208-3112, USA}}
\author{M.~Fritts} \affiliation{\mbox{School of Physics \& Astronomy, University of Minnesota, Minneapolis, MN 55455, USA}}
\author{G.~Gerbier} \affiliation{Department of Physics, Queen's University, Kingston, ON K7L 3N6, Canada}
\author{R.~Germond} \affiliation{Department of Physics, Queen's University, Kingston, ON K7L 3N6, Canada}
\author{M.~Ghaith} \affiliation{Department of Physics, Queen's University, Kingston, ON K7L 3N6, Canada}
%\author{G.L.~Godfrey} \affiliation{SLAC National Accelerator Laboratory/Kavli Institute for Particle Astrophysics and Cosmology, Menlo Park, CA 94025, USA}
\author{S.R.~Golwala} \affiliation{\mbox{Division of Physics, Mathematics, \& Astronomy, California Institute of Technology, Pasadena, CA 91125, USA}}
%\author{J.~Hall} \affiliation{Pacific Northwest National Laboratory, Richland, WA 99352, USA}
\author{H.R.~Harris} \affiliation{Department of Physics and Astronomy, and the Mitchell Institute for Fundamental Physics and Astronomy, Texas A\&M University, College Station, TX 77843, USA}
\author{Z.~Hong} \affiliation{\mbox{Department of Physics \& Astronomy, Northwestern University, Evanston, IL 60208-3112, USA}}
\author{E.W.~Hoppe} \affiliation{Pacific Northwest National Laboratory, Richland, WA 99352, USA}
\author{L.~Hsu} \affiliation{Fermi National Accelerator Laboratory, Batavia, IL 60510, USA}
\author{M.E.~Huber} \affiliation{\mbox{Department of Physics, University of Colorado Denver, Denver, CO 80217, USA}}\affiliation{\mbox{Department of Electrical Engineering, University of Colorado Denver, Denver, CO 80217, USA}}
\author{V.~Iyer} \affiliation{\mbox{School of Physical Sciences, National Institute of Science Education and Research, HBNI, Jatni - 752050, India}}
\author{D.~Jardin} \affiliation{\mbox{Department of Physics, Southern Methodist University, Dallas, TX 75275, USA}}
%\author{A.~Jastram} \affiliation{Department of Physics and Astronomy, and the Mitchell Institute for Fundamental Physics and Astronomy, Texas A\&M University, College Station, TX 77843, USA}
\author{C.~Jena} \affiliation{\mbox{School of Physical Sciences, National Institute of Science Education and Research, HBNI, Jatni - 752050, India}}
\author{M.H.~Kelsey} \affiliation{SLAC National Accelerator Laboratory/Kavli Institute for Particle Astrophysics and Cosmology, Menlo Park, CA 94025, USA}
\author{A.~Kennedy} \affiliation{\mbox{School of Physics \& Astronomy, University of Minnesota, Minneapolis, MN 55455, USA}}
\author{A.~Kubik} \affiliation{Department of Physics and Astronomy, and the Mitchell Institute for Fundamental Physics and Astronomy, Texas A\&M University, College Station, TX 77843, USA}
\author{N.A.~Kurinsky} \affiliation{SLAC National Accelerator Laboratory/Kavli Institute for Particle Astrophysics and Cosmology, Menlo Park, CA 94025, USA}
\author{R.E.~Lawrence} \affiliation{Department of Physics and Astronomy, and the Mitchell Institute for Fundamental Physics and Astronomy, Texas A\&M University, College Station, TX 77843, USA}
\author{B.~Loer} \affiliation{Pacific Northwest National Laboratory, Richland, WA 99352, USA}
\author{E.~Lopez~Asamar} \affiliation{Department of Physics, Durham University, Durham DH1 3LE, UK}
\author{P.~Lukens} \affiliation{Fermi National Accelerator Laboratory, Batavia, IL 60510, USA}
\author{D.~MacDonell} \affiliation{\mbox{Department of Physics \& Astronomy, University of British Columbia, Vancouver, BC V6T 1Z1, Canada}}\affiliation{TRIUMF, Vancouver, BC V6T 2A3, Canada}
\author{R.~Mahapatra} \affiliation{Department of Physics and Astronomy, and the Mitchell Institute for Fundamental Physics and Astronomy, Texas A\&M University, College Station, TX 77843, USA}
\author{V.~Mandic} \affiliation{\mbox{School of Physics \& Astronomy, University of Minnesota, Minneapolis, MN 55455, USA}}
\author{N.~Mast} \affiliation{\mbox{School of Physics \& Astronomy, University of Minnesota, Minneapolis, MN 55455, USA}}
\author{E.H.~Miller} \affiliation{\mbox{Department of Physics, South Dakota School of Mines and Technology, Rapid City, SD 57701, USA}}
\author{N.~Mirabolfathi} \affiliation{Department of Physics and Astronomy, and the Mitchell Institute for Fundamental Physics and Astronomy, Texas A\&M University, College Station, TX 77843, USA}
\author{B.~Mohanty} \affiliation{\mbox{School of Physical Sciences, National Institute of Science Education and Research, HBNI, Jatni - 752050, India}}
\author{J.D.~Morales~Mendoza} \affiliation{Department of Physics and Astronomy, and the Mitchell Institute for Fundamental Physics and Astronomy, Texas A\&M University, College Station, TX 77843, USA}
\author{J.~Nelson} \affiliation{\mbox{School of Physics \& Astronomy, University of Minnesota, Minneapolis, MN 55455, USA}}
\author{J.L.~Orrell} \affiliation{Pacific Northwest National Laboratory, Richland, WA 99352, USA}
\author{S.M.~Oser} \affiliation{\mbox{Department of Physics \& Astronomy, University of British Columbia, Vancouver, BC V6T 1Z1, Canada}}\affiliation{TRIUMF, Vancouver, BC V6T 2A3, Canada}
%\author{K.~Page} \affiliation{Department of Physics, Queen's University, Kingston, ON K7L 3N6, Canada}
\author{W.A.~Page} \affiliation{\mbox{Department of Physics \& Astronomy, University of British Columbia, Vancouver, BC V6T 1Z1, Canada}}\affiliation{TRIUMF, Vancouver, BC V6T 2A3, Canada}
\author{R.~Partridge} \affiliation{SLAC National Accelerator Laboratory/Kavli Institute for Particle Astrophysics and Cosmology, Menlo Park, CA 94025, USA}
%\author{M.~Penalver~Martinez} \affiliation{Department of Physics, Durham University, Durham DH1 3LE, UK}
\author{M.~Pepin} \affiliation{\mbox{School of Physics \& Astronomy, University of Minnesota, Minneapolis, MN 55455, USA}}
%\author{A.~Phipps} \affiliation{Department of Physics, University of California, Berkeley, CA 94720, USA}
\author{F.~Ponce} \affiliation{Department of Physics, Stanford University, Stanford, CA 94305, USA}
\author{S.~Poudel} \affiliation{\mbox{Department of Physics, University of South Dakota, Vermillion, SD 57069, USA}}
\author{M.~Pyle} \affiliation{Department of Physics, University of California, Berkeley, CA 94720, USA}
\author{H.~Qiu} \affiliation{\mbox{Department of Physics, Southern Methodist University, Dallas, TX 75275, USA}}
\author{W.~Rau} \affiliation{Department of Physics, Queen's University, Kingston, ON K7L 3N6, Canada}
%\author{P.~Redl} \affiliation{Department of Physics, Stanford University, Stanford, CA 94305, USA}
\author{A.~Reisetter} \affiliation{Department of Physics, University of Evansville, Evansville, IN 47722, USA}
\author{T.~Reynolds} \affiliation{Department of Physics, University of Florida, Gainesville, FL 32611, USA}
\author{A.~Roberts} \affiliation{\mbox{Department of Physics, University of Colorado Denver, Denver, CO 80217, USA}}
\author{A.E.~Robinson} \affiliation{Fermi National Accelerator Laboratory, Batavia, IL 60510, USA}
\author{H.E.~Rogers} \affiliation{\mbox{School of Physics \& Astronomy, University of Minnesota, Minneapolis, MN 55455, USA}}
\author{T.~Saab} \affiliation{Department of Physics, University of Florida, Gainesville, FL 32611, USA}
\author{B.~Sadoulet} \affiliation{Department of Physics, University of California, Berkeley, CA 94720, USA}\affiliation{Lawrence Berkeley National Laboratory, Berkeley, CA 94720, USA}
\author{J.~Sander} \affiliation{\mbox{Department of Physics, University of South Dakota, Vermillion, SD 57069, USA}}
\author{A.~Scarff} \affiliation{\mbox{Department of Physics \& Astronomy, University of British Columbia, Vancouver, BC V6T 1Z1, Canada}}\affiliation{TRIUMF, Vancouver, BC V6T 2A3, Canada}
%\author{K.~Schneck} \affiliation{SLAC National Accelerator Laboratory/Kavli Institute for Particle Astrophysics and Cosmology, Menlo Park, CA 94025, USA}
\author{R.W.~Schnee} \affiliation{\mbox{Department of Physics, South Dakota School of Mines and Technology, Rapid City, SD 57701, USA}}
\author{S.~Scorza} \affiliation{\mbox{SNOLAB, Creighton Mine \#9, 1039 Regional Road 24, Sudbury, ON P3Y 1N2, Canada}}
\author{K.~Senapati} \affiliation{\mbox{School of Physical Sciences, National Institute of Science Education and Research, HBNI, Jatni - 752050, India}}
\author{B.~Serfass} \affiliation{Department of Physics, University of California, Berkeley, CA 94720, USA}
\author{J.~So} \affiliation{\mbox{Department of Physics, South Dakota School of Mines and Technology, Rapid City, SD 57701, USA}}
\author{D.~Speller} \affiliation{Department of Physics, University of California, Berkeley, CA 94720, USA}
\author{M.~Stein} \email{mstein@smu.edu}\affiliation{\mbox{Department of Physics, Southern Methodist University, Dallas, TX 75275, USA}}

\author{J.~Street} \affiliation{\mbox{Department of Physics, South Dakota School of Mines and Technology, Rapid City, SD 57701, USA}}
\author{H.A.~Tanaka} \affiliation{Department of Physics, University of Toronto, Toronto, ON M5S 1A7, Canada}
\author{D.~Toback} \affiliation{Department of Physics and Astronomy, and the Mitchell Institute for Fundamental Physics and Astronomy, Texas A\&M University, College Station, TX 77843, USA}
\author{R.~Underwood} \affiliation{Department of Physics, Queen's University, Kingston, ON K7L 3N6, Canada}
\author{A.N.~Villano} \affiliation{\mbox{School of Physics \& Astronomy, University of Minnesota, Minneapolis, MN 55455, USA}}
\author{B.~von~Krosigk} \affiliation{\mbox{Department of Physics \& Astronomy, University of British Columbia, Vancouver, BC V6T 1Z1, Canada}}\affiliation{TRIUMF, Vancouver, BC V6T 2A3, Canada}
\author{S.L.~Watkins} \affiliation{Department of Physics, University of California, Berkeley, CA 94720, USA}
%\author{B.~Welliver} \affiliation{Department of Physics, University of Florida, Gainesville, FL 32611, USA}
\author{J.S.~Wilson} \affiliation{Department of Physics and Astronomy, and the Mitchell Institute for Fundamental Physics and Astronomy, Texas A\&M University, College Station, TX 77843, USA}
\author{M.J.~Wilson} \affiliation{Department of Physics, University of Toronto, Toronto, ON M5S 1A7, Canada}
\author{J.~Winchell} \affiliation{Department of Physics and Astronomy, and the Mitchell Institute for Fundamental Physics and Astronomy, Texas A\&M University, College Station, TX 77843, USA}
\author{D.H.~Wright} \affiliation{SLAC National Accelerator Laboratory/Kavli Institute for Particle Astrophysics and Cosmology, Menlo Park, CA 94025, USA}
\author{S.~Yellin} \affiliation{Department of Physics, Stanford University, Stanford, CA 94305, USA}
%\author{J.J.~Yen} \affiliation{Department of Physics, Stanford University, Stanford, CA 94305, USA}
\author{B.A.~Young} \affiliation{Department of Physics, Santa Clara University, Santa Clara, CA 95053, USA}
\author{X.~Zhang} \affiliation{Department of Physics, Queen's University, Kingston, ON K7L 3N6, Canada}
\author{X.~Zhao} \affiliation{Department of Physics and Astronomy, and the Mitchell Institute for Fundamental Physics and Astronomy, Texas A\&M University, College Station, TX 77843, USA}

\date{\today}

\begin{abstract}

%The Super Cryogenic Dark Matter Search experiment (SuperCDMS) at the Soudan Underground Laboratory employed \mbox{\textit{in situ}} $^{210}$Pb sources. Data from these sources were used to investigate energy loss associated with Frenkel defect formation in germanium crystals at mK temperatures. 
The Super Cryogenic Dark Matter Search experiment (SuperCDMS) at the Soudan Underground Laboratory studied energy loss associated with defect formation in germanium crystals at mK temperatures using \mbox{\textit{in situ}} $^{210}$Pb sources. 
We examine the spectrum of $^{206}$Pb nuclear recoils near its expected 103\,keV endpoint energy and determine an energy loss of \FDvalFin, which we attribute to defect formation. From this result and using TRIM simulations, we extract the first experimentally determined average displacement threshold energy of \GeThval for germanium. This has implications for the analysis thresholds of future germanium-based dark matter searches. 

\end{abstract}

%\pacs{61.80.-Jh, 61.82.Fk, 29.40.Wk, 95.35.+d, 85.25.Oj}
% 61.80.-Jh = Physical radiation effects, radiation damage ... from ions
% 61.82.Fk = radiation effects on semiconductors
% 29.40.Wk = solid-state radiation detectors
% 95.35.+d = dark matter
% 85.25.Oj = Superconducting detectors (including transition edge)

% RAB: SuperCDMS is not a good keyword and "threshold value" is too vague
%\keywords{Defect formation, Threshold value, SuperCDMS}
\keywords{SuperCDMS, crystal defect formation, displacement threshold energy, germanium detectors, dark matter, cryogenic}
\maketitle

\section{Introduction}

Crystal defects can occur when incident radiation recoils off of an atom transferring sufficient energy to displace the atom from its lattice site, thus creating a vacancy.  If the displaced atom remains in the crystal, it is referred to as an interstitial atom (or an ``interstitial'').  The combination of the vacancy and the interstitial are referred to as a Frenkel pair, or Frenkel defect\cite{Frenkel1926}.  Energy can also be lost through creation of defect clusters, dislocations and amorphous zones.  The creation of these defects permanently stores energy in the crystal, with the fraction of incident energy that goes into defect formation depending in part on the mass of the impinging particle, the deposited energy, and the crystal properties\cite{leroy_rancoita_2009}. Collectively, the total energy lost to formation of defects is referred to as the Wigner energy \cite{WignerEnergy}.

The energy required to displace an atom from its lattice site is the displacement threshold energy.  For germanium, previously determined displacement threshold energy values  from theory and various molecular dynamics simulations are inconsistent, ranging from 7 to 30~eV\cite{1956Vavilov,1977Vitovskii,PhysRevB.34.6987,1992Emtsev,PhysRevB.57.7556,0953-8984-14-11-302,1402-4896-81-3-035601}.

The value of the displacement threshold energy has implications for physics experiments that employ solid-state detectors to search for nuclear recoil events with sub-keV energy depositions.
%In this letter, we focus on the direct detection of dark matter with Ge detectors in the Super Cryogenic Dark Matter Search \mbox{(SuperCDMS)} experiment\cite{Surface2013,scdms2016,Agnese:2013jaa,Agnese:2015nto,Agnese:2017jvy,cdms2018,snolab2017}, which aims to detect nuclear recoils from weakly interacting massive particles (WIMPs)\cite{STEIGMAN1985375} by measuring the energy deposited when a WIMP scatters from an atom in a detector's crystal lattice.
In this letter, we focus on defect formation with data from the Super Cryogenic Dark Matter Search \mbox{(SuperCDMS)} experiment\cite{Surface2013,scdms2016,Agnese:2013jaa,Agnese:2015nto,Agnese:2017jvy,cdms2018,snolab2017}, which aims to detect nuclear recoils from weakly interacting massive particles (WIMPs)\cite{STEIGMAN1985375} by measuring the energy deposited when a WIMP scatters off of an atomic nucleus in a detector's crystal lattice. 
%Moving forward, t
The \mbox{SuperCDMS} program is targeting low-mass WIMPs\cite{snolab2017}---from a few hundred MeV/c$^{2}$ to several GeV/c$^{2}$---using advanced detector designs
%capable of resolving single electron-hole pairs\cite{ehpairs} and
having detection thresholds on the order of the Ge-atom displacement energy. Because the energy that goes into the formation of defects is not directly observable, an accurate determination of the energy loss to defect formation is important for understanding the low-energy detector response and thus for discerning the ultimate low-mass-WIMP sensitivity reach.

To measure the Wigner energy associated with   $^{206}$Pb-on-Ge interactions, we consider data from the most recent phase of the SuperCDMS experiment\cite{Surface2013,scdms2016,Agnese:2013jaa,Agnese:2015nto,Agnese:2017jvy,cdms2018}, when it was located in the Soudan Underground Laboratory. $^{210}$Pb sources were deployed adjacent to two detectors to evaluate their \textit{in situ} response to non-penetrating radiation from the decays of $^{210}$Pb and its daughters $^{210}$Bi and $^{210}$Po (see Ref.~\citenum{Surface2013} and Fig.~2 therein). These data include $^{206}$Pb-on-Ge recoils, for which a significant disagreement between the simulated and measured spectra is evident near the expected 103\,keV endpoint energy (\textit{cf.}\ Fig.~4 in Ref.~\citenum{Surface2013}). In this letter, we reconsider this discrepancy while allowing for the possibility that the measured recoil energy is effectively reduced due to formation of defects. 
The measured and simulated $^{206}$Pb spectra are compared using a $\chi^{2}$ statistic to find a best-fit energy-loss fraction that brings the two into agreement.
%The measured and simulated $^{206}$Pb spectra are  compared using a $\chi^{2}$ statistic, and a best-fit energy-loss percentage is obtained that brings the two into agreement.
The results from each detector are calibrated for events near the detector surface (\textit{vs}. in the bulk) to obtain a value for energy loss due to defect formation in Ge.

\section{Experimental Data \& Event Selection}\label{sec:data}

%This study utilized data from two of the 15 SuperCDMS detectors used in the SuperCDMS Soudan experiment\cite{CDMS2b}. All 15 detectors were interleaved Z-sensitive Ionization and Phonon (iZIP) cylindrical germanium detectors operated at approximately 50~mK\cite{Surface2013}.  This study used data from the top and bottom detectors of Tower 3 (T3Z1 and T3Z3, respectively).
The SuperCDMS Soudan experiment operated 15 cylindrical, interleaved Z-sensitive Ionization and Phonon (iZIP) Ge detectors at ${\sim}50$\,mK from 2012--2015\cite{Surface2013,scdms2016,cdms2018}, arranged in five stacks of three detectors each. Data from the top and bottom detectors of the third such stack---called T3Z1 and T3Z3, respectively---are used in this study.

%Ionization electrodes on the top and bottom faces of the detector were biased at $+2$~V and $-2$~V, respectively, creating a relatively uniform electric field ($\sim$0.5~V/cm) throughout the bulk of the crystal\cite{cdms2018}.
%The phonon channels remained grounded. Ionization measurements were made by drifting electrons and electron-holes through this field to the top and bottom faces of the detector. Events occurring within ${\sim}$1~mm of the surface will have asymmetric charge collection on opposing faces due to the non-uniform electric field near the biased ionization channels (see also Fig.\ 1(b) in Ref.\ \textcite{Surface2013}). This asymmetry in charge collection makes it possible to distinguish between surface events and bulk events.
Each iZIP detector had several independent phonon and ionization readout channels on both of its flat faces. The ionization electrodes on the top and bottom faces were biased respectively at $+$2\,V and $-$2\,V, while the interleaved phonon sensors were held at ground. The resulting electric field caused 
%electrons and electron-holes 
positive and negative charge carriers
from particle interactions in the detector bulk to drift to opposing faces, whereas within ${\sim}$1~mm of either face most of the 
%ionization 
charge carriers
were collected by the electrodes on that face of the detector. This asymmetry in charge collection between the two detector faces makes it possible to distinguish energy depositions near a detector face from those in its bulk (\textit{cf.}\ Fig.~3 in Ref.~\citenum{Surface2013}).

%During the operation of the SuperCDMS Soudan experiment, two $^{210}$Pb-implanted silicon wafers were installed facing T3Z1(T3Z3) with the source facing the +2(-2) electrode.
The SuperCDMS Soudan experiment had two \textit{in situ} $^{210}$Pb sources, with one installed facing the top side of T3Z1 and the other facing the bottom side of T3Z3. 
%Each source consisted of a $^{210}$Pb-implanted silicon wafer. 
The sources were produced by exposing silicon wafers to a 5~kBq $^{226}$Ra source (which produces $^{222}$Rn gas) for 12~days inside a sealed aluminum box.  After exposure, the wafers were surface etched to remove dust and radon daughters resting on the surface. This process resulted in a near-uniform implantation profile of $^{210}$Pb to a depth of approximately 58~nm\cite{Surface2013,redl2014}.
%From subsequent time of exposure to lab air, we estimate a $1.6\pm0.1$~nm oxide layer formed on the surface\cite{SiExp1,SiExp2}.  
Based on the subsequent time of exposure to lab air, we estimate a $(1.6\pm0.1)$\,nm oxide layer formed on the surface of each source wafer\cite{SiExp1,SiExp2}.  

\begin{figure}[tbph] 
	\includegraphics[width=\linewidth]{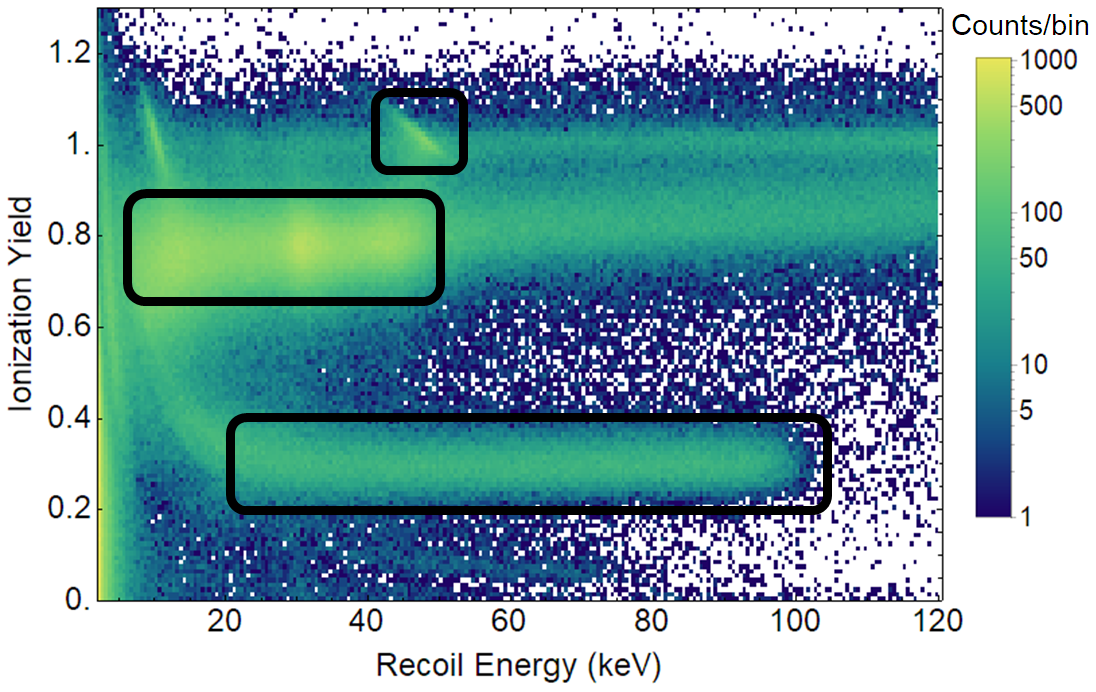}
    \caption{
    %Ionization yield vs.\ recoil energy for events from one detector adjacent to a $^{210}$Pb-implanted silicon plate. 46~keV gammas and surface betas from $^{210}$Pb and $^{210}$Bi decays appear in the top and middle boxes.  The $^{206}$Pb recoils are visible as the band near 0.3 on the vertical axis and have an obvious cut-off around 100~keV (bottom box).
    Ionization yield \textit{vs.}\ recoil energy for events from one of the $^{210}$Pb sources. Gamma rays and betas from $^{210}$Pb and $^{210}$Bi decays appear in the top and middle boxes, while $^{206}$Pb recoils from $^{210}$Po decays are visible as a band that cuts off at ${\sim}$100~keV near ionization yield of 0.3 (bottom box).
    }
	\label{im:T3Z1_YieldPlane}
\end{figure}

%The data used in this analysis were collected from the March 2012--July 2014 data-taking periods of the SuperCDMS experiment\cite{Surface2013,scdms2016}.  
The data used in this analysis were collected from March 2012--July 2014\cite{Surface2013,scdms2016}. 
%Since ionization charges and phonons are collected, ionization energy and recoil energy can be determined for each event. The ratio of these energies ("ionization yield") make it possible to identify different event types.
Ionization and phonon signals were measured for each event, and the ratio of these measurements (``ionization yield'') allowed for discrimination between event types.
%It is possible to identify different event types by examining the ratio of ionization to recoil energy (``ionization yield'').  
%The experiment is calibrated such that the ionization yield from gammas from a $^{133}$Ba source is 1, and the nuclear recoil band is defined by neutrons from a $^{252}$Cf source (see Fig.~4 in Ref.~\textcite{CDMS2b}). 
The detector responses were calibrated using $^{133}$Ba gamma rays such that electron recoils in the detector bulk have ionization yield equal to one.
%$^{206}$Pb recoils can be selected using their comparatively low ionization yield along with the surface event criteria developed in Ref. \textcite{Surface2013} (Fig.~\ref{im:T3Z1_YieldPlane}). We also select surface gammas and betas from $^{210}$Pb and $^{210}$Bi decays to create an initial estimate of the detector resolution for surface events (i.e. those with ionization yield $\sim$1). 
$^{206}$Pb recoils have comparatively low ionization yield; in Fig.~\ref{im:T3Z1_YieldPlane} they appear at a yield of $\sim$0.3 and they extend in energy to near the expected 103\,keV endpoint.

In this study, $^{206}$Pb recoils are selected based on their ionization yield and the surface-event criteria developed in Ref.~\citenum{Surface2013}. Similar criteria are used to select near-surface electron recoil events (highlighted in Fig.~\ref{im:T3Z1_YieldPlane}) that correspond to  gamma rays (top box) and betas (middle box) from decays of $^{210}$Pb and $^{210}$Bi. These event selections are used to estimate the detector resolution and energy scale for surface events, independent of the $^{206}$Pb recoils used to study energy loss from defect formation.

\section{Simulating $^{210}$P\lowercase{b} Decays}

The SuperCDMS Soudan experiment was simulated with G\textsc{eant}4\cite{Geant42003,Geant42006,Geant42016} version 10.1.p2 using the Screened Nuclear Recoil physics list\cite{MENDENHALL2005420}.  
%In simulation, the full experiment was created with detectors, shielding, and the $^{210}$Pb-implanted silicon source plates. The implantation profile models that expected from the exposure and latter etching of the plates\cite{redl2014,Surface2013}.
A detailed simulation geometry was used including the detectors, all surrounding materials, and the $^{210}$Pb source wafers. The source wafers were simulated with zero surface roughness. The full chain of $^{210}$Pb decays was simulated according to the source wafers' implantation profile.
One million primary $^{210}$Pb atoms were simulated, with G\textsc{eant}4 allowed to handle the full decay chain for each event.
%$^{206}$Pb recoils events were selected as those occuring within 1~$\mu${s} of a $^{210}$Po decay.  In this way, we do not discriminate against the species of particles depositing energy in the detector which should more closely resemble the real experiment.
%Pb-206 recoils events were identified as those occurring within 1~$\mu${s} of a $^{210}$Po decay.%  In this way, we do not discriminate against the species of particles depositing energy in the detector which should more closely resemble the real experiment.

\begin{figure}[tbph]
	\includegraphics[width=\linewidth]{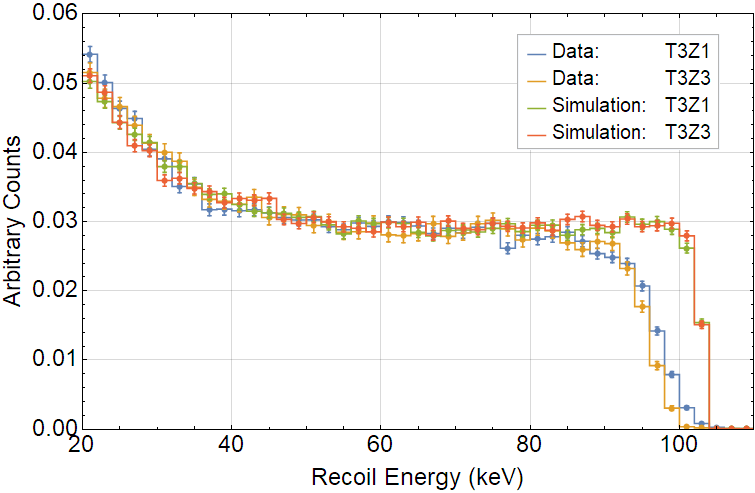}
    \caption{
    %Spectra of data taken from T3Z1 and T3Z3 compared to the Monte Carlo simulation of $^{206}$Pb recoils.  
    Measured $^{206}$Pb-recoil spectra for detectors T3Z1 (blue) and T3Z3 (yellow), compared to Monte Carlo simulations (green and red, respectively).
    The spectral shapes show approximate agreement up to $\sim$80\,keV.
    %with a noticeable divergence near the 103~keV end-point.  
    However, there is a clear discrepancy near the 103~keV $^{206}$Pb-recoil endpoint where  fewer counts are seen in the data compared to the Monte Carlo prediction.
    Error bars correspond to 1$\sigma$ statistical uncertainties.% on the number of counts in each bin.
    }
	\label{im:DataAndSim}
\end{figure}

Selecting simulated $^{206}$Pb events in the 80--110~keV region of interest yields a total of ${\sim}44,000$ simulated events, approximately twice the corresponding number of measured $^{206}$Pb recoils. The simulation results show good agreement with the shape of the measured spectra up to 80~keV. 
%and then showed a divergence as energies approached the 103~keV end-point (Fig.~\ref{im:DataAndSim}).
However, as shown in Fig.~\ref{im:DataAndSim}, for larger recoil energies there is a significant discrepancy between the measured and simulated spectra for each detector. The former are softer with substantially fewer events measuring the full 103~keV endpoint energy. This disagreement near the $^{206}$Pb-recoil endpoint is indicative of energy loss due to defect formation in the detector crystal, a process not taken into account in the simulation.

\section{Analysis Method}\label{sec:analysis}

\subsection{Fitting Method}\label{ssec:fitting}
%Two factors are considered in the analysis: Energy loss to account for defect formation, and energy smearing to account for detector resolution.
Two factors are considered to account for the discrepancy in Fig.~\ref{im:DataAndSim}: energy loss due to defect formation, and energy smearing due to detector resolution.
Each simulated event is first scaled to
\begin{equation}\label{eq:scale}
\hat{E}=E(1-f)\textrm{,}
\end{equation}
where $E$ is the deposited energy, $f$ is the fraction of energy lost to defect formation, and $\hat{E}$ is the remaining energy.  
$\hat{E}$ is then treated with a Gaussian smearing function that has a standard deviation corresponding to the 1$\sigma$ detector resolution at that energy. The resulting smeared event energy $\tilde{E}$ is thus representative of the actual energy measured by a detector.

%In the analysis region we consider, the resolution is expected to be linearly dependent on event energy \mbox{(i.e. $\sigma(E)=c_1+c2{E}$)} as demonstrated in Ref. \textcite{redl2014}. Surface resolution was determined by extracting the standard deviation $\sigma$ of a Gaussian peak fit to surface gamma events, and constructing a linear best fit to these points.  The \textit{simulated} events are then treated by a Gaussian smearing function with the mean at $\hat{E}$ and a standard deviation based on that found for surface gammas, but with one additional parameter $P_s$. 

%We assume that the resolution of $^{206}$Pb recoil events will have the same functional form as that of gammas, but the absolute value may slightly differ. To account for this, the absolute smearing strength is determined by multiplying the linear resolution function by a parameter that can float during analysis:
%\begin{equation}\label{eq:resolution}
%\sigma_{Pb}(E)=P_s\sigma(E)
%\end{equation}
%where $\sigma_{Pb}$ is the standard deviation used to smear events and $P_s$ is the floating parameter.

As demonstrated in Ref.~\citenum{redl2014}, the resolution is an approximately linear function of energy in the range of 80--110~keV: $\sigma(E)=0.63\,\textrm{keV}+0.024{E}$.   
%The parameters $c_1$ and $c_2$ are estimated by fitting to lines in surface-event gamma-ray spectra. 
The parameters are estimated by fitting to the 46 keV and 66.7 keV peaks in surface-event gamma-ray spectra. 
We assume that the energy resolution of $^{206}$Pb recoils has the same functional form, but the absolute value may differ slightly. To account for this difference, the resolution is scaled by a multiplicative factor $P_s$:
\begin{equation}\label{eq:resolution}
\sigma_{Pb}(E)=P_s\sigma(E)
\end{equation}
where $\sigma_{Pb}$ is the resolution function used to smear the simulated $^{206}$Pb recoil energies.  Both $f$ and $P_s$ represent free parameters that are allowed to float in the fitting method outlined below.

%Once all events are scaled and smeared, there are two sets of recoil energies ready for comparison, corresponding to data ($A$) and simulation ($B$):
After the simulated events are scaled and smeared, the resulting energies are compared directly to the measured energies as follows. Let $A$ and $B$ represent the set of measured and simulated event energies, respectively:
\begin{align*}
A=&\{E_1,E_2,\dots{E_N}\}\\
B=&\{\tilde{E}_1,\tilde{E}_2,\dots{\tilde{E}_M}\}
\end{align*}
where the sets are of size $N$ and $M$, respectively.
%The sets are of size $N$ and $M$, respectively.  
Each set is binned by energy into $q$ bins:
\begin{align*}
\text{Bins}_A=&\{a_1,a_2,\dots{a_q}\}\\
\text{Bins}_B=&\{b_1,b_2,\dots{b_q}\}
\end{align*}
where $a_i$ and $b_i$ indicate the number of events in the $i^{th}$ bin.  
%The $\chi^2$ value is determined by summing the residuals across all bins:
To gauge the level of agreement between these binned energy distributions, a $\chi^{2}$ statistic is calculated:
\begin{equation*}\label{eq:chi2}
\chi^2=\sum_{i=1}^{q}\frac{\left(\frac{a_i}{N}-\frac{b_i}{M}\right)^2}{\frac{a_i}{N^2}+\frac{b_i}{M^2}}
\end{equation*}
We generate approximately one million sets $B$ for each detector, corresponding to
%different smearing ($P_s$) and scaling ($f$) factors.  
different combinations of the scaling ($f$) and smearing ($P_s$) parameters.
A $\chi^{2}$ value is determined for each set, creating a well-defined parameter space from which a minimum can be found yielding the best-fit values for $f$ and $P_s$.

\subsection{Surface-Event Energy Scaling}

%The detectors were calibrated for analysis of events in the bulk of the crystal.  
The measured event energies are based on the detectors' default energy calibrations, which are developed using gamma rays in the bulk of the crystal.
%As such, there may be an energy-dependent scaling factor in the reconstructed recoil energies for surface events.  
The energy scale for surface events may be slightly different than for bulk events\cite{Surface2013}. Consequently, the measured $^{206}$Pb recoil energies may differ from their simulated counterparts by an additional energy scaling factor that represents an intrinsic miscalibration and therefore is independent of defect formation.
If present, a best-fit determination of the scaling factor $f$ in Equation~\ref{eq:scale} would account for both this miscalibration \textit{and} energy loss due to defect formation:  
\begin{equation}\label{eq:FDtotal}
\left(1-f\right)=\left(1-f_{DF}\right)\left(1-f_{sur}\right),
\end{equation}
where $f_{DF}$ is the scale factor from energy loss due to defect formation, and $f_{sur}$ is the surface-event scale factor.  Because the total energy loss to defect formation depends on the mass of the incident particle, surface events from gamma rays and betas should have $f_{DF}{\sim}0$ to within the precision of this study.  This allows for the determination of any intrinsic miscalibration via an independent examination of these alternate event classes. %where $f$ in Equation \ref{eq:scale} would be equal to $f_{int}$.
The energy loss to defect formation is thus
\begin{equation}\label{eq:FDfinal}
f_{DF}=1-\frac{\left(1-f\right)}{\left(1-f_{sur}\right)},
\end{equation}
with $f$ determined from $^{206}$Pb events, and $f_{sur}$ determined from surface gamma-ray and beta events.

\section{Results \& Discussion}

\subsection{Energy Loss to Defect Formation}

\begin{figure}[tbph]
\includegraphics[width=\linewidth]{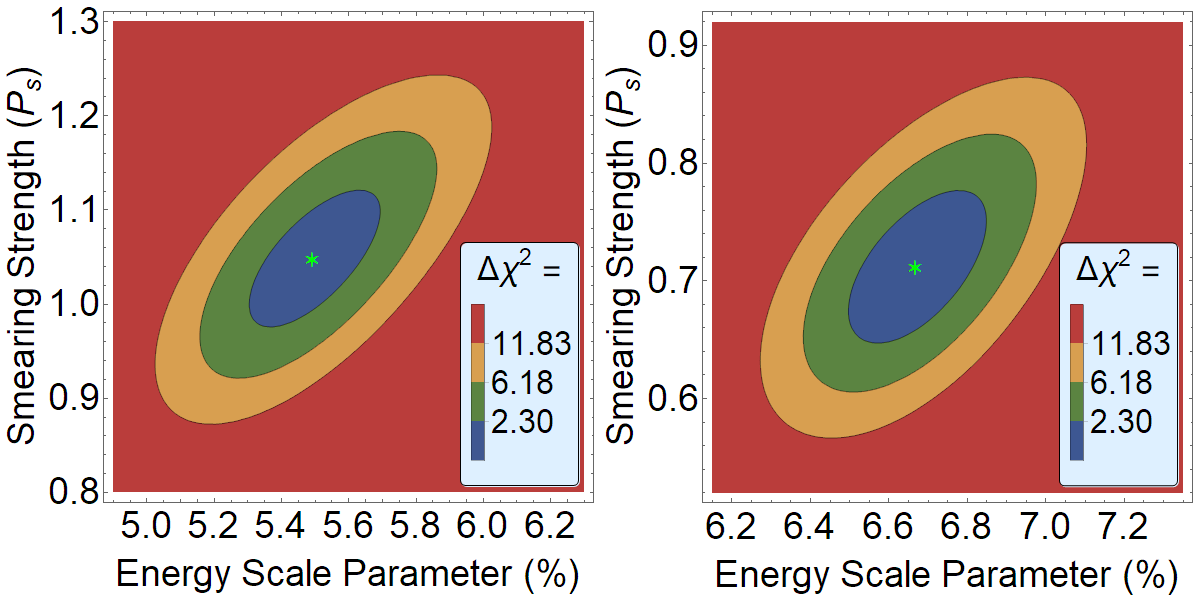}
    \caption{The $\chi^{2}$ statistic plotted versus the smearing factor $P_s$ and the energy-scale parameter $f$ for detectors T3Z1 (left) and T3Z3 (right).  The best-fit values are indicated by a green star.  The contours correspond to the 2-dimensional 1$\sigma$, 2$\sigma$ and 3$\sigma$ confidence intervals ($\Delta\chi^{2}$ = 2.3, 6.2, and 11.8, respectively).}
	\label{im:T3Both_Pb_CP}
\end{figure}

\begin{figure}[tbph]
	\includegraphics[width=\linewidth]{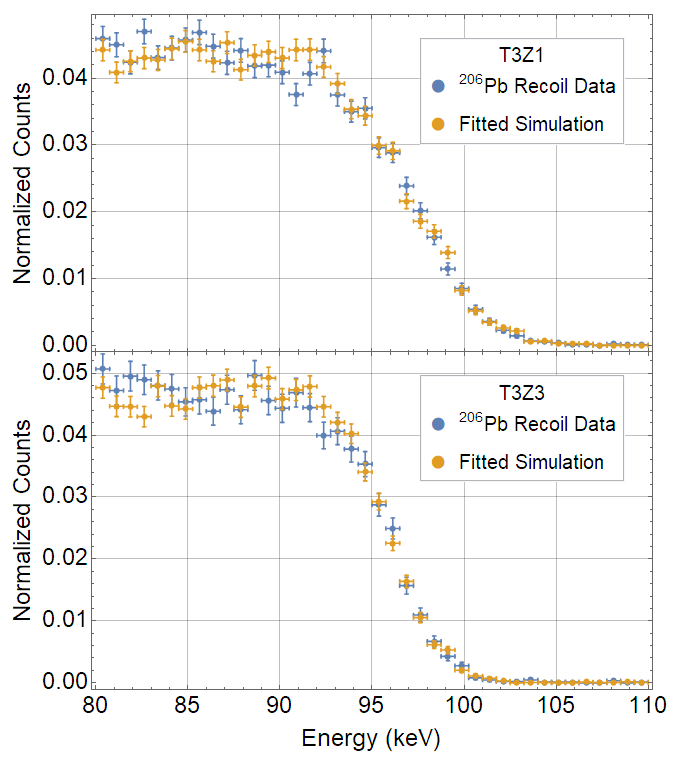}
    \caption{
    %$^{206}$Pb recoil data compared to the best fit values applied to simulation for T3Z1 (Top) and T3Z3 (Bottom).
    Measured $^{206}$Pb recoil spectrum (blue) compared to simulated $^{206}$Pb recoils after application of the best-fit energy scaling and smearing parameters (orange) for detectors T3Z1 (top) and T3Z3 (bottom).
    }
	\label{im:T3Both_Pb_Fit}
\end{figure}

%The best fit result for the energy scale parameter $f$ for $^{206}$Pb data is $6.35 \pm 0.14$~\% ($7.28 \pm 0.17$~\%) for T3Z1 (T3Z3). The parameter space for both detectors are illustrated in Figure~\ref{im:T3Both_Pb_CP}, and the uncertainty on $f$ is determined by projecting a $\Delta\chi^2=1$ contour in the parameter space to the ``Energy Loss'' axis. The best fit values applied to simulation are shown in Figure \ref{im:T3Both_Pb_Fit}. 
Application of the procedure outlined in Sec.~\ref{ssec:fitting} to the measured and simulated $^{206}$Pb recoil energies gives a best-fit energy-scale parameter of $f=$~\detSeven and \detNine for detectors T3Z1 and T3Z3, respectively. Figure~\ref{im:T3Both_Pb_CP} shows the $\chi^{2}$ statistic as a 2-dimensional function of the smearing strength $P_s$ and the energy loss parameter $f$. Statistical uncertainties (at 1$\sigma$ confidence) on the best-fit values of $f$ are determined by projecting the $\Delta\chi^{2} = 1$ contours onto the ``Energy Scale Parameter'' axis. After application of the best-fit parameters, the simulated and measured $^{206}$Pb recoil energy distributions are in good agreement, as shown in Fig.~\ref{im:T3Both_Pb_Fit}.

%The process is repeated identically for surface gamma and beta events with the results summarized in Table~\ref{table:IntScale}.  To determine $f_{sur}$ for each detector, a weighted mean between gamma and beta results is used.  The final preferred Frenkel defect value $f_{FD}$ is then calculated from Equation \ref{eq:FDfinal}, and the results are summarized in Table~\ref{table:PbScale}.
%The uncertainty of $f_{FD}$ is determined by the following error propagation formula:
%\begin{equation*}
%\sigma_{f_{FD}}=\sqrt{\left(\frac{\partial f_{FD}}{\partial f}\sigma{f}\right)^2+\left(\frac{\partial f_{FD}}{\partial f_{sur}}\sigma{f_{sur}}\right)^2}
%\end{equation*}
%With this parameter determined for each detector, a weighted mean is taken as the final value of \FDval where the uncertainty is a weighted sample variance. We use this method of determining uncertainty to account for the variance between each detector's results.

The same analysis procedure is applied to simulated and measured distributions of surface-event gamma rays and betas highlighted in Fig~\ref{im:T3Z1_YieldPlane}, with the results summarized in Table~\ref{table:IntScale}. Because $f_{DF}\sim0$ for these event classes, the values in Table~\ref{table:IntScale} are a direct measure of $f_{sur}$. A single value of $f_{sur}$ is obtained for each detector by taking a weighted mean of the gamma-ray and beta results, which is then used to determine $f_{DF}$  from Equation~\ref{eq:FDfinal}; these results are summarized in Table~\ref{table:PbScale}. 
%A final best estimate of the energy loss due to Frenkel defects from $^{206}$Pb recoils is obtained by taking a weighted mean of the two detectors, yielding \FDvalSolo. The uncertainty is taken from the biased weighted sample variance and is therefore a fair representation of the variance between the individual detector results.
The weighted mean of the two detectors is $\left(6.08\pm0.08\right)$\,\%, but because the individual measurements differ by 2.06 standard deviations (p-value 0.04), the uncertainty on the weighted mean is increased by a factor of 2.06 by increasing each uncertainty by that factor. This results in a more reasonable difference of one standard deviation, and gives a weighted mean of \FDvalSolo where the uncertainty is a combination of statistical and systematic uncertainty.
%we increased the variance of the weighted mean by a factor of the reduced chi-square\footnote{This is sometimes referred to as the biased weighted sample variance.}: $\chi^{2}_{\nu} = 4.2$. 
%This procedure yields a more reasonable uncertainty, giving a weighted mean of \FDvalSolo.
%There is an additional systematic uncertainty of 0.03\,\% from the native silicon dioxide on the surface of the source plate which we estimate to be $1.6\pm0.1$\,nm.
%There is an additional systematic uncertainty of $^{+0.04}_{-0.03}$\,\% from the native silicon dioxide on the surface of the source plate which we estimated to be 1.6\,$\pm$0.1\,nm as described in Section~\ref{sec:data}.  The best fit energy loss is then \mbox{\FDval}.  Additional potential sources of uncertainty are discussed in a following section.

There is an additional systematic uncertainty of $^{+0.04}_{-0.03}$\,\% from the $\left(1.6\pm0.1\right)$\,nm silicon dioxide layer on the surface of each source wafer (as described in Section~\ref{sec:data}).  
%We add this error in quadrature with our 
The best-fit energy loss is therefore \mbox{\FDvalFin} after adding the uncertainties in quadrature.  Additional potential sources of uncertainty are discussed in a following section.

% P-values:
%  Gammas:
%   T3Z1: 0.01
%   T3Z3: 0.01
%  Betas:
%   T3Z1: 0.11
%   T3Z3: 0.05

\begin{table}[tbph]
	\centering
	\small
    \caption{
    %The energy scale factor determined by examining gamma and beta events in T3Z1 and T3Z3.
    The best-fit energy-scale parameter $f$ obtained for surface-event gamma rays and betas with the associated reduced chi-square ($\chi^2_\nu$) and $p$-value for each detector.  Because $f_{DF}\sim0$ for these event classes, theses values provide a direct measure of the energy-scale correction factor for surface events.\label{table:IntScale}}
	\begin{tabular}{l
			>{\hspace{1pc}}c
			S[table-format=3.2]@{\,\( \pm \)\,}S[table-format=1.2]cc}
		\hline
		\Tstrut{Detector} & Population & \multicolumn{2}{c}{~~~$f$\,\%} & ~~~~$\chi^2_\nu$ & $p$-value\\
		\hline\Tstrut 
		\multirow{2}{*}{T3Z1} & Gamma Events & -0.74 & 0.07 & ~~~1.9 & 0.01\\
		& Beta Events  & -0.75 & 0.11 & ~~~1.3 & 0.11\\\hline\Tstrut
		%& Mean$_w$ & -0.75 & 0.06\\\hline
		\multirow{2}{*}{T3Z3} & Gamma Events & 0.84 & 0.09 & ~~~1.9 & 0.01\\
		& Beta Events  & 0.87 & 0.17 & ~~~1.4 & 0.05\\\hline
		%& Mean$_w$ & 0.85 & 0.08\\\hline
	\end{tabular}
	
\end{table}

% P-Values:
% T3Z1: 0.08
% T3Z3: 0.31

\begin{table}[tbph]
	\centering
    \caption{The energy scale factor $f$ determined by examining $^{206}$Pb recoils in detectors T3Z1 and T3Z3 and the corresponding reduced chi-square ($\chi_{\nu}^{2}$) and $p$-values.  The intrinsic scaling factor $f_{sur}$ is the weighted mean of the scale factors determined from gamma rays and betas (Table~\ref{table:IntScale}).  The energy loss to defect formation $f_{DF}$ is determined from Equation~\ref{eq:FDfinal}.\label{table:PbScale}}
	\begin{tabular}{lccc
			S[table-format=3.2]@{\,\( \pm \)\,}S[table-format=1.2]
			c}\hline
		\Tstrut{Detector} & $f$\,\% &$\chi^2_\nu$& $p$-value & \multicolumn{2}{c}{~~~~~$f_{sur}$\,\%} & $f_{DF}\,\%$\\
		%		Number & Number & Wall (m)\\
		\hline\Tstrut
		T3Z1 & $5.52\pm0.10$ & 1.3 & 0.08 & -0.75 & 0.06 & $6.22 \pm 0.11$\\
		T3Z3 & $6.67\pm0.11$ & 1.1 & 0.31 & 0.85 & 0.08 & $5.87 \pm 0.13$\\\hline
	\end{tabular}
	
\end{table}

\subsection{Displacement Threshold Energy}

%With a determined value for Frenkel defect energy loss, it is possible to further determine the germanium displacement threshold value.  
Using our best estimate of the value for  energy loss to the formation of defects, it is possible to determine the displacement threshold energy of a germanium atom.
This is the average displacement threshold energy over all lattice angles\cite{NORDLUND2006322} and is an important quantity for radiation detectors, WIMP-searches, and other applications\cite{2017arXiv170305371K,1402-4896-81-3-035601,2903}.  

For interactions involving the same species of incident and target atoms, the Kinchin-Pease equation estimates the number of defects formed\cite{1955RPPh...18....1K} (with further refinement by Norgett, Robinson and Torrens\cite{NORGETT197550}).  In the case of an incident $^{206}$Pb recoil on Ge, displaced Ge atoms may be liberated with enough energy to form yet more defects; so there are two types of interactions to consider.
\mbox{TRIM-2013}\cite{SRIM} simulations were used to model the entire defect formation process, for a range of user-defined values of the Ge displacement threshold energy from \mbox{15 to 23}\,eV.

The target material in the TRIM simulations was a solid mass of pure Ge with a thin layer of GeO$_2$ on top.  As with Si, pure Ge reacts with oxygen in the atmosphere to create GeO$_2$ with a thickness that logarithmically depends on exposure time\cite{GeExp1,GeExp2}.  We estimated a GeO$_2$ layer thickness of $\left(0.98\pm0.02\right)$\,nm.  %The Ge displacement threshold value was varied for each simulation from 15 to 23\,eV. 

%By simulating over one million incident $^{206}$Pb atoms, we developed a model to describe the relationship between the energy permanently lost to defects and the Ge displacement threshold value.  For our best estimate of the energy loss value of \FDval, this model suggests a threshold value of \GeThval.
TRIM predicts a monotonic, decreasing relationship between the percent energy lost to defects in the 80--110~keV energy range and the Ge displacement threshold energy.  To match our best estimate of the energy loss value of \FDvalFin, TRIM simulations suggest using a displacement threshold energy of \GeThval. The systematic error does not include modeling imperfections in G\textsc{eant}4 and TRIM.

%This value is in tension with some molecular dynamics simulations\cite{doi:10.1063/1.369709,1402-4896-81-3-035601,emtsev1992}.  
This value is somewhat in tension with some molecular dynamics calculations\cite{doi:10.1063/1.369709,1402-4896-81-3-035601,emtsev1992}. However TRIM uses simple potentials and includes tuned parameters to fit experimental implantation data. The more sophisticated potentials used in molecular dynamics simulations may yield different values. More experimental data are required to further investigate this.

%The simulations often rely on the Stillinger-Weber (SW) potential\cite{SW1985}, fitting as many as nine parameters to the energy function.  Small changes in the parameters to fit one feature (such as melting point or defect formation enthalpy) may yield large fluctuations in another\cite{Pizza2013}, which may account for the disagreement.  

\subsection{Other Sources of Uncertainty}

%The variance between results from each detector is caused by the polarization of each detector face.  The detectors carry a +2/-2~V bias on the top/bottom face which creates the electric field through which electrons and electron-holes drift.  Events used in this analysis are hitting the top of T3Z1 and the bottom of T3Z3 and so are dominated by electrons and electron-holes respectively.

%We examined the case of using an additional resolution parameter, allowing both $c_1$ and $c_2$ to float independently, as opposed to just $P_s$ in Equation \ref{eq:resolution}.
%This shifted the preferred energy loss result by less than two percent, and is small compared to the quoted uncertainty.
%Within the TRIM framework, we determine a displacement threshold energy of \GeThval where the systematic error does not include modeling imperfections in G\textsc{eant}4 and TRIM. This is somewhat in tension with some molecular dynamics calculations. However SRIM uses simple potentials and includes fudge factors to fit experimental implantation data and the more sophisticated potentials used in molecular dynamics may yield different values.

Other sources of uncertainty were considered, resulting in no significant increase in the quoted uncertainty.

%We investigated the dependence of the results on the shape of the energy resolution function by allowing both of the resolution parameters ($c_1$ and $c_2$) to float independently. The resulting shift in the energy loss parameter was less than two percent of our best-fit value and is thus small relative to the quoted statistical uncertainty.

We investigated the effects of varying the thickness of the germanium oxide layer on top of the detectors.  We estimated the thickness to be $\left(0.98\pm0.02\right)$\,nm, and varying the thickness by 1$\sigma$ did not change our results at the precision given.

In this analysis, we make the assumption that all recoils are $^{206}$Pb events.  However there are some events where sputtered silicon atoms from the source wafers might contribute to the total event energy.  
%By considering the incident energy of the Si atoms, Pb atoms, and the defect formation process for each species in Ge, the final result shifts by less than one percent from the Pb-only assumption. 
Considering the incident energies and formation of defects for both Si and Pb ions shifts our best-fit result by less than one percent of the value obtained with the Pb-only assumption.

Nevertheless, the defect energy loss parameters for the two detectors are not quite statistically consistent (\textit{cf.}\ $f_{DF}$ in Table~\ref{table:PbScale}). This inconsistency may represent a true physical difference due to differences in crystal properties between the two detectors.  It may also be a result of an operational difference.  The $^{206}$Pb recoils used in this analysis were incident on opposite faces of the two detectors (\textit{i.e.}\ top \textit{vs.}\ bottom), which were biased with opposite polarities and thus resulted in collection of predominantly positive or negative charge carriers by the ionization electrodes.  A corresponding difference in charge collection efficiency (electrons \textit{vs.}\ electron-holes) for $^{206}$Pb recoils relative to surface-event gamma rays and betas may explain the apparent inconsistency between the two detectors.

\section{Conclusion \& Outlook}

%The discriminating power of the SuperCDMS detectors are leveraged to find the Frenkel defect energy loss value for $^{206}$Pb recoils on Ge.  
The ability of SuperCDMS iZIP detectors to differentiate event types was leveraged to find the Wigner energy  following $^{206}$Pb  implantation on Ge.
We used this result with TRIM simulations to determine an average displacement threshold energy of \GeThval for germanium.
%Using this result, the average displacement threshold value for germanium is determined.  
%This value may play a critical role in future recoil-dependent experiments, especially as instruments move toward being able to resolve single electron-hole pairs\cite{ehpairs}. 
This value will play a critical role in understanding the sensitivity of future experiments designed to measure nuclear recoils (e.g.\ from dark matter interactions) in Ge detectors, especially as instruments move toward 
%being able to resolve single electron-hole pairs\cite{ehpairs}.
lower energy thresholds and better resolution.
Our results also provide another important, empirically determined value from which the Stillinger-Weber potential\cite{SW1985} or others could be fit. Future detectors with thresholds on the order of the displacement threshold energy could confirm this result more directly with low-energy neutron calibrations.

\section{Acknowledgments}

The SuperCDMS collaboration gratefully acknowledges technical assistance from the staff of the Soudan Underground Laboratory and the Minnesota Department of Natural Resources. The iZIP detectors were fabricated in the Stanford Nanofabrication Facility, which is a member of the National Nanofabrication Infrastructure Network, sponsored and supported by the NSF. We would like to thank François Schiettekatte for his feedback in reviewing an early manuscript. Funding and support were received from the National Science Foundation, the U.S.\ Department of Energy, Fermilab URA Visiting Scholar Grant No.\  15-S-33, NSERC Canada, the Canada First Research Excellence Fund, and MultiDark (Spanish MINECO).  This document was prepared by the SuperCDMS collaboration using the resources of the Fermi National Accelerator Laboratory (Fermilab), a U.S.\ Department of Energy, Office of Science, HEP User Facility. Fermilab is managed by Fermi Research Alliance, LLC (FRA), acting under Contract No.\ DE-AC02-07CH11359. Pacific Northwest National Laboratory is operated by Battelle Memorial Institute under Contract No.\ DE-AC05-76RL01830 for the U.S.\ Department of Energy. SLAC is operated under Contract No.\ DEAC02-76SF00515 with the U.S.\ Department of Energy.

\bibliography{references}

\begin{thebibliography}{36}%
\makeatletter
\providecommand \@ifxundefined [1]{%
 \@ifx{#1\undefined}
}%
\providecommand \@ifnum [1]{%
 \ifnum #1\expandafter \@firstoftwo
 \else \expandafter \@secondoftwo
 \fi
}%
\providecommand \@ifx [1]{%
 \ifx #1\expandafter \@firstoftwo
 \else \expandafter \@secondoftwo
 \fi
}%
\providecommand \natexlab [1]{#1}%
\providecommand \enquote  [1]{``#1''}%
\providecommand \bibnamefont  [1]{#1}%
\providecommand \bibfnamefont [1]{#1}%
\providecommand \citenamefont [1]{#1}%
\providecommand \href@noop [0]{\@secondoftwo}%
\providecommand \href [0]{\begingroup \@sanitize@url \@href}%
\providecommand \@href[1]{\@@startlink{#1}\@@href}%
\providecommand \@@href[1]{\endgroup#1\@@endlink}%
\providecommand \@sanitize@url [0]{\catcode `\\12\catcode `\$12\catcode
  `\&12\catcode `\#12\catcode `\^12\catcode `\_12\catcode `\%12\relax}%
\providecommand \@@startlink[1]{}%
\providecommand \@@endlink[0]{}%
\providecommand \url  [0]{\begingroup\@sanitize@url \@url }%
\providecommand \@url [1]{\endgroup\@href {#1}{\urlprefix }}%
\providecommand \urlprefix  [0]{URL }%
\providecommand \Eprint [0]{\href }%
\providecommand \doibase [0]{http://dx.doi.org/}%
\providecommand \selectlanguage [0]{\@gobble}%
\providecommand \bibinfo  [0]{\@secondoftwo}%
\providecommand \bibfield  [0]{\@secondoftwo}%
\providecommand \translation [1]{[#1]}%
\providecommand \BibitemOpen [0]{}%
\providecommand \bibitemStop [0]{}%
\providecommand \bibitemNoStop [0]{.\EOS\space}%
\providecommand \EOS [0]{\spacefactor3000\relax}%
\providecommand \BibitemShut  [1]{\csname bibitem#1\endcsname}%
\let\auto@bib@innerbib\@empty
%</preamble>
\bibitem [{\citenamefont {Frenkel}(1926)}]{Frenkel1926}%
  \BibitemOpen
  \bibfield  {author} {\bibinfo {author} {\bibfnamefont {J.}~\bibnamefont
  {Frenkel}},\ }\href {\doibase 10.1007/BF01379812} {\bibfield  {journal}
  {\bibinfo  {journal} {Zeitschrift f{\"u}r Physik}\ }\textbf {\bibinfo
  {volume} {35}},\ \bibinfo {pages} {652} (\bibinfo {year} {1926})}\BibitemShut
  {NoStop}%
\bibitem [{\citenamefont {Leroy}\ and\ \citenamefont
  {Rancoita}(2009)}]{leroy_rancoita_2009}%
  \BibitemOpen
  \bibfield  {author} {\bibinfo {author} {\bibfnamefont {C.}~\bibnamefont
  {Leroy}}\ and\ \bibinfo {author} {\bibfnamefont {P.~G.}\ \bibnamefont
  {Rancoita}},\ }\href@noop {} {\emph {\bibinfo {title} {Principles of
  radiation interaction in matter and detection}}},\ \bibinfo {edition} {2nd}\
  ed.\ (\bibinfo  {publisher} {World Scientific},\ \bibinfo {year}
  {2009})\BibitemShut {NoStop}%
\bibitem [{\citenamefont {Wigner}(1946)}]{WignerEnergy}%
  \BibitemOpen
  \bibfield  {author} {\bibinfo {author} {\bibfnamefont {E.~P.}\ \bibnamefont
  {Wigner}},\ }\href {\doibase 10.1063/1.1707653} {\bibfield  {journal}
  {\bibinfo  {journal} {Journal of Applied Physics}\ }\textbf {\bibinfo
  {volume} {17}},\ \bibinfo {pages} {857} (\bibinfo {year} {1946})},\ \Eprint
  {http://arxiv.org/abs/https://doi.org/10.1063/1.1707653}
  {https://doi.org/10.1063/1.1707653} \BibitemShut {NoStop}%
\bibitem [{\citenamefont {Vavilov}\ \emph {et~al.}(1956)\citenamefont
  {Vavilov}, \citenamefont {Smirnov}, \citenamefont {Galkin}, \citenamefont
  {Spitsyn},\ and\ \citenamefont {Patskevich}}]{1956Vavilov}%
  \BibitemOpen
  \bibfield  {author} {\bibinfo {author} {\bibnamefont {Vavilov}}, \bibinfo
  {author} {\bibnamefont {Smirnov}}, \bibinfo {author} {\bibnamefont {Galkin}},
  \bibinfo {author} {\bibnamefont {Spitsyn}}, \ and\ \bibinfo {author}
  {\bibnamefont {Patskevich}},\ }\href@noop {} {\bibfield  {journal} {\bibinfo
  {journal} {J. Tech. Phys. U.S.S.R.}\ }\textbf {\bibinfo {volume} {26}},\
  \bibinfo {pages} {1865} (\bibinfo {year} {1956})}\BibitemShut {NoStop}%
\bibitem [{\citenamefont {Vitovskii}, \citenamefont {Mustafakulov},\ and\
  \citenamefont {Chekmareva}(1977)}]{1977Vitovskii}%
  \BibitemOpen
  \bibfield  {author} {\bibinfo {author} {\bibfnamefont {N.}~\bibnamefont
  {Vitovskii}}, \bibinfo {author} {\bibfnamefont {D.}~\bibnamefont
  {Mustafakulov}}, \ and\ \bibinfo {author} {\bibfnamefont {A.}~\bibnamefont
  {Chekmareva}},\ }\href@noop {} {\bibfield  {journal} {\bibinfo  {journal}
  {Fiz. Tekh. Poluprovodn.}\ }\textbf {\bibinfo {volume} {11}},\ \bibinfo
  {pages} {1747} (\bibinfo {year} {1977})}\BibitemShut {NoStop}%
\bibitem [{\citenamefont {Ding}\ and\ \citenamefont
  {Andersen}(1986)}]{PhysRevB.34.6987}%
  \BibitemOpen
  \bibfield  {author} {\bibinfo {author} {\bibfnamefont {K.}~\bibnamefont
  {Ding}}\ and\ \bibinfo {author} {\bibfnamefont {H.~C.}\ \bibnamefont
  {Andersen}},\ }\href {\doibase 10.1103/PhysRevB.34.6987} {\bibfield
  {journal} {\bibinfo  {journal} {Phys. Rev. B}\ }\textbf {\bibinfo {volume}
  {34}},\ \bibinfo {pages} {6987} (\bibinfo {year} {1986})}\BibitemShut
  {NoStop}%
\bibitem [{\citenamefont {Emtsev}, \citenamefont {Mashovets},\ and\
  \citenamefont {Mikhnovich}(1992{\natexlab{a}})}]{1992Emtsev}%
  \BibitemOpen
  \bibfield  {author} {\bibinfo {author} {\bibfnamefont {V.}~\bibnamefont
  {Emtsev}}, \bibinfo {author} {\bibfnamefont {T.}~\bibnamefont {Mashovets}}, \
  and\ \bibinfo {author} {\bibfnamefont {V.}~\bibnamefont {Mikhnovich}},\
  }\href@noop {} {\bibfield  {journal} {\bibinfo  {journal} {Fiz. Tekh.
  Poluprovodn.}\ }\textbf {\bibinfo {volume} {26}},\ \bibinfo {pages} {20}
  (\bibinfo {year} {1992}{\natexlab{a}})}\BibitemShut {NoStop}%
\bibitem [{\citenamefont {Nordlund}\ \emph {et~al.}(1998)\citenamefont
  {Nordlund}, \citenamefont {Ghaly}, \citenamefont {Averback}, \citenamefont
  {Caturla}, \citenamefont {Diaz de~la Rubia},\ and\ \citenamefont
  {Tarus}}]{PhysRevB.57.7556}%
  \BibitemOpen
  \bibfield  {author} {\bibinfo {author} {\bibfnamefont {K.}~\bibnamefont
  {Nordlund}}, \bibinfo {author} {\bibfnamefont {M.}~\bibnamefont {Ghaly}},
  \bibinfo {author} {\bibfnamefont {R.~S.}\ \bibnamefont {Averback}}, \bibinfo
  {author} {\bibfnamefont {M.}~\bibnamefont {Caturla}}, \bibinfo {author}
  {\bibfnamefont {T.}~\bibnamefont {Diaz de~la Rubia}}, \ and\ \bibinfo
  {author} {\bibfnamefont {J.}~\bibnamefont {Tarus}},\ }\href {\doibase
  10.1103/PhysRevB.57.7556} {\bibfield  {journal} {\bibinfo  {journal} {Phys.
  Rev. B}\ }\textbf {\bibinfo {volume} {57}},\ \bibinfo {pages} {7556}
  (\bibinfo {year} {1998})}\BibitemShut {NoStop}%
\bibitem [{\citenamefont {Soler}\ \emph {et~al.}(2002)\citenamefont {Soler},
  \citenamefont {Artacho}, \citenamefont {Gale}, \citenamefont {García},
  \citenamefont {Junquera}, \citenamefont {Ordejón},\ and\ \citenamefont
  {Sánchez-Portal}}]{0953-8984-14-11-302}%
  \BibitemOpen
  \bibfield  {author} {\bibinfo {author} {\bibfnamefont {J.~M.}\ \bibnamefont
  {Soler}}, \bibinfo {author} {\bibfnamefont {E.}~\bibnamefont {Artacho}},
  \bibinfo {author} {\bibfnamefont {J.~D.}\ \bibnamefont {Gale}}, \bibinfo
  {author} {\bibfnamefont {A.}~\bibnamefont {García}}, \bibinfo {author}
  {\bibfnamefont {J.}~\bibnamefont {Junquera}}, \bibinfo {author}
  {\bibfnamefont {P.}~\bibnamefont {Ordejón}}, \ and\ \bibinfo {author}
  {\bibfnamefont {D.}~\bibnamefont {Sánchez-Portal}},\ }\href
  {http://stacks.iop.org/0953-8984/14/i=11/a=302} {\bibfield  {journal}
  {\bibinfo  {journal} {Journal of Physics: Condensed Matter}\ }\textbf
  {\bibinfo {volume} {14}},\ \bibinfo {pages} {2745} (\bibinfo {year}
  {2002})}\BibitemShut {NoStop}%
\bibitem [{\citenamefont {Holmström}, \citenamefont {Nordlund},\ and\
  \citenamefont {Kuronen}(2010)}]{1402-4896-81-3-035601}%
  \BibitemOpen
  \bibfield  {author} {\bibinfo {author} {\bibfnamefont {E.}~\bibnamefont
  {Holmström}}, \bibinfo {author} {\bibfnamefont {K.}~\bibnamefont
  {Nordlund}}, \ and\ \bibinfo {author} {\bibfnamefont {A.}~\bibnamefont
  {Kuronen}},\ }\href {http://stacks.iop.org/1402-4896/81/i=3/a=035601}
  {\bibfield  {journal} {\bibinfo  {journal} {Physica Scripta}\ }\textbf
  {\bibinfo {volume} {81}},\ \bibinfo {pages} {035601} (\bibinfo {year}
  {2010})}\BibitemShut {NoStop}%
\bibitem [{\citenamefont {Agnese}\ \emph {et~al.}(2013)\citenamefont {Agnese},
  \citenamefont {Anderson}, \citenamefont {Balakishiyeva}, \citenamefont
  {Thakur}, \citenamefont {Bauer}, \citenamefont {Borgland}, \citenamefont
  {Brandt}, \citenamefont {Brink}, \citenamefont {Bunker}, \citenamefont
  {Cabrera}, \citenamefont {Caldwell}, \citenamefont {Cerdeno}, \citenamefont
  {Chagani}, \citenamefont {Cherry}, \citenamefont {Cooley}, \citenamefont
  {Cornell}, \citenamefont {Crewdson}, \citenamefont {Cushman}, \citenamefont
  {Daal}, \citenamefont {Stefano}, \citenamefont {Silva}, \citenamefont
  {Doughty}, \citenamefont {Esteban}, \citenamefont {Fallows}, \citenamefont
  {Figueroa-Feliciano}, \citenamefont {Fox}, \citenamefont {Fritts},
  \citenamefont {Godfrey}, \citenamefont {Golwala}, \citenamefont {Hall},
  \citenamefont {Harris}, \citenamefont {Hasi}, \citenamefont {Hertel},
  \citenamefont {Hines}, \citenamefont {Hofer}, \citenamefont {Holmgren},
  \citenamefont {Hsu}, \citenamefont {Huber}, \citenamefont {Jastram},
  \citenamefont {Kamaev}, \citenamefont {Kara}, \citenamefont {Kelsey},
  \citenamefont {Kenany}, \citenamefont {Kennedy}, \citenamefont {Kenney},
  \citenamefont {Kiveni}, \citenamefont {Koch}, \citenamefont {Loer},
  \citenamefont {Asamar}, \citenamefont {Mahapatra}, \citenamefont {Mandic},
  \citenamefont {Martinez}, \citenamefont {McCarthy}, \citenamefont
  {Mirabolfathi}, \citenamefont {Moffatt}, \citenamefont {Moore}, \citenamefont
  {Nadeau}, \citenamefont {Nelson}, \citenamefont {Novak}, \citenamefont
  {Page}, \citenamefont {Partridge}, \citenamefont {Pepin}, \citenamefont
  {Phipps}, \citenamefont {Prasad}, \citenamefont {Pyle}, \citenamefont {Qiu},
  \citenamefont {Radpour}, \citenamefont {Rau}, \citenamefont {Redl},
  \citenamefont {Reisetter}, \citenamefont {Resch}, \citenamefont {Ricci},
  \citenamefont {Saab}, \citenamefont {Sadoulet}, \citenamefont {Sander},
  \citenamefont {Schmitt}, \citenamefont {Schneck}, \citenamefont {Schnee},
  \citenamefont {Scorza}, \citenamefont {Seitz}, \citenamefont {Serfass},
  \citenamefont {Shank}, \citenamefont {Speller}, \citenamefont {Tomada},
  \citenamefont {Villano}, \citenamefont {Welliver}, \citenamefont {Wright},
  \citenamefont {Yellin}, \citenamefont {Yen}, \citenamefont {Young},\ and\
  \citenamefont {Zhang}}]{Surface2013}%
  \BibitemOpen
  \bibfield  {author} {\bibinfo {author} {\bibfnamefont {R.}~\bibnamefont
  {Agnese}}, \bibinfo {author} {\bibfnamefont {A.~J.}\ \bibnamefont
  {Anderson}}, \bibinfo {author} {\bibfnamefont {D.}~\bibnamefont
  {Balakishiyeva}}, \bibinfo {author} {\bibfnamefont {R.~B.}\ \bibnamefont
  {Thakur}}, \bibinfo {author} {\bibfnamefont {D.~A.}\ \bibnamefont {Bauer}},
  \bibinfo {author} {\bibfnamefont {A.}~\bibnamefont {Borgland}}, \bibinfo
  {author} {\bibfnamefont {D.}~\bibnamefont {Brandt}}, \bibinfo {author}
  {\bibfnamefont {P.~L.}\ \bibnamefont {Brink}}, \bibinfo {author}
  {\bibfnamefont {R.}~\bibnamefont {Bunker}}, \bibinfo {author} {\bibfnamefont
  {B.}~\bibnamefont {Cabrera}}, \bibinfo {author} {\bibfnamefont {D.~O.}\
  \bibnamefont {Caldwell}}, \bibinfo {author} {\bibfnamefont {D.~G.}\
  \bibnamefont {Cerdeno}}, \bibinfo {author} {\bibfnamefont {H.}~\bibnamefont
  {Chagani}}, \bibinfo {author} {\bibfnamefont {M.}~\bibnamefont {Cherry}},
  \bibinfo {author} {\bibfnamefont {J.}~\bibnamefont {Cooley}}, \bibinfo
  {author} {\bibfnamefont {B.}~\bibnamefont {Cornell}}, \bibinfo {author}
  {\bibfnamefont {C.~H.}\ \bibnamefont {Crewdson}}, \bibinfo {author}
  {\bibfnamefont {P.}~\bibnamefont {Cushman}}, \bibinfo {author} {\bibfnamefont
  {M.}~\bibnamefont {Daal}}, \bibinfo {author} {\bibfnamefont {P.~C. F.~D.}\
  \bibnamefont {Stefano}}, \bibinfo {author} {\bibfnamefont {E.~D. C.~E.}\
  \bibnamefont {Silva}}, \bibinfo {author} {\bibfnamefont {T.}~\bibnamefont
  {Doughty}}, \bibinfo {author} {\bibfnamefont {L.}~\bibnamefont {Esteban}},
  \bibinfo {author} {\bibfnamefont {S.}~\bibnamefont {Fallows}}, \bibinfo
  {author} {\bibfnamefont {E.}~\bibnamefont {Figueroa-Feliciano}}, \bibinfo
  {author} {\bibfnamefont {J.}~\bibnamefont {Fox}}, \bibinfo {author}
  {\bibfnamefont {M.}~\bibnamefont {Fritts}}, \bibinfo {author} {\bibfnamefont
  {G.~L.}\ \bibnamefont {Godfrey}}, \bibinfo {author} {\bibfnamefont {S.~R.}\
  \bibnamefont {Golwala}}, \bibinfo {author} {\bibfnamefont {J.}~\bibnamefont
  {Hall}}, \bibinfo {author} {\bibfnamefont {H.~R.}\ \bibnamefont {Harris}},
  \bibinfo {author} {\bibfnamefont {J.}~\bibnamefont {Hasi}}, \bibinfo {author}
  {\bibfnamefont {S.~A.}\ \bibnamefont {Hertel}}, \bibinfo {author}
  {\bibfnamefont {B.~A.}\ \bibnamefont {Hines}}, \bibinfo {author}
  {\bibfnamefont {T.}~\bibnamefont {Hofer}}, \bibinfo {author} {\bibfnamefont
  {D.}~\bibnamefont {Holmgren}}, \bibinfo {author} {\bibfnamefont
  {L.}~\bibnamefont {Hsu}}, \bibinfo {author} {\bibfnamefont {M.~E.}\
  \bibnamefont {Huber}}, \bibinfo {author} {\bibfnamefont {A.}~\bibnamefont
  {Jastram}}, \bibinfo {author} {\bibfnamefont {O.}~\bibnamefont {Kamaev}},
  \bibinfo {author} {\bibfnamefont {B.}~\bibnamefont {Kara}}, \bibinfo {author}
  {\bibfnamefont {M.~H.}\ \bibnamefont {Kelsey}}, \bibinfo {author}
  {\bibfnamefont {S.~A.}\ \bibnamefont {Kenany}}, \bibinfo {author}
  {\bibfnamefont {A.}~\bibnamefont {Kennedy}}, \bibinfo {author} {\bibfnamefont
  {C.~J.}\ \bibnamefont {Kenney}}, \bibinfo {author} {\bibfnamefont
  {M.}~\bibnamefont {Kiveni}}, \bibinfo {author} {\bibfnamefont
  {K.}~\bibnamefont {Koch}}, \bibinfo {author} {\bibfnamefont {B.}~\bibnamefont
  {Loer}}, \bibinfo {author} {\bibfnamefont {E.~L.}\ \bibnamefont {Asamar}},
  \bibinfo {author} {\bibfnamefont {R.}~\bibnamefont {Mahapatra}}, \bibinfo
  {author} {\bibfnamefont {V.}~\bibnamefont {Mandic}}, \bibinfo {author}
  {\bibfnamefont {C.}~\bibnamefont {Martinez}}, \bibinfo {author}
  {\bibfnamefont {K.~A.}\ \bibnamefont {McCarthy}}, \bibinfo {author}
  {\bibfnamefont {N.}~\bibnamefont {Mirabolfathi}}, \bibinfo {author}
  {\bibfnamefont {R.~A.}\ \bibnamefont {Moffatt}}, \bibinfo {author}
  {\bibfnamefont {D.~C.}\ \bibnamefont {Moore}}, \bibinfo {author}
  {\bibfnamefont {P.}~\bibnamefont {Nadeau}}, \bibinfo {author} {\bibfnamefont
  {R.~H.}\ \bibnamefont {Nelson}}, \bibinfo {author} {\bibfnamefont
  {L.}~\bibnamefont {Novak}}, \bibinfo {author} {\bibfnamefont
  {K.}~\bibnamefont {Page}}, \bibinfo {author} {\bibfnamefont {R.}~\bibnamefont
  {Partridge}}, \bibinfo {author} {\bibfnamefont {M.}~\bibnamefont {Pepin}},
  \bibinfo {author} {\bibfnamefont {A.}~\bibnamefont {Phipps}}, \bibinfo
  {author} {\bibfnamefont {K.}~\bibnamefont {Prasad}}, \bibinfo {author}
  {\bibfnamefont {M.}~\bibnamefont {Pyle}}, \bibinfo {author} {\bibfnamefont
  {H.}~\bibnamefont {Qiu}}, \bibinfo {author} {\bibfnamefont {R.}~\bibnamefont
  {Radpour}}, \bibinfo {author} {\bibfnamefont {W.}~\bibnamefont {Rau}},
  \bibinfo {author} {\bibfnamefont {P.}~\bibnamefont {Redl}}, \bibinfo {author}
  {\bibfnamefont {A.}~\bibnamefont {Reisetter}}, \bibinfo {author}
  {\bibfnamefont {R.~W.}\ \bibnamefont {Resch}}, \bibinfo {author}
  {\bibfnamefont {Y.}~\bibnamefont {Ricci}}, \bibinfo {author} {\bibfnamefont
  {T.}~\bibnamefont {Saab}}, \bibinfo {author} {\bibfnamefont {B.}~\bibnamefont
  {Sadoulet}}, \bibinfo {author} {\bibfnamefont {J.}~\bibnamefont {Sander}},
  \bibinfo {author} {\bibfnamefont {R.}~\bibnamefont {Schmitt}}, \bibinfo
  {author} {\bibfnamefont {K.}~\bibnamefont {Schneck}}, \bibinfo {author}
  {\bibfnamefont {R.~W.}\ \bibnamefont {Schnee}}, \bibinfo {author}
  {\bibfnamefont {S.}~\bibnamefont {Scorza}}, \bibinfo {author} {\bibfnamefont
  {D.}~\bibnamefont {Seitz}}, \bibinfo {author} {\bibfnamefont
  {B.}~\bibnamefont {Serfass}}, \bibinfo {author} {\bibfnamefont
  {B.}~\bibnamefont {Shank}}, \bibinfo {author} {\bibfnamefont
  {D.}~\bibnamefont {Speller}}, \bibinfo {author} {\bibfnamefont
  {A.}~\bibnamefont {Tomada}}, \bibinfo {author} {\bibfnamefont {A.~N.}\
  \bibnamefont {Villano}}, \bibinfo {author} {\bibfnamefont {B.}~\bibnamefont
  {Welliver}}, \bibinfo {author} {\bibfnamefont {D.~H.}\ \bibnamefont
  {Wright}}, \bibinfo {author} {\bibfnamefont {S.}~\bibnamefont {Yellin}},
  \bibinfo {author} {\bibfnamefont {J.~J.}\ \bibnamefont {Yen}}, \bibinfo
  {author} {\bibfnamefont {B.~A.}\ \bibnamefont {Young}}, \ and\ \bibinfo
  {author} {\bibfnamefont {J.}~\bibnamefont {Zhang}},\ }\href {\doibase
  10.1063/1.4826093} {\bibfield  {journal} {\bibinfo  {journal} {Applied
  Physics Letters}\ }\textbf {\bibinfo {volume} {103}},\ \bibinfo {pages}
  {164105} (\bibinfo {year} {2013})},\ \Eprint
  {http://arxiv.org/abs/http://dx.doi.org/10.1063/1.4826093}
  {http://dx.doi.org/10.1063/1.4826093} \BibitemShut {NoStop}%
\bibitem [{\citenamefont {Agnese}\ \emph
  {et~al.}(2014{\natexlab{a}})\citenamefont {Agnese}, \citenamefont {Anderson},
  \citenamefont {Asai}, \citenamefont {Balakishiyeva}, \citenamefont
  {Basu~Thakur}, \citenamefont {Bauer}, \citenamefont {Beaty}, \citenamefont
  {Billard}, \citenamefont {Borgland}, \citenamefont {Bowles}, \citenamefont
  {Brandt}, \citenamefont {Brink}, \citenamefont {Bunker}, \citenamefont
  {Cabrera}, \citenamefont {Caldwell}, \citenamefont {Cerdeno}, \citenamefont
  {Chagani}, \citenamefont {Chen}, \citenamefont {Cherry}, \citenamefont
  {Cooley}, \citenamefont {Cornell}, \citenamefont {Crewdson}, \citenamefont
  {Cushman}, \citenamefont {Daal}, \citenamefont {DeVaney}, \citenamefont
  {Di~Stefano}, \citenamefont {Silva}, \citenamefont {Doughty}, \citenamefont
  {Esteban}, \citenamefont {Fallows}, \citenamefont {Figueroa-Feliciano},
  \citenamefont {Godfrey}, \citenamefont {Golwala}, \citenamefont {Hall},
  \citenamefont {Hansen}, \citenamefont {Harris}, \citenamefont {Hertel},
  \citenamefont {Hines}, \citenamefont {Hofer}, \citenamefont {Holmgren},
  \citenamefont {Hsu}, \citenamefont {Huber}, \citenamefont {Jastram},
  \citenamefont {Kamaev}, \citenamefont {Kara}, \citenamefont {Kelsey},
  \citenamefont {Kenany}, \citenamefont {Kennedy}, \citenamefont {Kiveni},
  \citenamefont {Koch}, \citenamefont {Leder}, \citenamefont {Loer},
  \citenamefont {Lopez~Asamar}, \citenamefont {Mahapatra}, \citenamefont
  {Mandic}, \citenamefont {Martinez}, \citenamefont {McCarthy}, \citenamefont
  {Mirabolfathi}, \citenamefont {Moffatt}, \citenamefont {Nelson},
  \citenamefont {Novak}, \citenamefont {Page}, \citenamefont {Partridge},
  \citenamefont {Pepin}, \citenamefont {Phipps}, \citenamefont {Platt},
  \citenamefont {Prasad}, \citenamefont {Pyle}, \citenamefont {Qiu},
  \citenamefont {Rau}, \citenamefont {Redl}, \citenamefont {Reisetter},
  \citenamefont {Resch}, \citenamefont {Ricci}, \citenamefont {Ruschman},
  \citenamefont {Saab}, \citenamefont {Sadoulet}, \citenamefont {Sander},
  \citenamefont {Schmitt}, \citenamefont {Schneck}, \citenamefont {Schnee},
  \citenamefont {Scorza}, \citenamefont {Seitz}, \citenamefont {Serfass},
  \citenamefont {Shank}, \citenamefont {Speller}, \citenamefont {Tomada},
  \citenamefont {Upadhyayula}, \citenamefont {Villano}, \citenamefont
  {Welliver}, \citenamefont {Wright}, \citenamefont {Yellin}, \citenamefont
  {Yen}, \citenamefont {Young},\ and\ \citenamefont {Zhang}}]{scdms2016}%
  \BibitemOpen
  \bibfield  {author} {\bibinfo {author} {\bibfnamefont {R.}~\bibnamefont
  {Agnese}}, \bibinfo {author} {\bibfnamefont {A.~J.}\ \bibnamefont
  {Anderson}}, \bibinfo {author} {\bibfnamefont {M.}~\bibnamefont {Asai}},
  \bibinfo {author} {\bibfnamefont {D.}~\bibnamefont {Balakishiyeva}}, \bibinfo
  {author} {\bibfnamefont {R.}~\bibnamefont {Basu~Thakur}}, \bibinfo {author}
  {\bibfnamefont {D.~A.}\ \bibnamefont {Bauer}}, \bibinfo {author}
  {\bibfnamefont {J.}~\bibnamefont {Beaty}}, \bibinfo {author} {\bibfnamefont
  {J.}~\bibnamefont {Billard}}, \bibinfo {author} {\bibfnamefont
  {A.}~\bibnamefont {Borgland}}, \bibinfo {author} {\bibfnamefont {M.~A.}\
  \bibnamefont {Bowles}}, \bibinfo {author} {\bibfnamefont {D.}~\bibnamefont
  {Brandt}}, \bibinfo {author} {\bibfnamefont {P.~L.}\ \bibnamefont {Brink}},
  \bibinfo {author} {\bibfnamefont {R.}~\bibnamefont {Bunker}}, \bibinfo
  {author} {\bibfnamefont {B.}~\bibnamefont {Cabrera}}, \bibinfo {author}
  {\bibfnamefont {D.~O.}\ \bibnamefont {Caldwell}}, \bibinfo {author}
  {\bibfnamefont {D.~G.}\ \bibnamefont {Cerdeno}}, \bibinfo {author}
  {\bibfnamefont {H.}~\bibnamefont {Chagani}}, \bibinfo {author} {\bibfnamefont
  {Y.}~\bibnamefont {Chen}}, \bibinfo {author} {\bibfnamefont {M.}~\bibnamefont
  {Cherry}}, \bibinfo {author} {\bibfnamefont {J.}~\bibnamefont {Cooley}},
  \bibinfo {author} {\bibfnamefont {B.}~\bibnamefont {Cornell}}, \bibinfo
  {author} {\bibfnamefont {C.~H.}\ \bibnamefont {Crewdson}}, \bibinfo {author}
  {\bibfnamefont {P.}~\bibnamefont {Cushman}}, \bibinfo {author} {\bibfnamefont
  {M.}~\bibnamefont {Daal}}, \bibinfo {author} {\bibfnamefont {D.}~\bibnamefont
  {DeVaney}}, \bibinfo {author} {\bibfnamefont {P.~C.~F.}\ \bibnamefont
  {Di~Stefano}}, \bibinfo {author} {\bibfnamefont {E.~D. C.~E.}\ \bibnamefont
  {Silva}}, \bibinfo {author} {\bibfnamefont {T.}~\bibnamefont {Doughty}},
  \bibinfo {author} {\bibfnamefont {L.}~\bibnamefont {Esteban}}, \bibinfo
  {author} {\bibfnamefont {S.}~\bibnamefont {Fallows}}, \bibinfo {author}
  {\bibfnamefont {E.}~\bibnamefont {Figueroa-Feliciano}}, \bibinfo {author}
  {\bibfnamefont {G.~L.}\ \bibnamefont {Godfrey}}, \bibinfo {author}
  {\bibfnamefont {S.~R.}\ \bibnamefont {Golwala}}, \bibinfo {author}
  {\bibfnamefont {J.}~\bibnamefont {Hall}}, \bibinfo {author} {\bibfnamefont
  {S.}~\bibnamefont {Hansen}}, \bibinfo {author} {\bibfnamefont {H.~R.}\
  \bibnamefont {Harris}}, \bibinfo {author} {\bibfnamefont {S.~A.}\
  \bibnamefont {Hertel}}, \bibinfo {author} {\bibfnamefont {B.~A.}\
  \bibnamefont {Hines}}, \bibinfo {author} {\bibfnamefont {T.}~\bibnamefont
  {Hofer}}, \bibinfo {author} {\bibfnamefont {D.}~\bibnamefont {Holmgren}},
  \bibinfo {author} {\bibfnamefont {L.}~\bibnamefont {Hsu}}, \bibinfo {author}
  {\bibfnamefont {M.~E.}\ \bibnamefont {Huber}}, \bibinfo {author}
  {\bibfnamefont {A.}~\bibnamefont {Jastram}}, \bibinfo {author} {\bibfnamefont
  {O.}~\bibnamefont {Kamaev}}, \bibinfo {author} {\bibfnamefont
  {B.}~\bibnamefont {Kara}}, \bibinfo {author} {\bibfnamefont {M.~H.}\
  \bibnamefont {Kelsey}}, \bibinfo {author} {\bibfnamefont {S.}~\bibnamefont
  {Kenany}}, \bibinfo {author} {\bibfnamefont {A.}~\bibnamefont {Kennedy}},
  \bibinfo {author} {\bibfnamefont {M.}~\bibnamefont {Kiveni}}, \bibinfo
  {author} {\bibfnamefont {K.}~\bibnamefont {Koch}}, \bibinfo {author}
  {\bibfnamefont {A.}~\bibnamefont {Leder}}, \bibinfo {author} {\bibfnamefont
  {B.}~\bibnamefont {Loer}}, \bibinfo {author} {\bibfnamefont {E.}~\bibnamefont
  {Lopez~Asamar}}, \bibinfo {author} {\bibfnamefont {R.}~\bibnamefont
  {Mahapatra}}, \bibinfo {author} {\bibfnamefont {V.}~\bibnamefont {Mandic}},
  \bibinfo {author} {\bibfnamefont {C.}~\bibnamefont {Martinez}}, \bibinfo
  {author} {\bibfnamefont {K.~A.}\ \bibnamefont {McCarthy}}, \bibinfo {author}
  {\bibfnamefont {N.}~\bibnamefont {Mirabolfathi}}, \bibinfo {author}
  {\bibfnamefont {R.~A.}\ \bibnamefont {Moffatt}}, \bibinfo {author}
  {\bibfnamefont {R.~H.}\ \bibnamefont {Nelson}}, \bibinfo {author}
  {\bibfnamefont {L.}~\bibnamefont {Novak}}, \bibinfo {author} {\bibfnamefont
  {K.}~\bibnamefont {Page}}, \bibinfo {author} {\bibfnamefont {R.}~\bibnamefont
  {Partridge}}, \bibinfo {author} {\bibfnamefont {M.}~\bibnamefont {Pepin}},
  \bibinfo {author} {\bibfnamefont {A.}~\bibnamefont {Phipps}}, \bibinfo
  {author} {\bibfnamefont {M.}~\bibnamefont {Platt}}, \bibinfo {author}
  {\bibfnamefont {K.}~\bibnamefont {Prasad}}, \bibinfo {author} {\bibfnamefont
  {M.}~\bibnamefont {Pyle}}, \bibinfo {author} {\bibfnamefont {H.}~\bibnamefont
  {Qiu}}, \bibinfo {author} {\bibfnamefont {W.}~\bibnamefont {Rau}}, \bibinfo
  {author} {\bibfnamefont {P.}~\bibnamefont {Redl}}, \bibinfo {author}
  {\bibfnamefont {A.}~\bibnamefont {Reisetter}}, \bibinfo {author}
  {\bibfnamefont {R.~W.}\ \bibnamefont {Resch}}, \bibinfo {author}
  {\bibfnamefont {Y.}~\bibnamefont {Ricci}}, \bibinfo {author} {\bibfnamefont
  {M.}~\bibnamefont {Ruschman}}, \bibinfo {author} {\bibfnamefont
  {T.}~\bibnamefont {Saab}}, \bibinfo {author} {\bibfnamefont {B.}~\bibnamefont
  {Sadoulet}}, \bibinfo {author} {\bibfnamefont {J.}~\bibnamefont {Sander}},
  \bibinfo {author} {\bibfnamefont {R.~L.}\ \bibnamefont {Schmitt}}, \bibinfo
  {author} {\bibfnamefont {K.}~\bibnamefont {Schneck}}, \bibinfo {author}
  {\bibfnamefont {R.~W.}\ \bibnamefont {Schnee}}, \bibinfo {author}
  {\bibfnamefont {S.}~\bibnamefont {Scorza}}, \bibinfo {author} {\bibfnamefont
  {D.~N.}\ \bibnamefont {Seitz}}, \bibinfo {author} {\bibfnamefont
  {B.}~\bibnamefont {Serfass}}, \bibinfo {author} {\bibfnamefont
  {B.}~\bibnamefont {Shank}}, \bibinfo {author} {\bibfnamefont
  {D.}~\bibnamefont {Speller}}, \bibinfo {author} {\bibfnamefont
  {A.}~\bibnamefont {Tomada}}, \bibinfo {author} {\bibfnamefont
  {S.}~\bibnamefont {Upadhyayula}}, \bibinfo {author} {\bibfnamefont {A.~N.}\
  \bibnamefont {Villano}}, \bibinfo {author} {\bibfnamefont {B.}~\bibnamefont
  {Welliver}}, \bibinfo {author} {\bibfnamefont {D.~H.}\ \bibnamefont
  {Wright}}, \bibinfo {author} {\bibfnamefont {S.}~\bibnamefont {Yellin}},
  \bibinfo {author} {\bibfnamefont {J.~J.}\ \bibnamefont {Yen}}, \bibinfo
  {author} {\bibfnamefont {B.~A.}\ \bibnamefont {Young}}, \ and\ \bibinfo
  {author} {\bibfnamefont {J.}~\bibnamefont {Zhang}} (\bibinfo {collaboration}
  {SuperCDMS Collaboration}),\ }\href {\doibase 10.1103/PhysRevLett.112.241302}
  {\bibfield  {journal} {\bibinfo  {journal} {Phys. Rev. Lett.}\ }\textbf
  {\bibinfo {volume} {112}},\ \bibinfo {pages} {241302} (\bibinfo {year}
  {2014}{\natexlab{a}})}\BibitemShut {NoStop}%
\bibitem [{\citenamefont {Agnese}\ \emph
  {et~al.}(2014{\natexlab{b}})\citenamefont {Agnese}, \citenamefont {Anderson},
  \citenamefont {Asai}, \citenamefont {Balakishiyeva}, \citenamefont
  {Basu~Thakur}, \citenamefont {Bauer}, \citenamefont {Billard}, \citenamefont
  {Borgland}, \citenamefont {Bowles}, \citenamefont {Brandt}, \citenamefont
  {Brink}, \citenamefont {Bunker}, \citenamefont {Cabrera}, \citenamefont
  {Caldwell}, \citenamefont {Cerdeno}, \citenamefont {Chagani}, \citenamefont
  {Cooley}, \citenamefont {Cornell}, \citenamefont {Crewdson}, \citenamefont
  {Cushman}, \citenamefont {Daal}, \citenamefont {Di~Stefano}, \citenamefont
  {Doughty}, \citenamefont {Esteban}, \citenamefont {Fallows}, \citenamefont
  {Figueroa-Feliciano}, \citenamefont {Godfrey}, \citenamefont {Golwala},
  \citenamefont {Hall}, \citenamefont {Harris}, \citenamefont {Hertel},
  \citenamefont {Hofer}, \citenamefont {Holmgren}, \citenamefont {Hsu},
  \citenamefont {Huber}, \citenamefont {Jastram}, \citenamefont {Kamaev},
  \citenamefont {Kara}, \citenamefont {Kelsey}, \citenamefont {Kennedy},
  \citenamefont {Kiveni}, \citenamefont {Koch}, \citenamefont {Loer},
  \citenamefont {Lopez~Asamar}, \citenamefont {Mahapatra}, \citenamefont
  {Mandic}, \citenamefont {Martinez}, \citenamefont {McCarthy}, \citenamefont
  {Mirabolfathi}, \citenamefont {Moffatt}, \citenamefont {Moore}, \citenamefont
  {Nadeau}, \citenamefont {Nelson}, \citenamefont {Page}, \citenamefont
  {Partridge}, \citenamefont {Pepin}, \citenamefont {Phipps}, \citenamefont
  {Prasad}, \citenamefont {Pyle}, \citenamefont {Qiu}, \citenamefont {Rau},
  \citenamefont {Redl}, \citenamefont {Reisetter}, \citenamefont {Ricci},
  \citenamefont {Saab}, \citenamefont {Sadoulet}, \citenamefont {Sander},
  \citenamefont {Schneck}, \citenamefont {Schnee}, \citenamefont {Scorza},
  \citenamefont {Serfass}, \citenamefont {Shank}, \citenamefont {Speller},
  \citenamefont {Villano}, \citenamefont {Welliver}, \citenamefont {Wright},
  \citenamefont {Yellin}, \citenamefont {Yen}, \citenamefont {Young},\ and\
  \citenamefont {Zhang}}]{Agnese:2013jaa}%
  \BibitemOpen
  \bibfield  {author} {\bibinfo {author} {\bibfnamefont {R.}~\bibnamefont
  {Agnese}}, \bibinfo {author} {\bibfnamefont {A.~J.}\ \bibnamefont
  {Anderson}}, \bibinfo {author} {\bibfnamefont {M.}~\bibnamefont {Asai}},
  \bibinfo {author} {\bibfnamefont {D.}~\bibnamefont {Balakishiyeva}}, \bibinfo
  {author} {\bibfnamefont {R.}~\bibnamefont {Basu~Thakur}}, \bibinfo {author}
  {\bibfnamefont {D.~A.}\ \bibnamefont {Bauer}}, \bibinfo {author}
  {\bibfnamefont {J.}~\bibnamefont {Billard}}, \bibinfo {author} {\bibfnamefont
  {A.}~\bibnamefont {Borgland}}, \bibinfo {author} {\bibfnamefont {M.~A.}\
  \bibnamefont {Bowles}}, \bibinfo {author} {\bibfnamefont {D.}~\bibnamefont
  {Brandt}}, \bibinfo {author} {\bibfnamefont {P.~L.}\ \bibnamefont {Brink}},
  \bibinfo {author} {\bibfnamefont {R.}~\bibnamefont {Bunker}}, \bibinfo
  {author} {\bibfnamefont {B.}~\bibnamefont {Cabrera}}, \bibinfo {author}
  {\bibfnamefont {D.~O.}\ \bibnamefont {Caldwell}}, \bibinfo {author}
  {\bibfnamefont {D.~G.}\ \bibnamefont {Cerdeno}}, \bibinfo {author}
  {\bibfnamefont {H.}~\bibnamefont {Chagani}}, \bibinfo {author} {\bibfnamefont
  {J.}~\bibnamefont {Cooley}}, \bibinfo {author} {\bibfnamefont
  {B.}~\bibnamefont {Cornell}}, \bibinfo {author} {\bibfnamefont {C.~H.}\
  \bibnamefont {Crewdson}}, \bibinfo {author} {\bibfnamefont {P.}~\bibnamefont
  {Cushman}}, \bibinfo {author} {\bibfnamefont {M.}~\bibnamefont {Daal}},
  \bibinfo {author} {\bibfnamefont {P.~C.~F.}\ \bibnamefont {Di~Stefano}},
  \bibinfo {author} {\bibfnamefont {T.}~\bibnamefont {Doughty}}, \bibinfo
  {author} {\bibfnamefont {L.}~\bibnamefont {Esteban}}, \bibinfo {author}
  {\bibfnamefont {S.}~\bibnamefont {Fallows}}, \bibinfo {author} {\bibfnamefont
  {E.}~\bibnamefont {Figueroa-Feliciano}}, \bibinfo {author} {\bibfnamefont
  {G.~L.}\ \bibnamefont {Godfrey}}, \bibinfo {author} {\bibfnamefont {S.~R.}\
  \bibnamefont {Golwala}}, \bibinfo {author} {\bibfnamefont {J.}~\bibnamefont
  {Hall}}, \bibinfo {author} {\bibfnamefont {H.~R.}\ \bibnamefont {Harris}},
  \bibinfo {author} {\bibfnamefont {S.~A.}\ \bibnamefont {Hertel}}, \bibinfo
  {author} {\bibfnamefont {T.}~\bibnamefont {Hofer}}, \bibinfo {author}
  {\bibfnamefont {D.}~\bibnamefont {Holmgren}}, \bibinfo {author}
  {\bibfnamefont {L.}~\bibnamefont {Hsu}}, \bibinfo {author} {\bibfnamefont
  {M.~E.}\ \bibnamefont {Huber}}, \bibinfo {author} {\bibfnamefont
  {A.}~\bibnamefont {Jastram}}, \bibinfo {author} {\bibfnamefont
  {O.}~\bibnamefont {Kamaev}}, \bibinfo {author} {\bibfnamefont
  {B.}~\bibnamefont {Kara}}, \bibinfo {author} {\bibfnamefont {M.~H.}\
  \bibnamefont {Kelsey}}, \bibinfo {author} {\bibfnamefont {A.}~\bibnamefont
  {Kennedy}}, \bibinfo {author} {\bibfnamefont {M.}~\bibnamefont {Kiveni}},
  \bibinfo {author} {\bibfnamefont {K.}~\bibnamefont {Koch}}, \bibinfo {author}
  {\bibfnamefont {B.}~\bibnamefont {Loer}}, \bibinfo {author} {\bibfnamefont
  {E.}~\bibnamefont {Lopez~Asamar}}, \bibinfo {author} {\bibfnamefont
  {R.}~\bibnamefont {Mahapatra}}, \bibinfo {author} {\bibfnamefont
  {V.}~\bibnamefont {Mandic}}, \bibinfo {author} {\bibfnamefont
  {C.}~\bibnamefont {Martinez}}, \bibinfo {author} {\bibfnamefont {K.~A.}\
  \bibnamefont {McCarthy}}, \bibinfo {author} {\bibfnamefont {N.}~\bibnamefont
  {Mirabolfathi}}, \bibinfo {author} {\bibfnamefont {R.~A.}\ \bibnamefont
  {Moffatt}}, \bibinfo {author} {\bibfnamefont {D.~C.}\ \bibnamefont {Moore}},
  \bibinfo {author} {\bibfnamefont {P.}~\bibnamefont {Nadeau}}, \bibinfo
  {author} {\bibfnamefont {R.~H.}\ \bibnamefont {Nelson}}, \bibinfo {author}
  {\bibfnamefont {K.}~\bibnamefont {Page}}, \bibinfo {author} {\bibfnamefont
  {R.}~\bibnamefont {Partridge}}, \bibinfo {author} {\bibfnamefont
  {M.}~\bibnamefont {Pepin}}, \bibinfo {author} {\bibfnamefont
  {A.}~\bibnamefont {Phipps}}, \bibinfo {author} {\bibfnamefont
  {K.}~\bibnamefont {Prasad}}, \bibinfo {author} {\bibfnamefont
  {M.}~\bibnamefont {Pyle}}, \bibinfo {author} {\bibfnamefont {H.}~\bibnamefont
  {Qiu}}, \bibinfo {author} {\bibfnamefont {W.}~\bibnamefont {Rau}}, \bibinfo
  {author} {\bibfnamefont {P.}~\bibnamefont {Redl}}, \bibinfo {author}
  {\bibfnamefont {A.}~\bibnamefont {Reisetter}}, \bibinfo {author}
  {\bibfnamefont {Y.}~\bibnamefont {Ricci}}, \bibinfo {author} {\bibfnamefont
  {T.}~\bibnamefont {Saab}}, \bibinfo {author} {\bibfnamefont {B.}~\bibnamefont
  {Sadoulet}}, \bibinfo {author} {\bibfnamefont {J.}~\bibnamefont {Sander}},
  \bibinfo {author} {\bibfnamefont {K.}~\bibnamefont {Schneck}}, \bibinfo
  {author} {\bibfnamefont {R.~W.}\ \bibnamefont {Schnee}}, \bibinfo {author}
  {\bibfnamefont {S.}~\bibnamefont {Scorza}}, \bibinfo {author} {\bibfnamefont
  {B.}~\bibnamefont {Serfass}}, \bibinfo {author} {\bibfnamefont
  {B.}~\bibnamefont {Shank}}, \bibinfo {author} {\bibfnamefont
  {D.}~\bibnamefont {Speller}}, \bibinfo {author} {\bibfnamefont {A.~N.}\
  \bibnamefont {Villano}}, \bibinfo {author} {\bibfnamefont {B.}~\bibnamefont
  {Welliver}}, \bibinfo {author} {\bibfnamefont {D.~H.}\ \bibnamefont
  {Wright}}, \bibinfo {author} {\bibfnamefont {S.}~\bibnamefont {Yellin}},
  \bibinfo {author} {\bibfnamefont {J.~J.}\ \bibnamefont {Yen}}, \bibinfo
  {author} {\bibfnamefont {B.~A.}\ \bibnamefont {Young}}, \ and\ \bibinfo
  {author} {\bibfnamefont {J.}~\bibnamefont {Zhang}} (\bibinfo {collaboration}
  {SuperCDMS collaboration}),\ }\href {\doibase 10.1103/PhysRevLett.112.041302}
  {\bibfield  {journal} {\bibinfo  {journal} {Phys. Rev. Lett.}\ }\textbf
  {\bibinfo {volume} {112}},\ \bibinfo {pages} {041302} (\bibinfo {year}
  {2014}{\natexlab{b}})}\BibitemShut {NoStop}%
\bibitem [{\citenamefont {Agnese}\ \emph {et~al.}(2016)\citenamefont {Agnese},
  \citenamefont {Anderson}, \citenamefont {Aramaki}, \citenamefont {Asai},
  \citenamefont {Baker}, \citenamefont {Balakishiyeva}, \citenamefont {Barker},
  \citenamefont {Basu~Thakur}, \citenamefont {Bauer}, \citenamefont {Billard},
  \citenamefont {Borgland}, \citenamefont {Bowles}, \citenamefont {Brink},
  \citenamefont {Bunker}, \citenamefont {Cabrera}, \citenamefont {Caldwell},
  \citenamefont {Calkins}, \citenamefont {Cerdeno}, \citenamefont {Chagani},
  \citenamefont {Chen}, \citenamefont {Cooley}, \citenamefont {Cornell},
  \citenamefont {Cushman}, \citenamefont {Daal}, \citenamefont {Di~Stefano},
  \citenamefont {Doughty}, \citenamefont {Esteban}, \citenamefont {Fallows},
  \citenamefont {Figueroa-Feliciano}, \citenamefont {Ghaith}, \citenamefont
  {Godfrey}, \citenamefont {Golwala}, \citenamefont {Hall}, \citenamefont
  {Harris}, \citenamefont {Hofer}, \citenamefont {Holmgren}, \citenamefont
  {Hsu}, \citenamefont {Huber}, \citenamefont {Jardin}, \citenamefont
  {Jastram}, \citenamefont {Kamaev}, \citenamefont {Kara}, \citenamefont
  {Kelsey}, \citenamefont {Kennedy}, \citenamefont {Leder}, \citenamefont
  {Loer}, \citenamefont {Lopez~Asamar}, \citenamefont {Lukens}, \citenamefont
  {Mahapatra}, \citenamefont {Mandic}, \citenamefont {Mast}, \citenamefont
  {Mirabolfathi}, \citenamefont {Moffatt}, \citenamefont {Morales~Mendoza},
  \citenamefont {Oser}, \citenamefont {Page}, \citenamefont {Page},
  \citenamefont {Partridge}, \citenamefont {Pepin}, \citenamefont {Phipps},
  \citenamefont {Prasad}, \citenamefont {Pyle}, \citenamefont {Qiu},
  \citenamefont {Rau}, \citenamefont {Redl}, \citenamefont {Reisetter},
  \citenamefont {Ricci}, \citenamefont {Roberts}, \citenamefont {Rogers},
  \citenamefont {Saab}, \citenamefont {Sadoulet}, \citenamefont {Sander},
  \citenamefont {Schneck}, \citenamefont {Schnee}, \citenamefont {Scorza},
  \citenamefont {Serfass}, \citenamefont {Shank}, \citenamefont {Speller},
  \citenamefont {Toback}, \citenamefont {Underwood}, \citenamefont
  {Upadhyayula}, \citenamefont {Villano}, \citenamefont {Welliver},
  \citenamefont {Wilson}, \citenamefont {Wright}, \citenamefont {Yellin},
  \citenamefont {Yen}, \citenamefont {Young},\ and\ \citenamefont
  {Zhang}}]{Agnese:2015nto}%
  \BibitemOpen
  \bibfield  {author} {\bibinfo {author} {\bibfnamefont {R.}~\bibnamefont
  {Agnese}}, \bibinfo {author} {\bibfnamefont {A.~J.}\ \bibnamefont
  {Anderson}}, \bibinfo {author} {\bibfnamefont {T.}~\bibnamefont {Aramaki}},
  \bibinfo {author} {\bibfnamefont {M.}~\bibnamefont {Asai}}, \bibinfo {author}
  {\bibfnamefont {W.}~\bibnamefont {Baker}}, \bibinfo {author} {\bibfnamefont
  {D.}~\bibnamefont {Balakishiyeva}}, \bibinfo {author} {\bibfnamefont
  {D.}~\bibnamefont {Barker}}, \bibinfo {author} {\bibfnamefont
  {R.}~\bibnamefont {Basu~Thakur}}, \bibinfo {author} {\bibfnamefont {D.~A.}\
  \bibnamefont {Bauer}}, \bibinfo {author} {\bibfnamefont {J.}~\bibnamefont
  {Billard}}, \bibinfo {author} {\bibfnamefont {A.}~\bibnamefont {Borgland}},
  \bibinfo {author} {\bibfnamefont {M.~A.}\ \bibnamefont {Bowles}}, \bibinfo
  {author} {\bibfnamefont {P.~L.}\ \bibnamefont {Brink}}, \bibinfo {author}
  {\bibfnamefont {R.}~\bibnamefont {Bunker}}, \bibinfo {author} {\bibfnamefont
  {B.}~\bibnamefont {Cabrera}}, \bibinfo {author} {\bibfnamefont {D.~O.}\
  \bibnamefont {Caldwell}}, \bibinfo {author} {\bibfnamefont {R.}~\bibnamefont
  {Calkins}}, \bibinfo {author} {\bibfnamefont {D.~G.}\ \bibnamefont
  {Cerdeno}}, \bibinfo {author} {\bibfnamefont {H.}~\bibnamefont {Chagani}},
  \bibinfo {author} {\bibfnamefont {Y.}~\bibnamefont {Chen}}, \bibinfo {author}
  {\bibfnamefont {J.}~\bibnamefont {Cooley}}, \bibinfo {author} {\bibfnamefont
  {B.}~\bibnamefont {Cornell}}, \bibinfo {author} {\bibfnamefont
  {P.}~\bibnamefont {Cushman}}, \bibinfo {author} {\bibfnamefont
  {M.}~\bibnamefont {Daal}}, \bibinfo {author} {\bibfnamefont {P.~C.~F.}\
  \bibnamefont {Di~Stefano}}, \bibinfo {author} {\bibfnamefont
  {T.}~\bibnamefont {Doughty}}, \bibinfo {author} {\bibfnamefont
  {L.}~\bibnamefont {Esteban}}, \bibinfo {author} {\bibfnamefont
  {S.}~\bibnamefont {Fallows}}, \bibinfo {author} {\bibfnamefont
  {E.}~\bibnamefont {Figueroa-Feliciano}}, \bibinfo {author} {\bibfnamefont
  {M.}~\bibnamefont {Ghaith}}, \bibinfo {author} {\bibfnamefont {G.~L.}\
  \bibnamefont {Godfrey}}, \bibinfo {author} {\bibfnamefont {S.~R.}\
  \bibnamefont {Golwala}}, \bibinfo {author} {\bibfnamefont {J.}~\bibnamefont
  {Hall}}, \bibinfo {author} {\bibfnamefont {H.~R.}\ \bibnamefont {Harris}},
  \bibinfo {author} {\bibfnamefont {T.}~\bibnamefont {Hofer}}, \bibinfo
  {author} {\bibfnamefont {D.}~\bibnamefont {Holmgren}}, \bibinfo {author}
  {\bibfnamefont {L.}~\bibnamefont {Hsu}}, \bibinfo {author} {\bibfnamefont
  {M.~E.}\ \bibnamefont {Huber}}, \bibinfo {author} {\bibfnamefont
  {D.}~\bibnamefont {Jardin}}, \bibinfo {author} {\bibfnamefont
  {A.}~\bibnamefont {Jastram}}, \bibinfo {author} {\bibfnamefont
  {O.}~\bibnamefont {Kamaev}}, \bibinfo {author} {\bibfnamefont
  {B.}~\bibnamefont {Kara}}, \bibinfo {author} {\bibfnamefont {M.~H.}\
  \bibnamefont {Kelsey}}, \bibinfo {author} {\bibfnamefont {A.}~\bibnamefont
  {Kennedy}}, \bibinfo {author} {\bibfnamefont {A.}~\bibnamefont {Leder}},
  \bibinfo {author} {\bibfnamefont {B.}~\bibnamefont {Loer}}, \bibinfo {author}
  {\bibfnamefont {E.}~\bibnamefont {Lopez~Asamar}}, \bibinfo {author}
  {\bibfnamefont {P.}~\bibnamefont {Lukens}}, \bibinfo {author} {\bibfnamefont
  {R.}~\bibnamefont {Mahapatra}}, \bibinfo {author} {\bibfnamefont
  {V.}~\bibnamefont {Mandic}}, \bibinfo {author} {\bibfnamefont
  {N.}~\bibnamefont {Mast}}, \bibinfo {author} {\bibfnamefont {N.}~\bibnamefont
  {Mirabolfathi}}, \bibinfo {author} {\bibfnamefont {R.~A.}\ \bibnamefont
  {Moffatt}}, \bibinfo {author} {\bibfnamefont {J.~D.}\ \bibnamefont
  {Morales~Mendoza}}, \bibinfo {author} {\bibfnamefont {S.~M.}\ \bibnamefont
  {Oser}}, \bibinfo {author} {\bibfnamefont {K.}~\bibnamefont {Page}}, \bibinfo
  {author} {\bibfnamefont {W.~A.}\ \bibnamefont {Page}}, \bibinfo {author}
  {\bibfnamefont {R.}~\bibnamefont {Partridge}}, \bibinfo {author}
  {\bibfnamefont {M.}~\bibnamefont {Pepin}}, \bibinfo {author} {\bibfnamefont
  {A.}~\bibnamefont {Phipps}}, \bibinfo {author} {\bibfnamefont
  {K.}~\bibnamefont {Prasad}}, \bibinfo {author} {\bibfnamefont
  {M.}~\bibnamefont {Pyle}}, \bibinfo {author} {\bibfnamefont {H.}~\bibnamefont
  {Qiu}}, \bibinfo {author} {\bibfnamefont {W.}~\bibnamefont {Rau}}, \bibinfo
  {author} {\bibfnamefont {P.}~\bibnamefont {Redl}}, \bibinfo {author}
  {\bibfnamefont {A.}~\bibnamefont {Reisetter}}, \bibinfo {author}
  {\bibfnamefont {Y.}~\bibnamefont {Ricci}}, \bibinfo {author} {\bibfnamefont
  {A.}~\bibnamefont {Roberts}}, \bibinfo {author} {\bibfnamefont {H.~E.}\
  \bibnamefont {Rogers}}, \bibinfo {author} {\bibfnamefont {T.}~\bibnamefont
  {Saab}}, \bibinfo {author} {\bibfnamefont {B.}~\bibnamefont {Sadoulet}},
  \bibinfo {author} {\bibfnamefont {J.}~\bibnamefont {Sander}}, \bibinfo
  {author} {\bibfnamefont {K.}~\bibnamefont {Schneck}}, \bibinfo {author}
  {\bibfnamefont {R.~W.}\ \bibnamefont {Schnee}}, \bibinfo {author}
  {\bibfnamefont {S.}~\bibnamefont {Scorza}}, \bibinfo {author} {\bibfnamefont
  {B.}~\bibnamefont {Serfass}}, \bibinfo {author} {\bibfnamefont
  {B.}~\bibnamefont {Shank}}, \bibinfo {author} {\bibfnamefont
  {D.}~\bibnamefont {Speller}}, \bibinfo {author} {\bibfnamefont
  {D.}~\bibnamefont {Toback}}, \bibinfo {author} {\bibfnamefont
  {R.}~\bibnamefont {Underwood}}, \bibinfo {author} {\bibfnamefont
  {S.}~\bibnamefont {Upadhyayula}}, \bibinfo {author} {\bibfnamefont {A.~N.}\
  \bibnamefont {Villano}}, \bibinfo {author} {\bibfnamefont {B.}~\bibnamefont
  {Welliver}}, \bibinfo {author} {\bibfnamefont {J.~S.}\ \bibnamefont
  {Wilson}}, \bibinfo {author} {\bibfnamefont {D.~H.}\ \bibnamefont {Wright}},
  \bibinfo {author} {\bibfnamefont {S.}~\bibnamefont {Yellin}}, \bibinfo
  {author} {\bibfnamefont {J.~J.}\ \bibnamefont {Yen}}, \bibinfo {author}
  {\bibfnamefont {B.~A.}\ \bibnamefont {Young}}, \ and\ \bibinfo {author}
  {\bibfnamefont {J.}~\bibnamefont {Zhang}} (\bibinfo {collaboration}
  {SuperCDMS Collaboration}),\ }\href {\doibase 10.1103/PhysRevLett.116.071301}
  {\bibfield  {journal} {\bibinfo  {journal} {Phys. Rev. Lett.}\ }\textbf
  {\bibinfo {volume} {116}},\ \bibinfo {pages} {071301} (\bibinfo {year}
  {2016})}\BibitemShut {NoStop}%
\bibitem [{\citenamefont {Agnese}\ \emph
  {et~al.}(2018{\natexlab{a}})\citenamefont {Agnese}, \citenamefont {Anderson},
  \citenamefont {Aralis}, \citenamefont {Aramaki}, \citenamefont {Arnquist},
  \citenamefont {Baker}, \citenamefont {Balakishiyeva}, \citenamefont {Barker},
  \citenamefont {Basu~Thakur}, \citenamefont {Bauer}, \citenamefont {Binder},
  \citenamefont {Bowles}, \citenamefont {Brink}, \citenamefont {Bunker},
  \citenamefont {Cabrera}, \citenamefont {Caldwell}, \citenamefont {Calkins},
  \citenamefont {Cartaro}, \citenamefont {Cerde\~no}, \citenamefont {Chang},
  \citenamefont {Chagani}, \citenamefont {Chen}, \citenamefont {Cooley},
  \citenamefont {Cornell}, \citenamefont {Cushman}, \citenamefont {Daal},
  \citenamefont {Di~Stefano}, \citenamefont {Doughty}, \citenamefont {Esteban},
  \citenamefont {Fascione}, \citenamefont {Figueroa-Feliciano}, \citenamefont
  {Fritts}, \citenamefont {Gerbier}, \citenamefont {Ghaith}, \citenamefont
  {Godfrey}, \citenamefont {Golwala}, \citenamefont {Hall}, \citenamefont
  {Harris}, \citenamefont {Hong}, \citenamefont {Hoppe}, \citenamefont {Hsu},
  \citenamefont {Huber}, \citenamefont {Iyer}, \citenamefont {Jardin},
  \citenamefont {Jastram}, \citenamefont {Jena}, \citenamefont {Kelsey},
  \citenamefont {Kennedy}, \citenamefont {Kubik}, \citenamefont {Kurinsky},
  \citenamefont {Leder}, \citenamefont {Loer}, \citenamefont {Lopez~Asamar},
  \citenamefont {Lukens}, \citenamefont {MacDonell}, \citenamefont {Mahapatra},
  \citenamefont {Mandic}, \citenamefont {Mast}, \citenamefont {Miller},
  \citenamefont {Mirabolfathi}, \citenamefont {Moffatt}, \citenamefont
  {Mohanty}, \citenamefont {Morales~Mendoza}, \citenamefont {Nelson},
  \citenamefont {Orrell}, \citenamefont {Oser}, \citenamefont {Page},
  \citenamefont {Page}, \citenamefont {Partridge}, \citenamefont {Pepin},
  \citenamefont {Pe\~nalver Martinez}, \citenamefont {Phipps}, \citenamefont
  {Poudel}, \citenamefont {Pyle}, \citenamefont {Qiu}, \citenamefont {Rau},
  \citenamefont {Redl}, \citenamefont {Reisetter}, \citenamefont {Reynolds},
  \citenamefont {Roberts}, \citenamefont {Robinson}, \citenamefont {Rogers},
  \citenamefont {Saab}, \citenamefont {Sadoulet}, \citenamefont {Sander},
  \citenamefont {Schneck}, \citenamefont {Schnee}, \citenamefont {Scorza},
  \citenamefont {Senapati}, \citenamefont {Serfass}, \citenamefont {Speller},
  \citenamefont {Stein}, \citenamefont {Street}, \citenamefont {Tanaka},
  \citenamefont {Toback}, \citenamefont {Underwood}, \citenamefont {Villano},
  \citenamefont {von Krosigk}, \citenamefont {Welliver}, \citenamefont
  {Wilson}, \citenamefont {Wilson}, \citenamefont {Wright}, \citenamefont
  {Yellin}, \citenamefont {Yen}, \citenamefont {Young}, \citenamefont {Zhang},\
  and\ \citenamefont {Zhao}}]{Agnese:2017jvy}%
  \BibitemOpen
  \bibfield  {author} {\bibinfo {author} {\bibfnamefont {R.}~\bibnamefont
  {Agnese}}, \bibinfo {author} {\bibfnamefont {A.~J.}\ \bibnamefont
  {Anderson}}, \bibinfo {author} {\bibfnamefont {T.}~\bibnamefont {Aralis}},
  \bibinfo {author} {\bibfnamefont {T.}~\bibnamefont {Aramaki}}, \bibinfo
  {author} {\bibfnamefont {I.~J.}\ \bibnamefont {Arnquist}}, \bibinfo {author}
  {\bibfnamefont {W.}~\bibnamefont {Baker}}, \bibinfo {author} {\bibfnamefont
  {D.}~\bibnamefont {Balakishiyeva}}, \bibinfo {author} {\bibfnamefont
  {D.}~\bibnamefont {Barker}}, \bibinfo {author} {\bibfnamefont
  {R.}~\bibnamefont {Basu~Thakur}}, \bibinfo {author} {\bibfnamefont {D.~A.}\
  \bibnamefont {Bauer}}, \bibinfo {author} {\bibfnamefont {T.}~\bibnamefont
  {Binder}}, \bibinfo {author} {\bibfnamefont {M.~A.}\ \bibnamefont {Bowles}},
  \bibinfo {author} {\bibfnamefont {P.~L.}\ \bibnamefont {Brink}}, \bibinfo
  {author} {\bibfnamefont {R.}~\bibnamefont {Bunker}}, \bibinfo {author}
  {\bibfnamefont {B.}~\bibnamefont {Cabrera}}, \bibinfo {author} {\bibfnamefont
  {D.~O.}\ \bibnamefont {Caldwell}}, \bibinfo {author} {\bibfnamefont
  {R.}~\bibnamefont {Calkins}}, \bibinfo {author} {\bibfnamefont
  {C.}~\bibnamefont {Cartaro}}, \bibinfo {author} {\bibfnamefont {D.~G.}\
  \bibnamefont {Cerde\~no}}, \bibinfo {author} {\bibfnamefont {Y.}~\bibnamefont
  {Chang}}, \bibinfo {author} {\bibfnamefont {H.}~\bibnamefont {Chagani}},
  \bibinfo {author} {\bibfnamefont {Y.}~\bibnamefont {Chen}}, \bibinfo {author}
  {\bibfnamefont {J.}~\bibnamefont {Cooley}}, \bibinfo {author} {\bibfnamefont
  {B.}~\bibnamefont {Cornell}}, \bibinfo {author} {\bibfnamefont
  {P.}~\bibnamefont {Cushman}}, \bibinfo {author} {\bibfnamefont
  {M.}~\bibnamefont {Daal}}, \bibinfo {author} {\bibfnamefont {P.~C.~F.}\
  \bibnamefont {Di~Stefano}}, \bibinfo {author} {\bibfnamefont
  {T.}~\bibnamefont {Doughty}}, \bibinfo {author} {\bibfnamefont
  {L.}~\bibnamefont {Esteban}}, \bibinfo {author} {\bibfnamefont
  {E.}~\bibnamefont {Fascione}}, \bibinfo {author} {\bibfnamefont
  {E.}~\bibnamefont {Figueroa-Feliciano}}, \bibinfo {author} {\bibfnamefont
  {M.}~\bibnamefont {Fritts}}, \bibinfo {author} {\bibfnamefont
  {G.}~\bibnamefont {Gerbier}}, \bibinfo {author} {\bibfnamefont
  {M.}~\bibnamefont {Ghaith}}, \bibinfo {author} {\bibfnamefont {G.~L.}\
  \bibnamefont {Godfrey}}, \bibinfo {author} {\bibfnamefont {S.~R.}\
  \bibnamefont {Golwala}}, \bibinfo {author} {\bibfnamefont {J.}~\bibnamefont
  {Hall}}, \bibinfo {author} {\bibfnamefont {H.~R.}\ \bibnamefont {Harris}},
  \bibinfo {author} {\bibfnamefont {Z.}~\bibnamefont {Hong}}, \bibinfo {author}
  {\bibfnamefont {E.~W.}\ \bibnamefont {Hoppe}}, \bibinfo {author}
  {\bibfnamefont {L.}~\bibnamefont {Hsu}}, \bibinfo {author} {\bibfnamefont
  {M.~E.}\ \bibnamefont {Huber}}, \bibinfo {author} {\bibfnamefont
  {V.}~\bibnamefont {Iyer}}, \bibinfo {author} {\bibfnamefont {D.}~\bibnamefont
  {Jardin}}, \bibinfo {author} {\bibfnamefont {A.}~\bibnamefont {Jastram}},
  \bibinfo {author} {\bibfnamefont {C.}~\bibnamefont {Jena}}, \bibinfo {author}
  {\bibfnamefont {M.~H.}\ \bibnamefont {Kelsey}}, \bibinfo {author}
  {\bibfnamefont {A.}~\bibnamefont {Kennedy}}, \bibinfo {author} {\bibfnamefont
  {A.}~\bibnamefont {Kubik}}, \bibinfo {author} {\bibfnamefont {N.~A.}\
  \bibnamefont {Kurinsky}}, \bibinfo {author} {\bibfnamefont {A.}~\bibnamefont
  {Leder}}, \bibinfo {author} {\bibfnamefont {B.}~\bibnamefont {Loer}},
  \bibinfo {author} {\bibfnamefont {E.}~\bibnamefont {Lopez~Asamar}}, \bibinfo
  {author} {\bibfnamefont {P.}~\bibnamefont {Lukens}}, \bibinfo {author}
  {\bibfnamefont {D.}~\bibnamefont {MacDonell}}, \bibinfo {author}
  {\bibfnamefont {R.}~\bibnamefont {Mahapatra}}, \bibinfo {author}
  {\bibfnamefont {V.}~\bibnamefont {Mandic}}, \bibinfo {author} {\bibfnamefont
  {N.}~\bibnamefont {Mast}}, \bibinfo {author} {\bibfnamefont {E.~H.}\
  \bibnamefont {Miller}}, \bibinfo {author} {\bibfnamefont {N.}~\bibnamefont
  {Mirabolfathi}}, \bibinfo {author} {\bibfnamefont {R.~A.}\ \bibnamefont
  {Moffatt}}, \bibinfo {author} {\bibfnamefont {B.}~\bibnamefont {Mohanty}},
  \bibinfo {author} {\bibfnamefont {J.~D.}\ \bibnamefont {Morales~Mendoza}},
  \bibinfo {author} {\bibfnamefont {J.}~\bibnamefont {Nelson}}, \bibinfo
  {author} {\bibfnamefont {J.~L.}\ \bibnamefont {Orrell}}, \bibinfo {author}
  {\bibfnamefont {S.~M.}\ \bibnamefont {Oser}}, \bibinfo {author}
  {\bibfnamefont {K.}~\bibnamefont {Page}}, \bibinfo {author} {\bibfnamefont
  {W.~A.}\ \bibnamefont {Page}}, \bibinfo {author} {\bibfnamefont
  {R.}~\bibnamefont {Partridge}}, \bibinfo {author} {\bibfnamefont
  {M.}~\bibnamefont {Pepin}}, \bibinfo {author} {\bibfnamefont
  {M.}~\bibnamefont {Pe\~nalver Martinez}}, \bibinfo {author} {\bibfnamefont
  {A.}~\bibnamefont {Phipps}}, \bibinfo {author} {\bibfnamefont
  {S.}~\bibnamefont {Poudel}}, \bibinfo {author} {\bibfnamefont
  {M.}~\bibnamefont {Pyle}}, \bibinfo {author} {\bibfnamefont {H.}~\bibnamefont
  {Qiu}}, \bibinfo {author} {\bibfnamefont {W.}~\bibnamefont {Rau}}, \bibinfo
  {author} {\bibfnamefont {P.}~\bibnamefont {Redl}}, \bibinfo {author}
  {\bibfnamefont {A.}~\bibnamefont {Reisetter}}, \bibinfo {author}
  {\bibfnamefont {T.}~\bibnamefont {Reynolds}}, \bibinfo {author}
  {\bibfnamefont {A.}~\bibnamefont {Roberts}}, \bibinfo {author} {\bibfnamefont
  {A.~E.}\ \bibnamefont {Robinson}}, \bibinfo {author} {\bibfnamefont {H.~E.}\
  \bibnamefont {Rogers}}, \bibinfo {author} {\bibfnamefont {T.}~\bibnamefont
  {Saab}}, \bibinfo {author} {\bibfnamefont {B.}~\bibnamefont {Sadoulet}},
  \bibinfo {author} {\bibfnamefont {J.}~\bibnamefont {Sander}}, \bibinfo
  {author} {\bibfnamefont {K.}~\bibnamefont {Schneck}}, \bibinfo {author}
  {\bibfnamefont {R.~W.}\ \bibnamefont {Schnee}}, \bibinfo {author}
  {\bibfnamefont {S.}~\bibnamefont {Scorza}}, \bibinfo {author} {\bibfnamefont
  {K.}~\bibnamefont {Senapati}}, \bibinfo {author} {\bibfnamefont
  {B.}~\bibnamefont {Serfass}}, \bibinfo {author} {\bibfnamefont
  {D.}~\bibnamefont {Speller}}, \bibinfo {author} {\bibfnamefont
  {M.}~\bibnamefont {Stein}}, \bibinfo {author} {\bibfnamefont
  {J.}~\bibnamefont {Street}}, \bibinfo {author} {\bibfnamefont {H.~A.}\
  \bibnamefont {Tanaka}}, \bibinfo {author} {\bibfnamefont {D.}~\bibnamefont
  {Toback}}, \bibinfo {author} {\bibfnamefont {R.}~\bibnamefont {Underwood}},
  \bibinfo {author} {\bibfnamefont {A.~N.}\ \bibnamefont {Villano}}, \bibinfo
  {author} {\bibfnamefont {B.}~\bibnamefont {von Krosigk}}, \bibinfo {author}
  {\bibfnamefont {B.}~\bibnamefont {Welliver}}, \bibinfo {author}
  {\bibfnamefont {J.~S.}\ \bibnamefont {Wilson}}, \bibinfo {author}
  {\bibfnamefont {M.~J.}\ \bibnamefont {Wilson}}, \bibinfo {author}
  {\bibfnamefont {D.~H.}\ \bibnamefont {Wright}}, \bibinfo {author}
  {\bibfnamefont {S.}~\bibnamefont {Yellin}}, \bibinfo {author} {\bibfnamefont
  {J.~J.}\ \bibnamefont {Yen}}, \bibinfo {author} {\bibfnamefont {B.~A.}\
  \bibnamefont {Young}}, \bibinfo {author} {\bibfnamefont {X.}~\bibnamefont
  {Zhang}}, \ and\ \bibinfo {author} {\bibfnamefont {X.}~\bibnamefont {Zhao}}
  (\bibinfo {collaboration} {SuperCDMS Collaboration}),\ }\href {\doibase
  10.1103/PhysRevD.97.022002} {\bibfield  {journal} {\bibinfo  {journal} {Phys.
  Rev. D}\ }\textbf {\bibinfo {volume} {97}},\ \bibinfo {pages} {022002}
  (\bibinfo {year} {2018}{\natexlab{a}})}\BibitemShut {NoStop}%
\bibitem [{\citenamefont {Agnese}\ \emph
  {et~al.}(2018{\natexlab{b}})\citenamefont {Agnese}, \citenamefont {Aramaki},
  \citenamefont {Arnquist}, \citenamefont {Baker}, \citenamefont
  {Balakishiyeva}, \citenamefont {Banik}, \citenamefont {Barker}, \citenamefont
  {Basu~Thakur}, \citenamefont {Bauer}, \citenamefont {Binder}, \citenamefont
  {Bowles}, \citenamefont {Brink}, \citenamefont {Bunker}, \citenamefont
  {Cabrera}, \citenamefont {Caldwell}, \citenamefont {Calkins}, \citenamefont
  {Cartaro}, \citenamefont {Cerde\~no}, \citenamefont {Chang}, \citenamefont
  {Chen}, \citenamefont {Cooley}, \citenamefont {Cornell}, \citenamefont
  {Cushman}, \citenamefont {Daal}, \citenamefont {Di~Stefano}, \citenamefont
  {Doughty}, \citenamefont {Fascione}, \citenamefont {Figueroa-Feliciano},
  \citenamefont {Fritts}, \citenamefont {Gerbier}, \citenamefont {Germond},
  \citenamefont {Ghaith}, \citenamefont {Godfrey}, \citenamefont {Golwala},
  \citenamefont {Hall}, \citenamefont {Harris}, \citenamefont {Hong},
  \citenamefont {Hoppe}, \citenamefont {Hsu}, \citenamefont {Huber},
  \citenamefont {Iyer}, \citenamefont {Jardin}, \citenamefont {Jastram},
  \citenamefont {Jena}, \citenamefont {Kelsey}, \citenamefont {Kennedy},
  \citenamefont {Kubik}, \citenamefont {Kurinsky}, \citenamefont {Loer},
  \citenamefont {Lopez~Asamar}, \citenamefont {Lukens}, \citenamefont
  {MacDonell}, \citenamefont {Mahapatra}, \citenamefont {Mandic}, \citenamefont
  {Mast}, \citenamefont {Miller}, \citenamefont {Mirabolfathi}, \citenamefont
  {Mohanty}, \citenamefont {Morales~Mendoza}, \citenamefont {Nelson},
  \citenamefont {Orrell}, \citenamefont {Oser}, \citenamefont {Page},
  \citenamefont {Page}, \citenamefont {Partridge}, \citenamefont
  {Penalver~Martinez}, \citenamefont {Pepin}, \citenamefont {Phipps},
  \citenamefont {Poudel}, \citenamefont {Pyle}, \citenamefont {Qiu},
  \citenamefont {Rau}, \citenamefont {Redl}, \citenamefont {Reisetter},
  \citenamefont {Reynolds}, \citenamefont {Roberts}, \citenamefont {Robinson},
  \citenamefont {Rogers}, \citenamefont {Saab}, \citenamefont {Sadoulet},
  \citenamefont {Sander}, \citenamefont {Schneck}, \citenamefont {Schnee},
  \citenamefont {Scorza}, \citenamefont {Senapati}, \citenamefont {Serfass},
  \citenamefont {Speller}, \citenamefont {Stein}, \citenamefont {Street},
  \citenamefont {Tanaka}, \citenamefont {Toback}, \citenamefont {Underwood},
  \citenamefont {Villano}, \citenamefont {von Krosigk}, \citenamefont
  {Welliver}, \citenamefont {Wilson}, \citenamefont {Wilson}, \citenamefont
  {Wright}, \citenamefont {Yellin}, \citenamefont {Yen}, \citenamefont {Young},
  \citenamefont {Zhang},\ and\ \citenamefont {Zhao}}]{cdms2018}%
  \BibitemOpen
  \bibfield  {author} {\bibinfo {author} {\bibfnamefont {R.}~\bibnamefont
  {Agnese}}, \bibinfo {author} {\bibfnamefont {T.}~\bibnamefont {Aramaki}},
  \bibinfo {author} {\bibfnamefont {I.~J.}\ \bibnamefont {Arnquist}}, \bibinfo
  {author} {\bibfnamefont {W.}~\bibnamefont {Baker}}, \bibinfo {author}
  {\bibfnamefont {D.}~\bibnamefont {Balakishiyeva}}, \bibinfo {author}
  {\bibfnamefont {S.}~\bibnamefont {Banik}}, \bibinfo {author} {\bibfnamefont
  {D.}~\bibnamefont {Barker}}, \bibinfo {author} {\bibfnamefont
  {R.}~\bibnamefont {Basu~Thakur}}, \bibinfo {author} {\bibfnamefont {D.~A.}\
  \bibnamefont {Bauer}}, \bibinfo {author} {\bibfnamefont {T.}~\bibnamefont
  {Binder}}, \bibinfo {author} {\bibfnamefont {M.~A.}\ \bibnamefont {Bowles}},
  \bibinfo {author} {\bibfnamefont {P.~L.}\ \bibnamefont {Brink}}, \bibinfo
  {author} {\bibfnamefont {R.}~\bibnamefont {Bunker}}, \bibinfo {author}
  {\bibfnamefont {B.}~\bibnamefont {Cabrera}}, \bibinfo {author} {\bibfnamefont
  {D.~O.}\ \bibnamefont {Caldwell}}, \bibinfo {author} {\bibfnamefont
  {R.}~\bibnamefont {Calkins}}, \bibinfo {author} {\bibfnamefont
  {C.}~\bibnamefont {Cartaro}}, \bibinfo {author} {\bibfnamefont {D.~G.}\
  \bibnamefont {Cerde\~no}}, \bibinfo {author} {\bibfnamefont {Y.}~\bibnamefont
  {Chang}}, \bibinfo {author} {\bibfnamefont {Y.}~\bibnamefont {Chen}},
  \bibinfo {author} {\bibfnamefont {J.}~\bibnamefont {Cooley}}, \bibinfo
  {author} {\bibfnamefont {B.}~\bibnamefont {Cornell}}, \bibinfo {author}
  {\bibfnamefont {P.}~\bibnamefont {Cushman}}, \bibinfo {author} {\bibfnamefont
  {M.}~\bibnamefont {Daal}}, \bibinfo {author} {\bibfnamefont {P.~C.~F.}\
  \bibnamefont {Di~Stefano}}, \bibinfo {author} {\bibfnamefont
  {T.}~\bibnamefont {Doughty}}, \bibinfo {author} {\bibfnamefont
  {E.}~\bibnamefont {Fascione}}, \bibinfo {author} {\bibfnamefont
  {E.}~\bibnamefont {Figueroa-Feliciano}}, \bibinfo {author} {\bibfnamefont
  {M.}~\bibnamefont {Fritts}}, \bibinfo {author} {\bibfnamefont
  {G.}~\bibnamefont {Gerbier}}, \bibinfo {author} {\bibfnamefont
  {R.}~\bibnamefont {Germond}}, \bibinfo {author} {\bibfnamefont
  {M.}~\bibnamefont {Ghaith}}, \bibinfo {author} {\bibfnamefont {G.~L.}\
  \bibnamefont {Godfrey}}, \bibinfo {author} {\bibfnamefont {S.~R.}\
  \bibnamefont {Golwala}}, \bibinfo {author} {\bibfnamefont {J.}~\bibnamefont
  {Hall}}, \bibinfo {author} {\bibfnamefont {H.~R.}\ \bibnamefont {Harris}},
  \bibinfo {author} {\bibfnamefont {Z.}~\bibnamefont {Hong}}, \bibinfo {author}
  {\bibfnamefont {E.~W.}\ \bibnamefont {Hoppe}}, \bibinfo {author}
  {\bibfnamefont {L.}~\bibnamefont {Hsu}}, \bibinfo {author} {\bibfnamefont
  {M.~E.}\ \bibnamefont {Huber}}, \bibinfo {author} {\bibfnamefont
  {V.}~\bibnamefont {Iyer}}, \bibinfo {author} {\bibfnamefont {D.}~\bibnamefont
  {Jardin}}, \bibinfo {author} {\bibfnamefont {A.}~\bibnamefont {Jastram}},
  \bibinfo {author} {\bibfnamefont {C.}~\bibnamefont {Jena}}, \bibinfo {author}
  {\bibfnamefont {M.~H.}\ \bibnamefont {Kelsey}}, \bibinfo {author}
  {\bibfnamefont {A.}~\bibnamefont {Kennedy}}, \bibinfo {author} {\bibfnamefont
  {A.}~\bibnamefont {Kubik}}, \bibinfo {author} {\bibfnamefont {N.~A.}\
  \bibnamefont {Kurinsky}}, \bibinfo {author} {\bibfnamefont {B.}~\bibnamefont
  {Loer}}, \bibinfo {author} {\bibfnamefont {E.}~\bibnamefont {Lopez~Asamar}},
  \bibinfo {author} {\bibfnamefont {P.}~\bibnamefont {Lukens}}, \bibinfo
  {author} {\bibfnamefont {D.}~\bibnamefont {MacDonell}}, \bibinfo {author}
  {\bibfnamefont {R.}~\bibnamefont {Mahapatra}}, \bibinfo {author}
  {\bibfnamefont {V.}~\bibnamefont {Mandic}}, \bibinfo {author} {\bibfnamefont
  {N.}~\bibnamefont {Mast}}, \bibinfo {author} {\bibfnamefont {E.~H.}\
  \bibnamefont {Miller}}, \bibinfo {author} {\bibfnamefont {N.}~\bibnamefont
  {Mirabolfathi}}, \bibinfo {author} {\bibfnamefont {B.}~\bibnamefont
  {Mohanty}}, \bibinfo {author} {\bibfnamefont {J.~D.}\ \bibnamefont
  {Morales~Mendoza}}, \bibinfo {author} {\bibfnamefont {J.}~\bibnamefont
  {Nelson}}, \bibinfo {author} {\bibfnamefont {J.~L.}\ \bibnamefont {Orrell}},
  \bibinfo {author} {\bibfnamefont {S.~M.}\ \bibnamefont {Oser}}, \bibinfo
  {author} {\bibfnamefont {K.}~\bibnamefont {Page}}, \bibinfo {author}
  {\bibfnamefont {W.~A.}\ \bibnamefont {Page}}, \bibinfo {author}
  {\bibfnamefont {R.}~\bibnamefont {Partridge}}, \bibinfo {author}
  {\bibfnamefont {M.}~\bibnamefont {Penalver~Martinez}}, \bibinfo {author}
  {\bibfnamefont {M.}~\bibnamefont {Pepin}}, \bibinfo {author} {\bibfnamefont
  {A.}~\bibnamefont {Phipps}}, \bibinfo {author} {\bibfnamefont
  {S.}~\bibnamefont {Poudel}}, \bibinfo {author} {\bibfnamefont
  {M.}~\bibnamefont {Pyle}}, \bibinfo {author} {\bibfnamefont {H.}~\bibnamefont
  {Qiu}}, \bibinfo {author} {\bibfnamefont {W.}~\bibnamefont {Rau}}, \bibinfo
  {author} {\bibfnamefont {P.}~\bibnamefont {Redl}}, \bibinfo {author}
  {\bibfnamefont {A.}~\bibnamefont {Reisetter}}, \bibinfo {author}
  {\bibfnamefont {T.}~\bibnamefont {Reynolds}}, \bibinfo {author}
  {\bibfnamefont {A.}~\bibnamefont {Roberts}}, \bibinfo {author} {\bibfnamefont
  {A.~E.}\ \bibnamefont {Robinson}}, \bibinfo {author} {\bibfnamefont {H.~E.}\
  \bibnamefont {Rogers}}, \bibinfo {author} {\bibfnamefont {T.}~\bibnamefont
  {Saab}}, \bibinfo {author} {\bibfnamefont {B.}~\bibnamefont {Sadoulet}},
  \bibinfo {author} {\bibfnamefont {J.}~\bibnamefont {Sander}}, \bibinfo
  {author} {\bibfnamefont {K.}~\bibnamefont {Schneck}}, \bibinfo {author}
  {\bibfnamefont {R.~W.}\ \bibnamefont {Schnee}}, \bibinfo {author}
  {\bibfnamefont {S.}~\bibnamefont {Scorza}}, \bibinfo {author} {\bibfnamefont
  {K.}~\bibnamefont {Senapati}}, \bibinfo {author} {\bibfnamefont
  {B.}~\bibnamefont {Serfass}}, \bibinfo {author} {\bibfnamefont
  {D.}~\bibnamefont {Speller}}, \bibinfo {author} {\bibfnamefont
  {M.}~\bibnamefont {Stein}}, \bibinfo {author} {\bibfnamefont
  {J.}~\bibnamefont {Street}}, \bibinfo {author} {\bibfnamefont {H.~A.}\
  \bibnamefont {Tanaka}}, \bibinfo {author} {\bibfnamefont {D.}~\bibnamefont
  {Toback}}, \bibinfo {author} {\bibfnamefont {R.}~\bibnamefont {Underwood}},
  \bibinfo {author} {\bibfnamefont {A.~N.}\ \bibnamefont {Villano}}, \bibinfo
  {author} {\bibfnamefont {B.}~\bibnamefont {von Krosigk}}, \bibinfo {author}
  {\bibfnamefont {B.}~\bibnamefont {Welliver}}, \bibinfo {author}
  {\bibfnamefont {J.~S.}\ \bibnamefont {Wilson}}, \bibinfo {author}
  {\bibfnamefont {M.~J.}\ \bibnamefont {Wilson}}, \bibinfo {author}
  {\bibfnamefont {D.~H.}\ \bibnamefont {Wright}}, \bibinfo {author}
  {\bibfnamefont {S.}~\bibnamefont {Yellin}}, \bibinfo {author} {\bibfnamefont
  {J.~J.}\ \bibnamefont {Yen}}, \bibinfo {author} {\bibfnamefont {B.~A.}\
  \bibnamefont {Young}}, \bibinfo {author} {\bibfnamefont {X.}~\bibnamefont
  {Zhang}}, \ and\ \bibinfo {author} {\bibfnamefont {X.}~\bibnamefont {Zhao}}
  (\bibinfo {collaboration} {SuperCDMS Collaboration}),\ }\href {\doibase
  10.1103/PhysRevLett.120.061802} {\bibfield  {journal} {\bibinfo  {journal}
  {Phys. Rev. Lett.}\ }\textbf {\bibinfo {volume} {120}},\ \bibinfo {pages}
  {061802} (\bibinfo {year} {2018}{\natexlab{b}})}\BibitemShut {NoStop}%
\bibitem [{\citenamefont {Agnese}\ \emph {et~al.}(2017)\citenamefont {Agnese},
  \citenamefont {Anderson}, \citenamefont {Aramaki}, \citenamefont {Arnquist},
  \citenamefont {Baker}, \citenamefont {Barker}, \citenamefont {Basu~Thakur},
  \citenamefont {Bauer}, \citenamefont {Borgland}, \citenamefont {Bowles},
  \citenamefont {Brink}, \citenamefont {Bunker}, \citenamefont {Cabrera},
  \citenamefont {Caldwell}, \citenamefont {Calkins}, \citenamefont {Cartaro},
  \citenamefont {Cerde\~no}, \citenamefont {Chagani}, \citenamefont {Chen},
  \citenamefont {Cooley}, \citenamefont {Cornell}, \citenamefont {Cushman},
  \citenamefont {Daal}, \citenamefont {Di~Stefano}, \citenamefont {Doughty},
  \citenamefont {Esteban}, \citenamefont {Fallows}, \citenamefont
  {Figueroa-Feliciano}, \citenamefont {Fritts}, \citenamefont {Gerbier},
  \citenamefont {Ghaith}, \citenamefont {Godfrey}, \citenamefont {Golwala},
  \citenamefont {Hall}, \citenamefont {Harris}, \citenamefont {Hofer},
  \citenamefont {Holmgren}, \citenamefont {Hong}, \citenamefont {Hoppe},
  \citenamefont {Hsu}, \citenamefont {Huber}, \citenamefont {Iyer},
  \citenamefont {Jardin}, \citenamefont {Jastram}, \citenamefont {Kelsey},
  \citenamefont {Kennedy}, \citenamefont {Kubik}, \citenamefont {Kurinsky},
  \citenamefont {Leder}, \citenamefont {Loer}, \citenamefont {Lopez~Asamar},
  \citenamefont {Lukens}, \citenamefont {Mahapatra}, \citenamefont {Mandic},
  \citenamefont {Mast}, \citenamefont {Mirabolfathi}, \citenamefont {Moffatt},
  \citenamefont {Morales~Mendoza}, \citenamefont {Orrell}, \citenamefont
  {Oser}, \citenamefont {Page}, \citenamefont {Page}, \citenamefont
  {Partridge}, \citenamefont {Pepin}, \citenamefont {Phipps}, \citenamefont
  {Poudel}, \citenamefont {Pyle}, \citenamefont {Qiu}, \citenamefont {Rau},
  \citenamefont {Redl}, \citenamefont {Reisetter}, \citenamefont {Roberts},
  \citenamefont {Robinson}, \citenamefont {Rogers}, \citenamefont {Saab},
  \citenamefont {Sadoulet}, \citenamefont {Sander}, \citenamefont {Schneck},
  \citenamefont {Schnee}, \citenamefont {Serfass}, \citenamefont {Speller},
  \citenamefont {Stein}, \citenamefont {Street}, \citenamefont {Tanaka},
  \citenamefont {Toback}, \citenamefont {Underwood}, \citenamefont {Villano},
  \citenamefont {von Krosigk}, \citenamefont {Welliver}, \citenamefont
  {Wilson}, \citenamefont {Wright}, \citenamefont {Yellin}, \citenamefont
  {Yen}, \citenamefont {Young}, \citenamefont {Zhang},\ and\ \citenamefont
  {Zhao}}]{snolab2017}%
  \BibitemOpen
  \bibfield  {author} {\bibinfo {author} {\bibfnamefont {R.}~\bibnamefont
  {Agnese}}, \bibinfo {author} {\bibfnamefont {A.~J.}\ \bibnamefont
  {Anderson}}, \bibinfo {author} {\bibfnamefont {T.}~\bibnamefont {Aramaki}},
  \bibinfo {author} {\bibfnamefont {I.}~\bibnamefont {Arnquist}}, \bibinfo
  {author} {\bibfnamefont {W.}~\bibnamefont {Baker}}, \bibinfo {author}
  {\bibfnamefont {D.}~\bibnamefont {Barker}}, \bibinfo {author} {\bibfnamefont
  {R.}~\bibnamefont {Basu~Thakur}}, \bibinfo {author} {\bibfnamefont {D.~A.}\
  \bibnamefont {Bauer}}, \bibinfo {author} {\bibfnamefont {A.}~\bibnamefont
  {Borgland}}, \bibinfo {author} {\bibfnamefont {M.~A.}\ \bibnamefont
  {Bowles}}, \bibinfo {author} {\bibfnamefont {P.~L.}\ \bibnamefont {Brink}},
  \bibinfo {author} {\bibfnamefont {R.}~\bibnamefont {Bunker}}, \bibinfo
  {author} {\bibfnamefont {B.}~\bibnamefont {Cabrera}}, \bibinfo {author}
  {\bibfnamefont {D.~O.}\ \bibnamefont {Caldwell}}, \bibinfo {author}
  {\bibfnamefont {R.}~\bibnamefont {Calkins}}, \bibinfo {author} {\bibfnamefont
  {C.}~\bibnamefont {Cartaro}}, \bibinfo {author} {\bibfnamefont {D.~G.}\
  \bibnamefont {Cerde\~no}}, \bibinfo {author} {\bibfnamefont {H.}~\bibnamefont
  {Chagani}}, \bibinfo {author} {\bibfnamefont {Y.}~\bibnamefont {Chen}},
  \bibinfo {author} {\bibfnamefont {J.}~\bibnamefont {Cooley}}, \bibinfo
  {author} {\bibfnamefont {B.}~\bibnamefont {Cornell}}, \bibinfo {author}
  {\bibfnamefont {P.}~\bibnamefont {Cushman}}, \bibinfo {author} {\bibfnamefont
  {M.}~\bibnamefont {Daal}}, \bibinfo {author} {\bibfnamefont {P.~C.~F.}\
  \bibnamefont {Di~Stefano}}, \bibinfo {author} {\bibfnamefont
  {T.}~\bibnamefont {Doughty}}, \bibinfo {author} {\bibfnamefont
  {L.}~\bibnamefont {Esteban}}, \bibinfo {author} {\bibfnamefont
  {S.}~\bibnamefont {Fallows}}, \bibinfo {author} {\bibfnamefont
  {E.}~\bibnamefont {Figueroa-Feliciano}}, \bibinfo {author} {\bibfnamefont
  {M.}~\bibnamefont {Fritts}}, \bibinfo {author} {\bibfnamefont
  {G.}~\bibnamefont {Gerbier}}, \bibinfo {author} {\bibfnamefont
  {M.}~\bibnamefont {Ghaith}}, \bibinfo {author} {\bibfnamefont {G.~L.}\
  \bibnamefont {Godfrey}}, \bibinfo {author} {\bibfnamefont {S.~R.}\
  \bibnamefont {Golwala}}, \bibinfo {author} {\bibfnamefont {J.}~\bibnamefont
  {Hall}}, \bibinfo {author} {\bibfnamefont {H.~R.}\ \bibnamefont {Harris}},
  \bibinfo {author} {\bibfnamefont {T.}~\bibnamefont {Hofer}}, \bibinfo
  {author} {\bibfnamefont {D.}~\bibnamefont {Holmgren}}, \bibinfo {author}
  {\bibfnamefont {Z.}~\bibnamefont {Hong}}, \bibinfo {author} {\bibfnamefont
  {E.}~\bibnamefont {Hoppe}}, \bibinfo {author} {\bibfnamefont
  {L.}~\bibnamefont {Hsu}}, \bibinfo {author} {\bibfnamefont {M.~E.}\
  \bibnamefont {Huber}}, \bibinfo {author} {\bibfnamefont {V.}~\bibnamefont
  {Iyer}}, \bibinfo {author} {\bibfnamefont {D.}~\bibnamefont {Jardin}},
  \bibinfo {author} {\bibfnamefont {A.}~\bibnamefont {Jastram}}, \bibinfo
  {author} {\bibfnamefont {M.~H.}\ \bibnamefont {Kelsey}}, \bibinfo {author}
  {\bibfnamefont {A.}~\bibnamefont {Kennedy}}, \bibinfo {author} {\bibfnamefont
  {A.}~\bibnamefont {Kubik}}, \bibinfo {author} {\bibfnamefont {N.~A.}\
  \bibnamefont {Kurinsky}}, \bibinfo {author} {\bibfnamefont {A.}~\bibnamefont
  {Leder}}, \bibinfo {author} {\bibfnamefont {B.}~\bibnamefont {Loer}},
  \bibinfo {author} {\bibfnamefont {E.}~\bibnamefont {Lopez~Asamar}}, \bibinfo
  {author} {\bibfnamefont {P.}~\bibnamefont {Lukens}}, \bibinfo {author}
  {\bibfnamefont {R.}~\bibnamefont {Mahapatra}}, \bibinfo {author}
  {\bibfnamefont {V.}~\bibnamefont {Mandic}}, \bibinfo {author} {\bibfnamefont
  {N.}~\bibnamefont {Mast}}, \bibinfo {author} {\bibfnamefont {N.}~\bibnamefont
  {Mirabolfathi}}, \bibinfo {author} {\bibfnamefont {R.~A.}\ \bibnamefont
  {Moffatt}}, \bibinfo {author} {\bibfnamefont {J.~D.}\ \bibnamefont
  {Morales~Mendoza}}, \bibinfo {author} {\bibfnamefont {J.~L.}\ \bibnamefont
  {Orrell}}, \bibinfo {author} {\bibfnamefont {S.~M.}\ \bibnamefont {Oser}},
  \bibinfo {author} {\bibfnamefont {K.}~\bibnamefont {Page}}, \bibinfo {author}
  {\bibfnamefont {W.~A.}\ \bibnamefont {Page}}, \bibinfo {author}
  {\bibfnamefont {R.}~\bibnamefont {Partridge}}, \bibinfo {author}
  {\bibfnamefont {M.}~\bibnamefont {Pepin}}, \bibinfo {author} {\bibfnamefont
  {A.}~\bibnamefont {Phipps}}, \bibinfo {author} {\bibfnamefont
  {S.}~\bibnamefont {Poudel}}, \bibinfo {author} {\bibfnamefont
  {M.}~\bibnamefont {Pyle}}, \bibinfo {author} {\bibfnamefont {H.}~\bibnamefont
  {Qiu}}, \bibinfo {author} {\bibfnamefont {W.}~\bibnamefont {Rau}}, \bibinfo
  {author} {\bibfnamefont {P.}~\bibnamefont {Redl}}, \bibinfo {author}
  {\bibfnamefont {A.}~\bibnamefont {Reisetter}}, \bibinfo {author}
  {\bibfnamefont {A.}~\bibnamefont {Roberts}}, \bibinfo {author} {\bibfnamefont
  {A.~E.}\ \bibnamefont {Robinson}}, \bibinfo {author} {\bibfnamefont {H.~E.}\
  \bibnamefont {Rogers}}, \bibinfo {author} {\bibfnamefont {T.}~\bibnamefont
  {Saab}}, \bibinfo {author} {\bibfnamefont {B.}~\bibnamefont {Sadoulet}},
  \bibinfo {author} {\bibfnamefont {J.}~\bibnamefont {Sander}}, \bibinfo
  {author} {\bibfnamefont {K.}~\bibnamefont {Schneck}}, \bibinfo {author}
  {\bibfnamefont {R.~W.}\ \bibnamefont {Schnee}}, \bibinfo {author}
  {\bibfnamefont {B.}~\bibnamefont {Serfass}}, \bibinfo {author} {\bibfnamefont
  {D.}~\bibnamefont {Speller}}, \bibinfo {author} {\bibfnamefont
  {M.}~\bibnamefont {Stein}}, \bibinfo {author} {\bibfnamefont
  {J.}~\bibnamefont {Street}}, \bibinfo {author} {\bibfnamefont {H.~A.}\
  \bibnamefont {Tanaka}}, \bibinfo {author} {\bibfnamefont {D.}~\bibnamefont
  {Toback}}, \bibinfo {author} {\bibfnamefont {R.}~\bibnamefont {Underwood}},
  \bibinfo {author} {\bibfnamefont {A.~N.}\ \bibnamefont {Villano}}, \bibinfo
  {author} {\bibfnamefont {B.}~\bibnamefont {von Krosigk}}, \bibinfo {author}
  {\bibfnamefont {B.}~\bibnamefont {Welliver}}, \bibinfo {author}
  {\bibfnamefont {J.~S.}\ \bibnamefont {Wilson}}, \bibinfo {author}
  {\bibfnamefont {D.~H.}\ \bibnamefont {Wright}}, \bibinfo {author}
  {\bibfnamefont {S.}~\bibnamefont {Yellin}}, \bibinfo {author} {\bibfnamefont
  {J.~J.}\ \bibnamefont {Yen}}, \bibinfo {author} {\bibfnamefont {B.~A.}\
  \bibnamefont {Young}}, \bibinfo {author} {\bibfnamefont {X.}~\bibnamefont
  {Zhang}}, \ and\ \bibinfo {author} {\bibfnamefont {X.}~\bibnamefont {Zhao}}
  (\bibinfo {collaboration} {SuperCDMS Collaboration}),\ }\href {\doibase
  10.1103/PhysRevD.95.082002} {\bibfield  {journal} {\bibinfo  {journal} {Phys.
  Rev. D}\ }\textbf {\bibinfo {volume} {95}},\ \bibinfo {pages} {082002}
  (\bibinfo {year} {2017})}\BibitemShut {NoStop}%
\bibitem [{\citenamefont {Steigman}\ and\ \citenamefont
  {Turner}(1985)}]{STEIGMAN1985375}%
  \BibitemOpen
  \bibfield  {author} {\bibinfo {author} {\bibfnamefont {G.}~\bibnamefont
  {Steigman}}\ and\ \bibinfo {author} {\bibfnamefont {M.~S.}\ \bibnamefont
  {Turner}},\ }\href {\doibase 10.1016/0550-3213(85)90537-1} {\bibfield
  {journal} {\bibinfo  {journal} {Nuclear Physics B}\ }\textbf {\bibinfo
  {volume} {253}},\ \bibinfo {pages} {375} (\bibinfo {year}
  {1985})}\BibitemShut {NoStop}%
\bibitem [{\citenamefont {{Redl}}(2014)}]{redl2014}%
  \BibitemOpen
  \bibfield  {author} {\bibinfo {author} {\bibfnamefont {P.}~\bibnamefont
  {{Redl}}},\ }\href {\doibase 10.1007/s10909-014-1102-z} {\bibfield  {journal}
  {\bibinfo  {journal} {Journal of Low Temperature Physics}\ }\textbf {\bibinfo
  {volume} {176}},\ \bibinfo {pages} {937} (\bibinfo {year}
  {2014})}\BibitemShut {NoStop}%
\bibitem [{\citenamefont {Raider}, \citenamefont {Flitsch},\ and\ \citenamefont
  {Palmer}(1975)}]{SiExp1}%
  \BibitemOpen
  \bibfield  {author} {\bibinfo {author} {\bibfnamefont {S.}~\bibnamefont
  {Raider}}, \bibinfo {author} {\bibfnamefont {R.}~\bibnamefont {Flitsch}}, \
  and\ \bibinfo {author} {\bibfnamefont {M.}~\bibnamefont {Palmer}},\ }\href
  {\doibase https://doi.org/10.1149/1.2134225} {\bibfield  {journal} {\bibinfo
  {journal} {Journal of the Electrochemical Society}\ }\textbf {\bibinfo
  {volume} {122}},\ \bibinfo {pages} {413 } (\bibinfo {year}
  {1975})}\BibitemShut {NoStop}%
\bibitem [{\citenamefont {Morita}\ \emph {et~al.}(1990)\citenamefont {Morita},
  \citenamefont {Ohmi}, \citenamefont {Hasegawa}, \citenamefont {Kawakami},\
  and\ \citenamefont {Ohwada}}]{SiExp2}%
  \BibitemOpen
  \bibfield  {author} {\bibinfo {author} {\bibfnamefont {M.}~\bibnamefont
  {Morita}}, \bibinfo {author} {\bibfnamefont {T.}~\bibnamefont {Ohmi}},
  \bibinfo {author} {\bibfnamefont {E.}~\bibnamefont {Hasegawa}}, \bibinfo
  {author} {\bibfnamefont {M.}~\bibnamefont {Kawakami}}, \ and\ \bibinfo
  {author} {\bibfnamefont {M.}~\bibnamefont {Ohwada}},\ }\href {\doibase
  10.1063/1.347181} {\bibfield  {journal} {\bibinfo  {journal} {Journal of
  Applied Physics}\ }\textbf {\bibinfo {volume} {68}},\ \bibinfo {pages} {1272}
  (\bibinfo {year} {1990})},\ \Eprint
  {http://arxiv.org/abs/https://doi.org/10.1063/1.347181}
  {https://doi.org/10.1063/1.347181} \BibitemShut {NoStop}%
\bibitem [{\citenamefont {Agostinelli}\ \emph {et~al.}(2003)\citenamefont
  {Agostinelli}, \citenamefont {Allison}, \citenamefont {Amako}, \citenamefont
  {Apostolakis}, \citenamefont {Araujo}, \citenamefont {Arce}, \citenamefont
  {Asai}, \citenamefont {Axen}, \citenamefont {Banerjee}, \citenamefont
  {Barrand}, \citenamefont {Behner}, \citenamefont {Bellagamba}, \citenamefont
  {Boudreau}, \citenamefont {Broglia}, \citenamefont {Brunengo}, \citenamefont
  {Burkhardt}, \citenamefont {Chauvie}, \citenamefont {Chuma}, \citenamefont
  {Chytracek}, \citenamefont {Cooperman}, \citenamefont {Cosmo}, \citenamefont
  {Degtyarenko}, \citenamefont {Dell'Acqua}, \citenamefont {Depaola},
  \citenamefont {Dietrich}, \citenamefont {Enami}, \citenamefont {Feliciello},
  \citenamefont {Ferguson}, \citenamefont {Fesefeldt}, \citenamefont {Folger},
  \citenamefont {Foppiano}, \citenamefont {Forti}, \citenamefont {Garelli},
  \citenamefont {Giani}, \citenamefont {Giannitrapani}, \citenamefont {Gibin},
  \citenamefont {Cadenas}, \citenamefont {González}, \citenamefont {Abril},
  \citenamefont {Greeniaus}, \citenamefont {Greiner}, \citenamefont {Grichine},
  \citenamefont {Grossheim}, \citenamefont {Guatelli}, \citenamefont
  {Gumplinger}, \citenamefont {Hamatsu}, \citenamefont {Hashimoto},
  \citenamefont {Hasui}, \citenamefont {Heikkinen}, \citenamefont {Howard},
  \citenamefont {Ivanchenko}, \citenamefont {Johnson}, \citenamefont {Jones},
  \citenamefont {Kallenbach}, \citenamefont {Kanaya}, \citenamefont {Kawabata},
  \citenamefont {Kawabata}, \citenamefont {Kawaguti}, \citenamefont {Kelner},
  \citenamefont {Kent}, \citenamefont {Kimura}, \citenamefont {Kodama},
  \citenamefont {Kokoulin}, \citenamefont {Kossov}, \citenamefont {Kurashige},
  \citenamefont {Lamanna}, \citenamefont {Lampén}, \citenamefont {Lara},
  \citenamefont {Lefebure}, \citenamefont {Lei}, \citenamefont {Liendl},
  \citenamefont {Lockman}, \citenamefont {Longo}, \citenamefont {Magni},
  \citenamefont {Maire}, \citenamefont {Medernach}, \citenamefont {Minamimoto},
  \citenamefont {de~Freitas}, \citenamefont {Morita}, \citenamefont {Murakami},
  \citenamefont {Nagamatu}, \citenamefont {Nartallo}, \citenamefont {Nieminen},
  \citenamefont {Nishimura}, \citenamefont {Ohtsubo}, \citenamefont {Okamura},
  \citenamefont {O'Neale}, \citenamefont {Oohata}, \citenamefont {Paech},
  \citenamefont {Perl}, \citenamefont {Pfeiffer}, \citenamefont {Pia},
  \citenamefont {Ranjard}, \citenamefont {Rybin}, \citenamefont {Sadilov},
  \citenamefont {Salvo}, \citenamefont {Santin}, \citenamefont {Sasaki},
  \citenamefont {Savvas}, \citenamefont {Sawada}, \citenamefont {Scherer},
  \citenamefont {Sei}, \citenamefont {Sirotenko}, \citenamefont {Smith},
  \citenamefont {Starkov}, \citenamefont {Stoecker}, \citenamefont {Sulkimo},
  \citenamefont {Takahata}, \citenamefont {Tanaka}, \citenamefont {Tcherniaev},
  \citenamefont {Tehrani}, \citenamefont {Tropeano}, \citenamefont {Truscott},
  \citenamefont {Uno}, \citenamefont {Urban}, \citenamefont {Urban},
  \citenamefont {Verderi}, \citenamefont {Walkden}, \citenamefont {Wander},
  \citenamefont {Weber}, \citenamefont {Wellisch}, \citenamefont {Wenaus},
  \citenamefont {Williams}, \citenamefont {Wright}, \citenamefont {Yamada},
  \citenamefont {Yoshida},\ and\ \citenamefont {Zschiesche}}]{Geant42003}%
  \BibitemOpen
  \bibfield  {author} {\bibinfo {author} {\bibfnamefont {S.}~\bibnamefont
  {Agostinelli}}, \bibinfo {author} {\bibfnamefont {J.}~\bibnamefont
  {Allison}}, \bibinfo {author} {\bibfnamefont {K.}~\bibnamefont {Amako}},
  \bibinfo {author} {\bibfnamefont {J.}~\bibnamefont {Apostolakis}}, \bibinfo
  {author} {\bibfnamefont {H.}~\bibnamefont {Araujo}}, \bibinfo {author}
  {\bibfnamefont {P.}~\bibnamefont {Arce}}, \bibinfo {author} {\bibfnamefont
  {M.}~\bibnamefont {Asai}}, \bibinfo {author} {\bibfnamefont {D.}~\bibnamefont
  {Axen}}, \bibinfo {author} {\bibfnamefont {S.}~\bibnamefont {Banerjee}},
  \bibinfo {author} {\bibfnamefont {G.}~\bibnamefont {Barrand}}, \bibinfo
  {author} {\bibfnamefont {F.}~\bibnamefont {Behner}}, \bibinfo {author}
  {\bibfnamefont {L.}~\bibnamefont {Bellagamba}}, \bibinfo {author}
  {\bibfnamefont {J.}~\bibnamefont {Boudreau}}, \bibinfo {author}
  {\bibfnamefont {L.}~\bibnamefont {Broglia}}, \bibinfo {author} {\bibfnamefont
  {A.}~\bibnamefont {Brunengo}}, \bibinfo {author} {\bibfnamefont
  {H.}~\bibnamefont {Burkhardt}}, \bibinfo {author} {\bibfnamefont
  {S.}~\bibnamefont {Chauvie}}, \bibinfo {author} {\bibfnamefont
  {J.}~\bibnamefont {Chuma}}, \bibinfo {author} {\bibfnamefont
  {R.}~\bibnamefont {Chytracek}}, \bibinfo {author} {\bibfnamefont
  {G.}~\bibnamefont {Cooperman}}, \bibinfo {author} {\bibfnamefont
  {G.}~\bibnamefont {Cosmo}}, \bibinfo {author} {\bibfnamefont
  {P.}~\bibnamefont {Degtyarenko}}, \bibinfo {author} {\bibfnamefont
  {A.}~\bibnamefont {Dell'Acqua}}, \bibinfo {author} {\bibfnamefont
  {G.}~\bibnamefont {Depaola}}, \bibinfo {author} {\bibfnamefont
  {D.}~\bibnamefont {Dietrich}}, \bibinfo {author} {\bibfnamefont
  {R.}~\bibnamefont {Enami}}, \bibinfo {author} {\bibfnamefont
  {A.}~\bibnamefont {Feliciello}}, \bibinfo {author} {\bibfnamefont
  {C.}~\bibnamefont {Ferguson}}, \bibinfo {author} {\bibfnamefont
  {H.}~\bibnamefont {Fesefeldt}}, \bibinfo {author} {\bibfnamefont
  {G.}~\bibnamefont {Folger}}, \bibinfo {author} {\bibfnamefont
  {F.}~\bibnamefont {Foppiano}}, \bibinfo {author} {\bibfnamefont
  {A.}~\bibnamefont {Forti}}, \bibinfo {author} {\bibfnamefont
  {S.}~\bibnamefont {Garelli}}, \bibinfo {author} {\bibfnamefont
  {S.}~\bibnamefont {Giani}}, \bibinfo {author} {\bibfnamefont
  {R.}~\bibnamefont {Giannitrapani}}, \bibinfo {author} {\bibfnamefont
  {D.}~\bibnamefont {Gibin}}, \bibinfo {author} {\bibfnamefont {J.~G.}\
  \bibnamefont {Cadenas}}, \bibinfo {author} {\bibfnamefont {I.}~\bibnamefont
  {González}}, \bibinfo {author} {\bibfnamefont {G.~G.}\ \bibnamefont
  {Abril}}, \bibinfo {author} {\bibfnamefont {G.}~\bibnamefont {Greeniaus}},
  \bibinfo {author} {\bibfnamefont {W.}~\bibnamefont {Greiner}}, \bibinfo
  {author} {\bibfnamefont {V.}~\bibnamefont {Grichine}}, \bibinfo {author}
  {\bibfnamefont {A.}~\bibnamefont {Grossheim}}, \bibinfo {author}
  {\bibfnamefont {S.}~\bibnamefont {Guatelli}}, \bibinfo {author}
  {\bibfnamefont {P.}~\bibnamefont {Gumplinger}}, \bibinfo {author}
  {\bibfnamefont {R.}~\bibnamefont {Hamatsu}}, \bibinfo {author} {\bibfnamefont
  {K.}~\bibnamefont {Hashimoto}}, \bibinfo {author} {\bibfnamefont
  {H.}~\bibnamefont {Hasui}}, \bibinfo {author} {\bibfnamefont
  {A.}~\bibnamefont {Heikkinen}}, \bibinfo {author} {\bibfnamefont
  {A.}~\bibnamefont {Howard}}, \bibinfo {author} {\bibfnamefont
  {V.}~\bibnamefont {Ivanchenko}}, \bibinfo {author} {\bibfnamefont
  {A.}~\bibnamefont {Johnson}}, \bibinfo {author} {\bibfnamefont
  {F.}~\bibnamefont {Jones}}, \bibinfo {author} {\bibfnamefont
  {J.}~\bibnamefont {Kallenbach}}, \bibinfo {author} {\bibfnamefont
  {N.}~\bibnamefont {Kanaya}}, \bibinfo {author} {\bibfnamefont
  {M.}~\bibnamefont {Kawabata}}, \bibinfo {author} {\bibfnamefont
  {Y.}~\bibnamefont {Kawabata}}, \bibinfo {author} {\bibfnamefont
  {M.}~\bibnamefont {Kawaguti}}, \bibinfo {author} {\bibfnamefont
  {S.}~\bibnamefont {Kelner}}, \bibinfo {author} {\bibfnamefont
  {P.}~\bibnamefont {Kent}}, \bibinfo {author} {\bibfnamefont {A.}~\bibnamefont
  {Kimura}}, \bibinfo {author} {\bibfnamefont {T.}~\bibnamefont {Kodama}},
  \bibinfo {author} {\bibfnamefont {R.}~\bibnamefont {Kokoulin}}, \bibinfo
  {author} {\bibfnamefont {M.}~\bibnamefont {Kossov}}, \bibinfo {author}
  {\bibfnamefont {H.}~\bibnamefont {Kurashige}}, \bibinfo {author}
  {\bibfnamefont {E.}~\bibnamefont {Lamanna}}, \bibinfo {author} {\bibfnamefont
  {T.}~\bibnamefont {Lampén}}, \bibinfo {author} {\bibfnamefont
  {V.}~\bibnamefont {Lara}}, \bibinfo {author} {\bibfnamefont {V.}~\bibnamefont
  {Lefebure}}, \bibinfo {author} {\bibfnamefont {F.}~\bibnamefont {Lei}},
  \bibinfo {author} {\bibfnamefont {M.}~\bibnamefont {Liendl}}, \bibinfo
  {author} {\bibfnamefont {W.}~\bibnamefont {Lockman}}, \bibinfo {author}
  {\bibfnamefont {F.}~\bibnamefont {Longo}}, \bibinfo {author} {\bibfnamefont
  {S.}~\bibnamefont {Magni}}, \bibinfo {author} {\bibfnamefont
  {M.}~\bibnamefont {Maire}}, \bibinfo {author} {\bibfnamefont
  {E.}~\bibnamefont {Medernach}}, \bibinfo {author} {\bibfnamefont
  {K.}~\bibnamefont {Minamimoto}}, \bibinfo {author} {\bibfnamefont {P.~M.}\
  \bibnamefont {de~Freitas}}, \bibinfo {author} {\bibfnamefont
  {Y.}~\bibnamefont {Morita}}, \bibinfo {author} {\bibfnamefont
  {K.}~\bibnamefont {Murakami}}, \bibinfo {author} {\bibfnamefont
  {M.}~\bibnamefont {Nagamatu}}, \bibinfo {author} {\bibfnamefont
  {R.}~\bibnamefont {Nartallo}}, \bibinfo {author} {\bibfnamefont
  {P.}~\bibnamefont {Nieminen}}, \bibinfo {author} {\bibfnamefont
  {T.}~\bibnamefont {Nishimura}}, \bibinfo {author} {\bibfnamefont
  {K.}~\bibnamefont {Ohtsubo}}, \bibinfo {author} {\bibfnamefont
  {M.}~\bibnamefont {Okamura}}, \bibinfo {author} {\bibfnamefont
  {S.}~\bibnamefont {O'Neale}}, \bibinfo {author} {\bibfnamefont
  {Y.}~\bibnamefont {Oohata}}, \bibinfo {author} {\bibfnamefont
  {K.}~\bibnamefont {Paech}}, \bibinfo {author} {\bibfnamefont
  {J.}~\bibnamefont {Perl}}, \bibinfo {author} {\bibfnamefont {A.}~\bibnamefont
  {Pfeiffer}}, \bibinfo {author} {\bibfnamefont {M.}~\bibnamefont {Pia}},
  \bibinfo {author} {\bibfnamefont {F.}~\bibnamefont {Ranjard}}, \bibinfo
  {author} {\bibfnamefont {A.}~\bibnamefont {Rybin}}, \bibinfo {author}
  {\bibfnamefont {S.}~\bibnamefont {Sadilov}}, \bibinfo {author} {\bibfnamefont
  {E.~D.}\ \bibnamefont {Salvo}}, \bibinfo {author} {\bibfnamefont
  {G.}~\bibnamefont {Santin}}, \bibinfo {author} {\bibfnamefont
  {T.}~\bibnamefont {Sasaki}}, \bibinfo {author} {\bibfnamefont
  {N.}~\bibnamefont {Savvas}}, \bibinfo {author} {\bibfnamefont
  {Y.}~\bibnamefont {Sawada}}, \bibinfo {author} {\bibfnamefont
  {S.}~\bibnamefont {Scherer}}, \bibinfo {author} {\bibfnamefont
  {S.}~\bibnamefont {Sei}}, \bibinfo {author} {\bibfnamefont {V.}~\bibnamefont
  {Sirotenko}}, \bibinfo {author} {\bibfnamefont {D.}~\bibnamefont {Smith}},
  \bibinfo {author} {\bibfnamefont {N.}~\bibnamefont {Starkov}}, \bibinfo
  {author} {\bibfnamefont {H.}~\bibnamefont {Stoecker}}, \bibinfo {author}
  {\bibfnamefont {J.}~\bibnamefont {Sulkimo}}, \bibinfo {author} {\bibfnamefont
  {M.}~\bibnamefont {Takahata}}, \bibinfo {author} {\bibfnamefont
  {S.}~\bibnamefont {Tanaka}}, \bibinfo {author} {\bibfnamefont
  {E.}~\bibnamefont {Tcherniaev}}, \bibinfo {author} {\bibfnamefont {E.~S.}\
  \bibnamefont {Tehrani}}, \bibinfo {author} {\bibfnamefont {M.}~\bibnamefont
  {Tropeano}}, \bibinfo {author} {\bibfnamefont {P.}~\bibnamefont {Truscott}},
  \bibinfo {author} {\bibfnamefont {H.}~\bibnamefont {Uno}}, \bibinfo {author}
  {\bibfnamefont {L.}~\bibnamefont {Urban}}, \bibinfo {author} {\bibfnamefont
  {P.}~\bibnamefont {Urban}}, \bibinfo {author} {\bibfnamefont
  {M.}~\bibnamefont {Verderi}}, \bibinfo {author} {\bibfnamefont
  {A.}~\bibnamefont {Walkden}}, \bibinfo {author} {\bibfnamefont
  {W.}~\bibnamefont {Wander}}, \bibinfo {author} {\bibfnamefont
  {H.}~\bibnamefont {Weber}}, \bibinfo {author} {\bibfnamefont
  {J.}~\bibnamefont {Wellisch}}, \bibinfo {author} {\bibfnamefont
  {T.}~\bibnamefont {Wenaus}}, \bibinfo {author} {\bibfnamefont
  {D.}~\bibnamefont {Williams}}, \bibinfo {author} {\bibfnamefont
  {D.}~\bibnamefont {Wright}}, \bibinfo {author} {\bibfnamefont
  {T.}~\bibnamefont {Yamada}}, \bibinfo {author} {\bibfnamefont
  {H.}~\bibnamefont {Yoshida}}, \ and\ \bibinfo {author} {\bibfnamefont
  {D.}~\bibnamefont {Zschiesche}},\ }\href {\doibase
  http://dx.doi.org/10.1016/S0168-9002(03)01368-8} {\bibfield  {journal}
  {\bibinfo  {journal} {Nuclear Instruments and Methods in Physics Research
  Section A: Accelerators, Spectrometers, Detectors and Associated Equipment}\
  }\textbf {\bibinfo {volume} {506}},\ \bibinfo {pages} {250 } (\bibinfo {year}
  {2003})}\BibitemShut {NoStop}%
\bibitem [{\citenamefont {Allison}\ \emph {et~al.}(2006)\citenamefont
  {Allison}, \citenamefont {Amako}, \citenamefont {Apostolakis}, \citenamefont
  {Araujo}, \citenamefont {Dubois}, \citenamefont {Asai}, \citenamefont
  {Barrand}, \citenamefont {Capra}, \citenamefont {Chauvie}, \citenamefont
  {Chytracek}, \citenamefont {Cirrone}, \citenamefont {Cooperman},
  \citenamefont {Cosmo}, \citenamefont {Cuttone}, \citenamefont {Daquino},
  \citenamefont {Donszelmann}, \citenamefont {Dressel}, \citenamefont {Folger},
  \citenamefont {Foppiano}, \citenamefont {Generowicz}, \citenamefont
  {Grichine}, \citenamefont {Guatelli}, \citenamefont {Gumplinger},
  \citenamefont {Heikkinen}, \citenamefont {Hrivnacova}, \citenamefont
  {Howard}, \citenamefont {Incerti}, \citenamefont {Ivanchenko}, \citenamefont
  {Johnson}, \citenamefont {Jones}, \citenamefont {Koi}, \citenamefont
  {Kokoulin}, \citenamefont {Kossov}, \citenamefont {Kurashige}, \citenamefont
  {Lara}, \citenamefont {Larsson}, \citenamefont {Lei}, \citenamefont {Link},
  \citenamefont {Longo}, \citenamefont {Maire}, \citenamefont {Mantero},
  \citenamefont {Mascialino}, \citenamefont {McLaren}, \citenamefont {Lorenzo},
  \citenamefont {Minamimoto}, \citenamefont {Murakami}, \citenamefont
  {Nieminen}, \citenamefont {Pandola}, \citenamefont {Parlati}, \citenamefont
  {Peralta}, \citenamefont {Perl}, \citenamefont {Pfeiffer}, \citenamefont
  {Pia}, \citenamefont {Ribon}, \citenamefont {Rodrigues}, \citenamefont
  {Russo}, \citenamefont {Sadilov}, \citenamefont {Santin}, \citenamefont
  {Sasaki}, \citenamefont {Smith}, \citenamefont {Starkov}, \citenamefont
  {Tanaka}, \citenamefont {Tcherniaev}, \citenamefont {Tome}, \citenamefont
  {Trindade}, \citenamefont {Truscott}, \citenamefont {Urban}, \citenamefont
  {Verderi}, \citenamefont {Walkden}, \citenamefont {Wellisch}, \citenamefont
  {Williams}, \citenamefont {Wright},\ and\ \citenamefont
  {Yoshida}}]{Geant42006}%
  \BibitemOpen
  \bibfield  {author} {\bibinfo {author} {\bibfnamefont {J.}~\bibnamefont
  {Allison}}, \bibinfo {author} {\bibfnamefont {K.}~\bibnamefont {Amako}},
  \bibinfo {author} {\bibfnamefont {J.}~\bibnamefont {Apostolakis}}, \bibinfo
  {author} {\bibfnamefont {H.}~\bibnamefont {Araujo}}, \bibinfo {author}
  {\bibfnamefont {P.~A.}\ \bibnamefont {Dubois}}, \bibinfo {author}
  {\bibfnamefont {M.}~\bibnamefont {Asai}}, \bibinfo {author} {\bibfnamefont
  {G.}~\bibnamefont {Barrand}}, \bibinfo {author} {\bibfnamefont
  {R.}~\bibnamefont {Capra}}, \bibinfo {author} {\bibfnamefont
  {S.}~\bibnamefont {Chauvie}}, \bibinfo {author} {\bibfnamefont
  {R.}~\bibnamefont {Chytracek}}, \bibinfo {author} {\bibfnamefont {G.~A.~P.}\
  \bibnamefont {Cirrone}}, \bibinfo {author} {\bibfnamefont {G.}~\bibnamefont
  {Cooperman}}, \bibinfo {author} {\bibfnamefont {G.}~\bibnamefont {Cosmo}},
  \bibinfo {author} {\bibfnamefont {G.}~\bibnamefont {Cuttone}}, \bibinfo
  {author} {\bibfnamefont {G.~G.}\ \bibnamefont {Daquino}}, \bibinfo {author}
  {\bibfnamefont {M.}~\bibnamefont {Donszelmann}}, \bibinfo {author}
  {\bibfnamefont {M.}~\bibnamefont {Dressel}}, \bibinfo {author} {\bibfnamefont
  {G.}~\bibnamefont {Folger}}, \bibinfo {author} {\bibfnamefont
  {F.}~\bibnamefont {Foppiano}}, \bibinfo {author} {\bibfnamefont
  {J.}~\bibnamefont {Generowicz}}, \bibinfo {author} {\bibfnamefont
  {V.}~\bibnamefont {Grichine}}, \bibinfo {author} {\bibfnamefont
  {S.}~\bibnamefont {Guatelli}}, \bibinfo {author} {\bibfnamefont
  {P.}~\bibnamefont {Gumplinger}}, \bibinfo {author} {\bibfnamefont
  {A.}~\bibnamefont {Heikkinen}}, \bibinfo {author} {\bibfnamefont
  {I.}~\bibnamefont {Hrivnacova}}, \bibinfo {author} {\bibfnamefont
  {A.}~\bibnamefont {Howard}}, \bibinfo {author} {\bibfnamefont
  {S.}~\bibnamefont {Incerti}}, \bibinfo {author} {\bibfnamefont
  {V.}~\bibnamefont {Ivanchenko}}, \bibinfo {author} {\bibfnamefont
  {T.}~\bibnamefont {Johnson}}, \bibinfo {author} {\bibfnamefont
  {F.}~\bibnamefont {Jones}}, \bibinfo {author} {\bibfnamefont
  {T.}~\bibnamefont {Koi}}, \bibinfo {author} {\bibfnamefont {R.}~\bibnamefont
  {Kokoulin}}, \bibinfo {author} {\bibfnamefont {M.}~\bibnamefont {Kossov}},
  \bibinfo {author} {\bibfnamefont {H.}~\bibnamefont {Kurashige}}, \bibinfo
  {author} {\bibfnamefont {V.}~\bibnamefont {Lara}}, \bibinfo {author}
  {\bibfnamefont {S.}~\bibnamefont {Larsson}}, \bibinfo {author} {\bibfnamefont
  {F.}~\bibnamefont {Lei}}, \bibinfo {author} {\bibfnamefont {O.}~\bibnamefont
  {Link}}, \bibinfo {author} {\bibfnamefont {F.}~\bibnamefont {Longo}},
  \bibinfo {author} {\bibfnamefont {M.}~\bibnamefont {Maire}}, \bibinfo
  {author} {\bibfnamefont {A.}~\bibnamefont {Mantero}}, \bibinfo {author}
  {\bibfnamefont {B.}~\bibnamefont {Mascialino}}, \bibinfo {author}
  {\bibfnamefont {I.}~\bibnamefont {McLaren}}, \bibinfo {author} {\bibfnamefont
  {P.~M.}\ \bibnamefont {Lorenzo}}, \bibinfo {author} {\bibfnamefont
  {K.}~\bibnamefont {Minamimoto}}, \bibinfo {author} {\bibfnamefont
  {K.}~\bibnamefont {Murakami}}, \bibinfo {author} {\bibfnamefont
  {P.}~\bibnamefont {Nieminen}}, \bibinfo {author} {\bibfnamefont
  {L.}~\bibnamefont {Pandola}}, \bibinfo {author} {\bibfnamefont
  {S.}~\bibnamefont {Parlati}}, \bibinfo {author} {\bibfnamefont
  {L.}~\bibnamefont {Peralta}}, \bibinfo {author} {\bibfnamefont
  {J.}~\bibnamefont {Perl}}, \bibinfo {author} {\bibfnamefont {A.}~\bibnamefont
  {Pfeiffer}}, \bibinfo {author} {\bibfnamefont {M.~G.}\ \bibnamefont {Pia}},
  \bibinfo {author} {\bibfnamefont {A.}~\bibnamefont {Ribon}}, \bibinfo
  {author} {\bibfnamefont {P.}~\bibnamefont {Rodrigues}}, \bibinfo {author}
  {\bibfnamefont {G.}~\bibnamefont {Russo}}, \bibinfo {author} {\bibfnamefont
  {S.}~\bibnamefont {Sadilov}}, \bibinfo {author} {\bibfnamefont
  {G.}~\bibnamefont {Santin}}, \bibinfo {author} {\bibfnamefont
  {T.}~\bibnamefont {Sasaki}}, \bibinfo {author} {\bibfnamefont
  {D.}~\bibnamefont {Smith}}, \bibinfo {author} {\bibfnamefont
  {N.}~\bibnamefont {Starkov}}, \bibinfo {author} {\bibfnamefont
  {S.}~\bibnamefont {Tanaka}}, \bibinfo {author} {\bibfnamefont
  {E.}~\bibnamefont {Tcherniaev}}, \bibinfo {author} {\bibfnamefont
  {B.}~\bibnamefont {Tome}}, \bibinfo {author} {\bibfnamefont {A.}~\bibnamefont
  {Trindade}}, \bibinfo {author} {\bibfnamefont {P.}~\bibnamefont {Truscott}},
  \bibinfo {author} {\bibfnamefont {L.}~\bibnamefont {Urban}}, \bibinfo
  {author} {\bibfnamefont {M.}~\bibnamefont {Verderi}}, \bibinfo {author}
  {\bibfnamefont {A.}~\bibnamefont {Walkden}}, \bibinfo {author} {\bibfnamefont
  {J.~P.}\ \bibnamefont {Wellisch}}, \bibinfo {author} {\bibfnamefont {D.~C.}\
  \bibnamefont {Williams}}, \bibinfo {author} {\bibfnamefont {D.}~\bibnamefont
  {Wright}}, \ and\ \bibinfo {author} {\bibfnamefont {H.}~\bibnamefont
  {Yoshida}},\ }\href {\doibase 10.1109/TNS.2006.869826} {\bibfield  {journal}
  {\bibinfo  {journal} {IEEE Transactions on Nuclear Science}\ }\textbf
  {\bibinfo {volume} {53}},\ \bibinfo {pages} {270} (\bibinfo {year}
  {2006})}\BibitemShut {NoStop}%
\bibitem [{\citenamefont {Allison}\ \emph {et~al.}(2016)\citenamefont
  {Allison}, \citenamefont {Amako}, \citenamefont {Apostolakis}, \citenamefont
  {Arce}, \citenamefont {Asai}, \citenamefont {Aso}, \citenamefont {Bagli},
  \citenamefont {Bagulya}, \citenamefont {Banerjee}, \citenamefont {Barrand},
  \citenamefont {Beck}, \citenamefont {Bogdanov}, \citenamefont {Brandt},
  \citenamefont {Brown}, \citenamefont {Burkhardt}, \citenamefont {Canal},
  \citenamefont {Cano-Ott}, \citenamefont {Chauvie}, \citenamefont {Cho},
  \citenamefont {Cirrone}, \citenamefont {Cooperman}, \citenamefont
  {Cortés-Giraldo}, \citenamefont {Cosmo}, \citenamefont {Cuttone},
  \citenamefont {Depaola}, \citenamefont {Desorgher}, \citenamefont {Dong},
  \citenamefont {Dotti}, \citenamefont {Elvira}, \citenamefont {Folger},
  \citenamefont {Francis}, \citenamefont {Galoyan}, \citenamefont {Garnier},
  \citenamefont {Gayer}, \citenamefont {Genser}, \citenamefont {Grichine},
  \citenamefont {Guatelli}, \citenamefont {Guèye}, \citenamefont {Gumplinger},
  \citenamefont {Howard}, \citenamefont {Hřivnáčová}, \citenamefont
  {Hwang}, \citenamefont {Incerti}, \citenamefont {Ivanchenko}, \citenamefont
  {Ivanchenko}, \citenamefont {Jones}, \citenamefont {Jun}, \citenamefont
  {Kaitaniemi}, \citenamefont {Karakatsanis}, \citenamefont {Karamitros},
  \citenamefont {Kelsey}, \citenamefont {Kimura}, \citenamefont {Koi},
  \citenamefont {Kurashige}, \citenamefont {Lechner}, \citenamefont {Lee},
  \citenamefont {Longo}, \citenamefont {Maire}, \citenamefont {Mancusi},
  \citenamefont {Mantero}, \citenamefont {Mendoza}, \citenamefont {Morgan},
  \citenamefont {Murakami}, \citenamefont {Nikitina}, \citenamefont {Pandola},
  \citenamefont {Paprocki}, \citenamefont {Perl}, \citenamefont {Petrović},
  \citenamefont {Pia}, \citenamefont {Pokorski}, \citenamefont {Quesada},
  \citenamefont {Raine}, \citenamefont {Reis}, \citenamefont {Ribon},
  \citenamefont {Fira}, \citenamefont {Romano}, \citenamefont {Russo},
  \citenamefont {Santin}, \citenamefont {Sasaki}, \citenamefont {Sawkey},
  \citenamefont {Shin}, \citenamefont {Strakovsky}, \citenamefont {Taborda},
  \citenamefont {Tanaka}, \citenamefont {Tomé}, \citenamefont {Toshito},
  \citenamefont {Tran}, \citenamefont {Truscott}, \citenamefont {Urban},
  \citenamefont {Uzhinsky}, \citenamefont {Verbeke}, \citenamefont {Verderi},
  \citenamefont {Wendt}, \citenamefont {Wenzel}, \citenamefont {Wright},
  \citenamefont {Wright}, \citenamefont {Yamashita}, \citenamefont {Yarba},\
  and\ \citenamefont {Yoshida}}]{Geant42016}%
  \BibitemOpen
  \bibfield  {author} {\bibinfo {author} {\bibfnamefont {J.}~\bibnamefont
  {Allison}}, \bibinfo {author} {\bibfnamefont {K.}~\bibnamefont {Amako}},
  \bibinfo {author} {\bibfnamefont {J.}~\bibnamefont {Apostolakis}}, \bibinfo
  {author} {\bibfnamefont {P.}~\bibnamefont {Arce}}, \bibinfo {author}
  {\bibfnamefont {M.}~\bibnamefont {Asai}}, \bibinfo {author} {\bibfnamefont
  {T.}~\bibnamefont {Aso}}, \bibinfo {author} {\bibfnamefont {E.}~\bibnamefont
  {Bagli}}, \bibinfo {author} {\bibfnamefont {A.}~\bibnamefont {Bagulya}},
  \bibinfo {author} {\bibfnamefont {S.}~\bibnamefont {Banerjee}}, \bibinfo
  {author} {\bibfnamefont {G.}~\bibnamefont {Barrand}}, \bibinfo {author}
  {\bibfnamefont {B.}~\bibnamefont {Beck}}, \bibinfo {author} {\bibfnamefont
  {A.}~\bibnamefont {Bogdanov}}, \bibinfo {author} {\bibfnamefont
  {D.}~\bibnamefont {Brandt}}, \bibinfo {author} {\bibfnamefont
  {J.}~\bibnamefont {Brown}}, \bibinfo {author} {\bibfnamefont
  {H.}~\bibnamefont {Burkhardt}}, \bibinfo {author} {\bibfnamefont
  {P.}~\bibnamefont {Canal}}, \bibinfo {author} {\bibfnamefont
  {D.}~\bibnamefont {Cano-Ott}}, \bibinfo {author} {\bibfnamefont
  {S.}~\bibnamefont {Chauvie}}, \bibinfo {author} {\bibfnamefont
  {K.}~\bibnamefont {Cho}}, \bibinfo {author} {\bibfnamefont {G.}~\bibnamefont
  {Cirrone}}, \bibinfo {author} {\bibfnamefont {G.}~\bibnamefont {Cooperman}},
  \bibinfo {author} {\bibfnamefont {M.}~\bibnamefont {Cortés-Giraldo}},
  \bibinfo {author} {\bibfnamefont {G.}~\bibnamefont {Cosmo}}, \bibinfo
  {author} {\bibfnamefont {G.}~\bibnamefont {Cuttone}}, \bibinfo {author}
  {\bibfnamefont {G.}~\bibnamefont {Depaola}}, \bibinfo {author} {\bibfnamefont
  {L.}~\bibnamefont {Desorgher}}, \bibinfo {author} {\bibfnamefont
  {X.}~\bibnamefont {Dong}}, \bibinfo {author} {\bibfnamefont {A.}~\bibnamefont
  {Dotti}}, \bibinfo {author} {\bibfnamefont {V.}~\bibnamefont {Elvira}},
  \bibinfo {author} {\bibfnamefont {G.}~\bibnamefont {Folger}}, \bibinfo
  {author} {\bibfnamefont {Z.}~\bibnamefont {Francis}}, \bibinfo {author}
  {\bibfnamefont {A.}~\bibnamefont {Galoyan}}, \bibinfo {author} {\bibfnamefont
  {L.}~\bibnamefont {Garnier}}, \bibinfo {author} {\bibfnamefont
  {M.}~\bibnamefont {Gayer}}, \bibinfo {author} {\bibfnamefont
  {K.}~\bibnamefont {Genser}}, \bibinfo {author} {\bibfnamefont
  {V.}~\bibnamefont {Grichine}}, \bibinfo {author} {\bibfnamefont
  {S.}~\bibnamefont {Guatelli}}, \bibinfo {author} {\bibfnamefont
  {P.}~\bibnamefont {Guèye}}, \bibinfo {author} {\bibfnamefont
  {P.}~\bibnamefont {Gumplinger}}, \bibinfo {author} {\bibfnamefont
  {A.}~\bibnamefont {Howard}}, \bibinfo {author} {\bibfnamefont
  {I.}~\bibnamefont {Hřivnáčová}}, \bibinfo {author} {\bibfnamefont
  {S.}~\bibnamefont {Hwang}}, \bibinfo {author} {\bibfnamefont
  {S.}~\bibnamefont {Incerti}}, \bibinfo {author} {\bibfnamefont
  {A.}~\bibnamefont {Ivanchenko}}, \bibinfo {author} {\bibfnamefont
  {V.}~\bibnamefont {Ivanchenko}}, \bibinfo {author} {\bibfnamefont
  {F.}~\bibnamefont {Jones}}, \bibinfo {author} {\bibfnamefont
  {S.}~\bibnamefont {Jun}}, \bibinfo {author} {\bibfnamefont {P.}~\bibnamefont
  {Kaitaniemi}}, \bibinfo {author} {\bibfnamefont {N.}~\bibnamefont
  {Karakatsanis}}, \bibinfo {author} {\bibfnamefont {M.}~\bibnamefont
  {Karamitros}}, \bibinfo {author} {\bibfnamefont {M.}~\bibnamefont {Kelsey}},
  \bibinfo {author} {\bibfnamefont {A.}~\bibnamefont {Kimura}}, \bibinfo
  {author} {\bibfnamefont {T.}~\bibnamefont {Koi}}, \bibinfo {author}
  {\bibfnamefont {H.}~\bibnamefont {Kurashige}}, \bibinfo {author}
  {\bibfnamefont {A.}~\bibnamefont {Lechner}}, \bibinfo {author} {\bibfnamefont
  {S.}~\bibnamefont {Lee}}, \bibinfo {author} {\bibfnamefont {F.}~\bibnamefont
  {Longo}}, \bibinfo {author} {\bibfnamefont {M.}~\bibnamefont {Maire}},
  \bibinfo {author} {\bibfnamefont {D.}~\bibnamefont {Mancusi}}, \bibinfo
  {author} {\bibfnamefont {A.}~\bibnamefont {Mantero}}, \bibinfo {author}
  {\bibfnamefont {E.}~\bibnamefont {Mendoza}}, \bibinfo {author} {\bibfnamefont
  {B.}~\bibnamefont {Morgan}}, \bibinfo {author} {\bibfnamefont
  {K.}~\bibnamefont {Murakami}}, \bibinfo {author} {\bibfnamefont
  {T.}~\bibnamefont {Nikitina}}, \bibinfo {author} {\bibfnamefont
  {L.}~\bibnamefont {Pandola}}, \bibinfo {author} {\bibfnamefont
  {P.}~\bibnamefont {Paprocki}}, \bibinfo {author} {\bibfnamefont
  {J.}~\bibnamefont {Perl}}, \bibinfo {author} {\bibfnamefont {I.}~\bibnamefont
  {Petrović}}, \bibinfo {author} {\bibfnamefont {M.}~\bibnamefont {Pia}},
  \bibinfo {author} {\bibfnamefont {W.}~\bibnamefont {Pokorski}}, \bibinfo
  {author} {\bibfnamefont {J.}~\bibnamefont {Quesada}}, \bibinfo {author}
  {\bibfnamefont {M.}~\bibnamefont {Raine}}, \bibinfo {author} {\bibfnamefont
  {M.}~\bibnamefont {Reis}}, \bibinfo {author} {\bibfnamefont {A.}~\bibnamefont
  {Ribon}}, \bibinfo {author} {\bibfnamefont {A.~R.}\ \bibnamefont {Fira}},
  \bibinfo {author} {\bibfnamefont {F.}~\bibnamefont {Romano}}, \bibinfo
  {author} {\bibfnamefont {G.}~\bibnamefont {Russo}}, \bibinfo {author}
  {\bibfnamefont {G.}~\bibnamefont {Santin}}, \bibinfo {author} {\bibfnamefont
  {T.}~\bibnamefont {Sasaki}}, \bibinfo {author} {\bibfnamefont
  {D.}~\bibnamefont {Sawkey}}, \bibinfo {author} {\bibfnamefont
  {J.}~\bibnamefont {Shin}}, \bibinfo {author} {\bibfnamefont {I.}~\bibnamefont
  {Strakovsky}}, \bibinfo {author} {\bibfnamefont {A.}~\bibnamefont {Taborda}},
  \bibinfo {author} {\bibfnamefont {S.}~\bibnamefont {Tanaka}}, \bibinfo
  {author} {\bibfnamefont {B.}~\bibnamefont {Tomé}}, \bibinfo {author}
  {\bibfnamefont {T.}~\bibnamefont {Toshito}}, \bibinfo {author} {\bibfnamefont
  {H.}~\bibnamefont {Tran}}, \bibinfo {author} {\bibfnamefont {P.}~\bibnamefont
  {Truscott}}, \bibinfo {author} {\bibfnamefont {L.}~\bibnamefont {Urban}},
  \bibinfo {author} {\bibfnamefont {V.}~\bibnamefont {Uzhinsky}}, \bibinfo
  {author} {\bibfnamefont {J.}~\bibnamefont {Verbeke}}, \bibinfo {author}
  {\bibfnamefont {M.}~\bibnamefont {Verderi}}, \bibinfo {author} {\bibfnamefont
  {B.}~\bibnamefont {Wendt}}, \bibinfo {author} {\bibfnamefont
  {H.}~\bibnamefont {Wenzel}}, \bibinfo {author} {\bibfnamefont
  {D.}~\bibnamefont {Wright}}, \bibinfo {author} {\bibfnamefont
  {D.}~\bibnamefont {Wright}}, \bibinfo {author} {\bibfnamefont
  {T.}~\bibnamefont {Yamashita}}, \bibinfo {author} {\bibfnamefont
  {J.}~\bibnamefont {Yarba}}, \ and\ \bibinfo {author} {\bibfnamefont
  {H.}~\bibnamefont {Yoshida}},\ }\href {\doibase
  https://doi.org/10.1016/j.nima.2016.06.125} {\bibfield  {journal} {\bibinfo
  {journal} {Nuclear Instruments and Methods in Physics Research Section A:
  Accelerators, Spectrometers, Detectors and Associated Equipment}\ }\textbf
  {\bibinfo {volume} {835}},\ \bibinfo {pages} {186 } (\bibinfo {year}
  {2016})}\BibitemShut {NoStop}%
\bibitem [{\citenamefont {Mendenhall}\ and\ \citenamefont
  {Weller}(2005)}]{MENDENHALL2005420}%
  \BibitemOpen
  \bibfield  {author} {\bibinfo {author} {\bibfnamefont {M.~H.}\ \bibnamefont
  {Mendenhall}}\ and\ \bibinfo {author} {\bibfnamefont {R.~A.}\ \bibnamefont
  {Weller}},\ }\href {\doibase https://doi.org/10.1016/j.nimb.2004.08.014}
  {\bibfield  {journal} {\bibinfo  {journal} {Nuclear Instruments and Methods
  in Physics Research Section B: Beam Interactions with Materials and Atoms}\
  }\textbf {\bibinfo {volume} {227}},\ \bibinfo {pages} {420 } (\bibinfo {year}
  {2005})}\BibitemShut {NoStop}%
\bibitem [{\citenamefont {Nordlund}, \citenamefont {Wallenius},\ and\
  \citenamefont {Malerba}(2006)}]{NORDLUND2006322}%
  \BibitemOpen
  \bibfield  {author} {\bibinfo {author} {\bibfnamefont {K.}~\bibnamefont
  {Nordlund}}, \bibinfo {author} {\bibfnamefont {J.}~\bibnamefont {Wallenius}},
  \ and\ \bibinfo {author} {\bibfnamefont {L.}~\bibnamefont {Malerba}},\ }\href
  {\doibase https://doi.org/10.1016/j.nimb.2006.01.003} {\bibfield  {journal}
  {\bibinfo  {journal} {Nuclear Instruments and Methods in Physics Research
  Section B: Beam Interactions with Materials and Atoms}\ }\textbf {\bibinfo
  {volume} {246}},\ \bibinfo {pages} {322 } (\bibinfo {year}
  {2006})}\BibitemShut {NoStop}%
\bibitem [{\citenamefont {Kadribasic}\ \emph {et~al.}(2018)\citenamefont
  {Kadribasic}, \citenamefont {Mirabolfathi}, \citenamefont {Nordlund},
  \citenamefont {Sand}, \citenamefont {Holmstr\"om},\ and\ \citenamefont
  {Djurabekova}}]{2017arXiv170305371K}%
  \BibitemOpen
  \bibfield  {author} {\bibinfo {author} {\bibfnamefont {F.}~\bibnamefont
  {Kadribasic}}, \bibinfo {author} {\bibfnamefont {N.}~\bibnamefont
  {Mirabolfathi}}, \bibinfo {author} {\bibfnamefont {K.}~\bibnamefont
  {Nordlund}}, \bibinfo {author} {\bibfnamefont {A.~E.}\ \bibnamefont {Sand}},
  \bibinfo {author} {\bibfnamefont {E.}~\bibnamefont {Holmstr\"om}}, \ and\
  \bibinfo {author} {\bibfnamefont {F.}~\bibnamefont {Djurabekova}},\ }\href
  {\doibase 10.1103/PhysRevLett.120.111301} {\bibfield  {journal} {\bibinfo
  {journal} {Phys. Rev. Lett.}\ }\textbf {\bibinfo {volume} {120}},\ \bibinfo
  {pages} {111301} (\bibinfo {year} {2018})}\BibitemShut {NoStop}%
\bibitem [{\citenamefont {Lazanu}\ and\ \citenamefont {Lazanu}(2010)}]{2903}%
  \BibitemOpen
  \bibfield  {author} {\bibinfo {author} {\bibfnamefont {I.}~\bibnamefont
  {Lazanu}}\ and\ \bibinfo {author} {\bibfnamefont {S.}~\bibnamefont
  {Lazanu}},\ }\href@noop {} {\bibfield  {journal} {\bibinfo  {journal}
  {Romanian Reports in Physics}\ }\textbf {\bibinfo {volume} {62}},\ \bibinfo
  {pages} {309} (\bibinfo {year} {2010})}\BibitemShut {NoStop}%
\bibitem [{\citenamefont {{Kinchin}}\ and\ \citenamefont
  {{Pease}}(1955)}]{1955RPPh...18....1K}%
  \BibitemOpen
  \bibfield  {author} {\bibinfo {author} {\bibfnamefont {G.~H.}\ \bibnamefont
  {{Kinchin}}}\ and\ \bibinfo {author} {\bibfnamefont {R.~S.}\ \bibnamefont
  {{Pease}}},\ }\href {\doibase 10.1088/0034-4885/18/1/301} {\bibfield
  {journal} {\bibinfo  {journal} {Reports on Progress in Physics}\ }\textbf
  {\bibinfo {volume} {18}},\ \bibinfo {pages} {1} (\bibinfo {year}
  {1955})}\BibitemShut {NoStop}%
\bibitem [{\citenamefont {Norgett}, \citenamefont {Robinson},\ and\
  \citenamefont {Torrens}(1975)}]{NORGETT197550}%
  \BibitemOpen
  \bibfield  {author} {\bibinfo {author} {\bibfnamefont {M.}~\bibnamefont
  {Norgett}}, \bibinfo {author} {\bibfnamefont {M.}~\bibnamefont {Robinson}}, \
  and\ \bibinfo {author} {\bibfnamefont {I.}~\bibnamefont {Torrens}},\ }\href
  {\doibase https://doi.org/10.1016/0029-5493(75)90035-7} {\bibfield  {journal}
  {\bibinfo  {journal} {Nuclear Engineering and Design}\ }\textbf {\bibinfo
  {volume} {33}},\ \bibinfo {pages} {50 } (\bibinfo {year} {1975})}\BibitemShut
  {NoStop}%
\bibitem [{\citenamefont {{Ziegler}}, \citenamefont {{Ziegler}},\ and\
  \citenamefont {{Biersack}}(2010)}]{SRIM}%
  \BibitemOpen
  \bibfield  {author} {\bibinfo {author} {\bibfnamefont {J.~F.}\ \bibnamefont
  {{Ziegler}}}, \bibinfo {author} {\bibfnamefont {M.~D.}\ \bibnamefont
  {{Ziegler}}}, \ and\ \bibinfo {author} {\bibfnamefont {J.~P.}\ \bibnamefont
  {{Biersack}}},\ }\href {\doibase 10.1016/j.nimb.2010.02.091} {\bibfield
  {journal} {\bibinfo  {journal} {Nuclear Instruments and Methods in Physics
  Research B}\ }\textbf {\bibinfo {volume} {268}},\ \bibinfo {pages} {1818}
  (\bibinfo {year} {2010})}\BibitemShut {NoStop}%
\bibitem [{\citenamefont {Bodlaki}\ \emph {et~al.}(2003)\citenamefont
  {Bodlaki}, \citenamefont {Yamamoto}, \citenamefont {Waldeck},\ and\
  \citenamefont {Borguet}}]{GeExp1}%
  \BibitemOpen
  \bibfield  {author} {\bibinfo {author} {\bibfnamefont {D.}~\bibnamefont
  {Bodlaki}}, \bibinfo {author} {\bibfnamefont {H.}~\bibnamefont {Yamamoto}},
  \bibinfo {author} {\bibfnamefont {D.}~\bibnamefont {Waldeck}}, \ and\
  \bibinfo {author} {\bibfnamefont {E.}~\bibnamefont {Borguet}},\ }\href
  {\doibase https://doi.org/10.1016/S0039-6028(03)00958-0} {\bibfield
  {journal} {\bibinfo  {journal} {Surface Science}\ }\textbf {\bibinfo {volume}
  {543}},\ \bibinfo {pages} {63 } (\bibinfo {year} {2003})}\BibitemShut
  {NoStop}%
\bibitem [{\citenamefont {Sahari}\ \emph {et~al.}(2011)\citenamefont {Sahari},
  \citenamefont {Murakami}, \citenamefont {Fujioka}, \citenamefont {Bando},
  \citenamefont {Ohta}, \citenamefont {Makihara}, \citenamefont {Higashi},\
  and\ \citenamefont {Miyazaki}}]{GeExp2}%
  \BibitemOpen
  \bibfield  {author} {\bibinfo {author} {\bibfnamefont {S.~K.}\ \bibnamefont
  {Sahari}}, \bibinfo {author} {\bibfnamefont {H.}~\bibnamefont {Murakami}},
  \bibinfo {author} {\bibfnamefont {T.}~\bibnamefont {Fujioka}}, \bibinfo
  {author} {\bibfnamefont {T.}~\bibnamefont {Bando}}, \bibinfo {author}
  {\bibfnamefont {A.}~\bibnamefont {Ohta}}, \bibinfo {author} {\bibfnamefont
  {K.}~\bibnamefont {Makihara}}, \bibinfo {author} {\bibfnamefont
  {S.}~\bibnamefont {Higashi}}, \ and\ \bibinfo {author} {\bibfnamefont
  {S.}~\bibnamefont {Miyazaki}},\ }\href
  {http://stacks.iop.org/1347-4065/50/i=4S/a=04DA12} {\bibfield  {journal}
  {\bibinfo  {journal} {Japanese Journal of Applied Physics}\ }\textbf
  {\bibinfo {volume} {50}},\ \bibinfo {pages} {04DA12} (\bibinfo {year}
  {2011})}\BibitemShut {NoStop}%
\bibitem [{\citenamefont {Ehrhart}\ and\ \citenamefont
  {Zillgen}(1999)}]{doi:10.1063/1.369709}%
  \BibitemOpen
  \bibfield  {author} {\bibinfo {author} {\bibfnamefont {P.}~\bibnamefont
  {Ehrhart}}\ and\ \bibinfo {author} {\bibfnamefont {H.}~\bibnamefont
  {Zillgen}},\ }\href {\doibase 10.1063/1.369709} {\bibfield  {journal}
  {\bibinfo  {journal} {Journal of Applied Physics}\ }\textbf {\bibinfo
  {volume} {85}},\ \bibinfo {pages} {3503} (\bibinfo {year} {1999})},\ \Eprint
  {http://arxiv.org/abs/https://doi.org/10.1063/1.369709}
  {https://doi.org/10.1063/1.369709} \BibitemShut {NoStop}%
\bibitem [{\citenamefont {Emtsev}, \citenamefont {Mashovets},\ and\
  \citenamefont {Mikhnovich}(1992{\natexlab{b}})}]{emtsev1992}%
  \BibitemOpen
  \bibfield  {author} {\bibinfo {author} {\bibfnamefont {V.}~\bibnamefont
  {Emtsev}}, \bibinfo {author} {\bibfnamefont {T.}~\bibnamefont {Mashovets}}, \
  and\ \bibinfo {author} {\bibfnamefont {V.}~\bibnamefont {Mikhnovich}},\
  }\href {https://inis.iaea.org/search/search.aspx?orig_q=RN:24060482}
  {\bibfield  {journal} {\bibinfo  {journal} {Soviet Physics - Semiconductors}\
  }\textbf {\bibinfo {volume} {26}},\ \bibinfo {pages} {12 } (\bibinfo {year}
  {1992}{\natexlab{b}})}\BibitemShut {NoStop}%
\bibitem [{\citenamefont {Stillinger}\ and\ \citenamefont
  {Weber}(1985)}]{SW1985}%
  \BibitemOpen
  \bibfield  {author} {\bibinfo {author} {\bibfnamefont {F.~H.}\ \bibnamefont
  {Stillinger}}\ and\ \bibinfo {author} {\bibfnamefont {T.~A.}\ \bibnamefont
  {Weber}},\ }\href {\doibase 10.1103/PhysRevB.31.5262} {\bibfield  {journal}
  {\bibinfo  {journal} {Phys. Rev. B}\ }\textbf {\bibinfo {volume} {31}},\
  \bibinfo {pages} {5262} (\bibinfo {year} {1985})}\BibitemShut {NoStop}%
\end{thebibliography}%
	
\end{document}